\documentclass[a4paper,11pt]{article}
\usepackage{jheppub} 
\usepackage[T1]{fontenc} 
\usepackage{amsmath,amssymb,graphicx,bbm,mathrsfs,nicefrac}
\usepackage{braket} % bra and ket notation
\usepackage{wasysym} % permil symbol
\usepackage{slashed} % feynman slash notation
\DeclareMathOperator{\Tr}{Tr}
\def\nlo{\mathrm{NLO}}
\def\nlops{\mathrm{NLO+PS}}
\def\helacnlo{\textsc{Helac-Nlo}}
\def\dipoles{\textsc{Helac-Dipoles}}
\def\OneLoop{\textsc{Helac-1Loop}}
\def\sherpa{\textsc{Sherpa}}
\def\amcnlo{a\textsc{Mc@Nlo}}
\def\powhegbox{\textsc{Powheg-Box}}
\def\mcnlo{\textsc{Mc@Nlo}}
\def\powheg{\textsc{Powheg}}
\def\deductor{\textsc{Deductor}}
\def\pyQ{\textsc{Pythia6Q}}
\def\pythia{\textsc{Pythia8}}
\def\herwig{\textsc{Herwig}}
\def\amcnlopyQ{\amcnlo+\pyQ}
\def\amcnlopy{\amcnlo+\pythia}
\def\powhegpy{\powheg+\pythia}
\def\heldeductor{\helacnlo+\deductor}
\title{Matching the Nagy-Soper parton shower at next-to-leading order}

\author[]{M.~Czakon, H.~B.~Hartanto, M.~Kraus, M.~Worek}
\affiliation[]{Institut f\"{u}r Theoretische Teilchenphysik und Kosmologie,
 RWTH Aachen University, D-52056}
\emailAdd{mczakon@physik.rwth-aachen.de}
\emailAdd{hartanto@physik.rwth-aachen.de}
\emailAdd{kraus@physik.rwth-aachen.de}
\emailAdd{worek@physik.rwth-aachen.de}
 
\abstract{We present an \mcnlo{}-like matching of
  next-to-leading order  QCD calculations with the Nagy-Soper parton
  shower.  An implementation of the algorithm within the
  \textsc{Helac-Dipoles} Monte Carlo  generator is used to address 
   the uncertainties
  and ambiguities of the matching scheme.  First results obtained using the
  Nagy-Soper parton shower implementation in \deductor{} in
  conjunction with the \textsc{Helac-Nlo} framework are given for the
  $pp\to t\bar{t}j+X$  process at the LHC with $\sqrt{s}=8$ TeV.
  Effects of resummation are discussed for various observables.  }

\dedicated{\rm TTK-15-07}

\begin{document}
\maketitle

\section{Introduction}

The simulation of scattering events according to field theoretical
models is at the core of particle physics, as it allows for both
discoveries of particles and the validation of theories. The source of
practical difficulties lies primarily in the presence of strong
interactions. Indeed, if the initial and/or final states involve
hadrons, then non-perturbative effects will have to be included at
some point. It is only due to factorization that it is possible to
obtain an accurate description using perturbative methods. For some
observables, the evolution of the state of the system between the high
scale, where fixed order perturbation theory is adequate, and the
non-perturbative scale, where it is not, requires resummation of
logarithmic enhancements due to soft and collinear singularities. This
can be achieved either by analytic resummation methods, or by means of
parton showers. The latter are the versatile tool of choice in the
communication between theory and experiment.

Parton showers have a long history~\cite{Marchesini:1983bm,Sjostrand:1985xi,
Bengtsson:1986gz,Bengtsson:1986et,Marchesini:1987cf,Lonnblad:1992tz}. 
They are based on a semi-classical picture of partons splitting into pairs 
of other partons. This picture corresponds to the collinear limit of QCD
amplitudes in next-to-leading order (NLO) configurations. Thus, the leading
singularity of collinear-soft origin may be correctly reproduced. In
consequence, the approach has leading logarithmic accuracy in the most
general case of many partons in the high scale process. Including pure
soft effects is necessary to correctly reproduce the first subleading
logarithms. Recent years have seen quite some activity in this
direction~\cite{Dinsdale:2007mf,Schumann:2007mg,Platzer:2009jq,
GehrmannDeRidder:2011dm,Platzer:2012np,Ritzmann:2012ca,Hartgring:2013jma}. 
In particular, it has been possible to include
leading colour soft effects using parton showers constructed around
known subtraction schemes, such as the one of Catani and 
Seymour~\cite{Catani:1996vz,Catani:2002hc}. In a series of 
papers~\cite{Nagy:2007ty,Nagy:2008ns,Nagy:2008eq,Nagy:2009vg,Nagy:2009re,
Nagy:2012bt,Nagy:2014nqa,Nagy:2014oqa,Nagy:2014mqa,Nagy:2015hwa}, 
Nagy and Soper have proposed a
different concept for a parton shower. Their construction should be
able to include soft effects at subleading colour. Recently, they have
also provided a public code~\cite{Nagy:2014mqa}, \deductor{}, that implements a
large part of their idea, albeit with the exlusion of exact colour
treatment, where an extension of the leading colour approximation is
used, and spin correlations.

The purpose of the present publication is to specify a matching
procedure between the Nagy-Soper parton shower and a fixed order
calculation at NLO. This can be achieved using at least two different
ideas: \mcnlo{} \cite{Frixione:2002ik,Frixione:2003ei}  and \powheg{}
\cite{Nason:2004rx,Frixione:2007vw,Alioli:2010xd}.  The first method,
which is the one we have chosen, consists in removing double counting
contributions by expanding the parton shower to first order in the
strong coupling and compensating for the terms, which are already
present at fixed order. This is closely related to the possibility of
obtaining a subtraction scheme from the parton shower, which was
exploited in
Refs.~\cite{Chung:2010fx,Chung:2012rq,Bevilacqua:2013iha,Robens:2013wga}
for the case of the Nagy-Soper construction. In particular,  the
implementation within the \dipoles{}  package~\cite{Czakon:2009ss}
has proven invaluable as the basis for the present work.

Having a working tool to produce parton shower matched event samples
with inclusive next-to-leading order accuracy is necessary in order to
assess the differences of the new shower concept and more established
alternatives. This is a second goal of our publication. Clearly, there
will be differences just as there are even in analytic resummation,
where a much higher logarithmic accuracy can be
achieved. Nevertheless, studying particular examples allows to
quantify the deviations and convince oneself that they do not point to
real errors either within the concept or the implementation. For this
first study we have chosen a process, which involves non-trivial colour
exchange and massive partons, and requires cuts in order to even
define the Born approximation cross section: top-quark pair production
in association with at least one jet in hadron collisions at the
Large Hadron Collider (LHC). 
There is one drawback of having top-quarks in the final state. A
realistic description would require the inclusion of decays. We did
not want to consider off-shell effects and the narrow width
approximation was also out of the question for technical reasons. In
such cases, one relies on the shower program to decay the quarks. This
is a crude approximation, which neglects spin
correlations. Unfortunately, \deductor{} does not even provide
that. Thus, we are simulating a process with stable top-quarks in the
final state. On the other hand, in our comparisons we have only taken
the perturbative evolution into account. In other words, we have
switched off the hadronization and multiple-interaction models available in
common Monte Carlo generators (though not in \deductor{}). This allows
us to quantify differences at the same level of approximation.

We should point out that our matching implementation is restricted to
leading colour. \deductor{} in the version we have used contains a more
advanced colour approximation. Nevertheless, as long as full colour
functionality is not available, we have decided to simplify our
work. On the other hand, working at leading colour is appropriate for
comparisons with other systems, which are also only certified to have
this accuracy.

Ours is the first phenomenological study performed using the
Nagy-Soper shower, which does not involve the authors of the
concept. We believe that this independence is important and actually
proves that \deductor{} is ready to use by outsiders. It also gave us the
motivation to prepare a summary of the main components of the
Nagy-Soper parton shower including a list of ambiguities, which
correspond to the places, where modifications might be expected once
more experience in practical applications is gathered. We hope that
this part of our paper will be useful to those, who do not necessarily
want to read the hundreds of pages of the original publications to
gain a basic understanding of the concept.

The present text is organised as follows. We start by a summary of the
Nagy-Soper parton shower concept. We are explicit as far as those
parts are concerned, which have an impact on the construction of the
matching procedure. We also point out some special features like
the modified parton distribution function (PDF) 
evolution for example. The next Section deals with the
matching itself. We consider both the case of processes, which have
well-defined Born approximation total cross sections, and the case of
processes, which require the specification of cuts already at this
level. Subsequently, we describe the details of our implementation
within \dipoles{} and the interface to \deductor{}. Finally, we show
some results for $t\bar{t}j$ production at the LHC and compare
them to results obtained with other systems. We close the main text
with some short conclusions. An appendix contains a more thorough
discussion of the matching in the leading colour approximation.

%------------------------X------------------------X------------------------X---%
\section{A parton shower with quantum interference}
\label{sec:shower}
%------------------------X------------------------X------------------------X---%
In this Section we review the Nagy-Soper parton shower concept, which allows
for parton state evolution to include both spin and colour
correlations.  Neverthless, the main focus of our discussion will lie
on the colour evolution. In fact, the exponentiation of non-diagonal
colour matrices is a long-standing problem. Notice that an independent
attempt to include full colour evolution in a parton shower was
already made in Ref.~\cite{Platzer:2012np}.

%------------------------X------------------------X------------------------X---%
\subsection{Quantum configuration space}
\label{subsec:confspace}
%------------------------X------------------------X------------------------X---%
A generic $2 \to  m$ process is defined by two initial state partons
$a$ and $b$\footnote{We treat the case of incoming hadrons. Colour
  neutral particles imply the usual simplifications, i.e. lack of
  parton distribution functions and no initial state evolution.}
and $1,...,m$ final state particles. Each particle is described by a set of 
quantum numbers to define the flavour $f_i$, spin $s_i$ and colour $c_i$ of the 
particle and its momentum $p_i$. Initial state partons are described by 
momentum fractions $\eta_a$ and $\eta_b$. Thus, a 
complete $m$-parton ensemble, i.e. the probability distribution over possible 
quantum states, can be parametrized by the following set of quantum 
numbers\footnote{The minus sign for the initial state flavour is only 
a convention, since all partons are considered as outgoing from 
the hard interaction.}
\begin{equation}
 \{p,f,s,c\}_m \equiv \{ [\eta_a,-f_a, s_a,c_a], [\eta_b,-f_b, s_b,c_b],
 [p_1,f_1, s_1,c_1],...,[p_m,f_m, s_m,c_m] \}\;.
\end{equation}
Special emphasis should be given to initial state partons. In the
Nagy-Soper shower,
charm and bottom quarks are allowed to be massive. In the presence of parton 
masses, the initial state momenta are parametrized by the hadron momenta $p_A$ 
and $p_B$, and the momentum fractions $\eta_a$ and $\eta_b$
\begin{align}
&p_a = \eta_a p_A + \frac{m^2_{f_a}}{\eta_a s}p_B\;, \\
&p_b = \eta_b p_B + \frac{m^2_{f_b}}{\eta_b s}p_A\;,
\end{align}
where $p_A^2 = p_B^2 =0$\footnote{Even though parton masses are kept 
non-vanishing, one assumes that the proton mass is negligible with respect to 
the hadronic center-of-mass energy.} and $s=2(p_A\cdot p_B)$. In the following, 
we will describe the treatment of spin and colour in the parton shower. 
A matrix element $\mathcal{M}$ can be viewed as a vector 
$\ket{\mathcal{M}(\{p,f\}_m)}$ in colour $\otimes$ spin space and can be 
resolved into basis vectors $\ket{\{s\}_m}$ and $\ket{\{c\}_m}$ with complex 
expansion coefficients $\mathcal{M}(\{p,f,s,c\}_m)$
\begin{equation}
  \begin{split}
     \ket{\mathcal{M}(\{p,f\}_m)} &= \sum_{\{s\}_m} \sum_{\{c\}_m} 
     \mathcal{M}(\{p,f,s,c\}_m) \ket{\{s\}_m} \otimes \ket{\{c\}_m} \\
	&= \sum_{\{s,c\}_m}  \mathcal{M}(\{p,f,s,c\}_m) \ket{\{s,c\}_m}\;.
  \end{split}
  \label{expandbasis}
\end{equation}
The basis in the spin space is orthonormal. Here, $\{s\}_m = \{ s_a,s_b, s_1,
...,s_m\}$ represents the physical helicities of all  particles.

The colour basis is defined in terms of colour string configurations
\cite{Paton:1969je,Berends:1987cv,Berends:1987me,
Mangano:1987xk,Mangano:1988kk}, which allows 
for a straightforward connection with a hadronization model based on colour 
strings like the one described in Ref.~\cite{Andersson:1983ia}. In practice, a 
colour state $\ket{\{c\}_m}$ is represented by several strings $\{ 
\mathcal{S}_1,...,\mathcal{S}_n \}$, where $\mathcal{S}_i$ can be an open or
a closed string. An open string is denoted by $\mathcal{S} = [l_1,l_2,...,
l_{n-1},l_n]$, where $l_1$ is the label of a quark, $l_n$ of an anti-quark and 
$l_2,...,l_{n-1}$ are gluon labels. Furthermore, a closed string is given by 
$\mathcal{S} = (l_1,...,l_n)$, where $l_i$ labels only gluons this time. 
For brevity we will write $l_i=i$ in the following. The label $l$ of a parton 
refers to its colour index $a_l$. In case $l$ labels a gluon, the colour index 
$a_l$ has values $1,...,8$. In case of quarks or anti-quarks $a_l$ only takes 
the values $1,2,3$. The set of colour indices $\{a_1,a_2,\cdots,a_n\}$ is
abbreviated in the following with $\{a\}$ and allows to define the colour
basis structures.

The colour structure of an open string can be defined by
\begin{equation}
  \Psi^{\{a\}}(\mathcal{S}=[1,2,...,n-1,n]) = \frac{1}{\sqrt{N_c C_F^{n-2}}}
  [T^{a_2}T^{a_3}\cdots T^{a_{n-1}}]_{a_1a_n}\;,
  \label{openstring}
\end{equation}
where $T^a$ is an SU($N_c$) generator. Open strings are normalized to
\begin{equation}
  \braket{\mathcal{S}|\mathcal{S}} \equiv \sum_{\{a\}} 
  |\Psi^{\{a\}}(\mathcal{S})|^2 = 1\;.
\end{equation}
For closed strings we define
\begin{equation}
  \Psi^{\{a\}}(\mathcal{S}=(1,...,n)) = \frac{1}{\sqrt{C_F^n}} \text{Tr}
  [T^{a_1}T^{a_2}\cdots T^{a_n}]\;,
  \label{closedstring}
\end{equation}
with normalization
\begin{equation}
  \braket{\mathcal{S}|\mathcal{S}} \equiv \sum_{\{a\}} 
  |\Psi^{\{a\}}(\mathcal{S})|^2 = 1 - \left(\frac{-1}{2N_cC_F}\right)^{n-1}\;.
\end{equation}
A colour state $\ket{\{c\}_m}$ is represented by a product of these
colour structures. Thus a state is normalized to
\begin{equation}
\braket{\{c\}_m|\{c\}_m} = \prod_k \braket{\mathcal{S}_k|\mathcal{S}_k}\;.
\end{equation}
With these definitions the basis is not orthonormal in general. For instance
\begin{equation}
  \braket{\{c^\prime\}_m|\{c\}_m} = \delta\left(\{c^\prime\}_m ; \{c\}_m
  \right)  + \mathcal{O}(1/N_c^2)\;.
\end{equation}
Therefore, the basis is only orthonormal in the leading colour approximation,
 $N_c \to \infty$.

%------------------------X------------------------X------------------------X---%
\subsection{Quantum density matrix}
%------------------------X------------------------X------------------------X---%
The basic object describing the parton shower evolution is the quantum density
matrix $\rho$, which gives the ``probability''\footnote{Since $\rho$ contains
the full colour information, it can become negative for subleading colour 
configurations. Thus, strictly speaking, one cannot interpret $\rho$ as a 
probability distribution. Nevertheless, the standard concepts from statistical 
mechanics apply here. } to find a certain parton ensemble $\{p,f\}_m$.
 In this Section we give the relation of the 
quantum density to matrix elements. Using the notation of the 
previous Section, one can write the expectation value for a completely inclusive
observable $F$, as
\begin{equation}
\begin{split}
  \sigma[F] &= \sum_m \frac{1}{m!}\int [d\{p,f\}_m] \bra{\mathcal{M}(
  \{p,f\}_m)}F(\{p,f\}_m)\ket{\mathcal{M}(\{p,f\}_m)} \frac{f_a(\eta_a,
  \mu_F^2)f_b(\eta_b,\mu_F^2) }{4n_c(a) n_c(b) \times \text{flux}} \\
  &\equiv\sum_m \frac{1}{m!}\int [d\{p,f\}_m]\Tr[\rho(\{p,f\}_m) 
  F(\{p,f\}_m)]\;,
  \label{rho_intro}
\end{split}
\end{equation}
where the sum runs over all final state multiplicities. Here, $[d\{p,f\}_m]$ 
is the sum of all $m$-particle phase space measures for different flavour 
sequences $\{f\}_m$. The factor $1/m!$ is necessary to account for 
identical contributions. The parton density functions evaluated 
at the momentum fraction $\eta$ and factorization scale $\mu_F^2$ are denoted 
by $f_{a/b}(\eta,\mu_F^2)$. 
We average over initial state spins (factor $4$ in the denominator) 
and colour (factor $n_c(i)$, with $n_c(q) = 3$ for quarks and $n_c(g)=8$ 
for gluons). The trace in the 
second line corresponds to a sum over indices in the colour $\otimes$ spin 
space. The quantum density $\rho$ is, therefore, given by
\begin{equation}
\begin{split}
  \rho(\{p,f\}_m) &= \ket{\mathcal{M}(\{p,f\}_m)}\bra{\mathcal{M}
  (\{p,f\}_m)}\frac{f_a(\eta_a,\mu_F^2)f_b(\eta_b,\mu_F^2) }{4n_c(a) n_c(b) 
  \times \text{flux}} \\
  &\equiv\sum_{s,c} \sum_{s^\prime,c^\prime} \ket{\{s,c\}_m} 
  \rho(\{p,f,s^\prime,c^\prime,s,c\}_m) \bra{\{s^\prime,c^\prime\}_m}\;,
\end{split}
\label{rho_def}
\end{equation}
where we used the expansion of $\ket{\mathcal{M}(\{p,f\}_m)}$ according to 
Eq.~\eqref{expandbasis}. The quantum density matrix is a projector onto the 
different helicity and gauge invariant subamplitudes of the full quantum 
amplitude
\begin{equation}
  \rho(\{p,f,s^\prime,c^\prime,s,c\}_m) =\mathcal{M}^*(\{p,f,s^\prime,
  c^\prime\}_m) \mathcal{M}(\{p,f,s,c\}_m)\frac{f_a(\eta_a,\mu_F^2)f_b(
  \eta_b,\mu_F^2) }{4n_c(a) n_c(b) \times \text{flux}}\;.
\end{equation}
Using the quantum density matrix, the expectation value of the observable $F$ 
is given by\footnote{We assume that $F$ is a unit operator in colour 
$\otimes$ spin space.}
\begin{equation}
  \sigma[F] = \sum_m \frac{1}{m!}\int [d\{p,f,s^\prime,c^\prime,s,c\}_m] 
  F(\{p,f\}_m) \braket{\{s^\prime,c^\prime\}_m|\{s,c\}_m} \rho(\{p,f,
  s^\prime,c^\prime,s,c\}_m)\;,
  \label{rhoFexpectation}
\end{equation}
where the integration measure has been extended by the inclusion of the 
summation over spin and colour. It is useful to define basis vectors
or statistical states $|\{p,f,s^\prime,c^\prime,s,c\}_m)$ 
(see Ref.~\cite{Nagy:2007ty}), such that
\begin{equation}
  \rho(\{p,f,s^\prime,c^\prime,s,c\}_m)=(\{p,f,s^\prime,c^\prime,s,c\}_m
  |\rho)\;.
\end{equation}
Defining additionally an abstract state $(F|$ according to
\begin{equation}
  (F|\{p,f,s^\prime,c^\prime,s,c\}_m) = F(\{p,f\}_m) 
  \braket{\{s^\prime,c^\prime\}_m|\{s,c\}_m}
\end{equation}
and using the decomposition of the identity
\begin{equation}
  1 = \sum_m\frac{1}{m!}\int [d(\{p,f,s^\prime,c^\prime,s,c\}_m)] 
  |\{p,f,s^\prime,c^\prime,s,c\}_m)(\{p,f,s^\prime,c^\prime,s,c\}_m|\;,
\end{equation}
one can reduce Eq.~\eqref{rhoFexpectation} to a scalar product
\begin{equation}
 \sigma[F] = (F|\rho)\;.
\end{equation}
Finally, we define the total cross section measurement function $(1|$ as
\begin{equation}
  (1|\{p,f,s^\prime,c^\prime,s,c\}_m) = \braket{\{s^\prime\}_m| \{s\}_m} 
  \braket{\{c^\prime\}_m| \{c\}_m}\;.
\end{equation}

%------------------------X------------------------X------------------------X---%
\subsection{Evolution equation}
%------------------------X------------------------X------------------------X---%
The evolution equation describes the propagation of the quantum
density matrix  $\rho$ from some initial shower time $t_0$ to some
final time $t_F$.  The initial time corresponds to the hard
interaction, while the final to the  formation of hadrons.  Therefore,
$t_F$  characterizes the physical scale at which parton emissions
cannot be described  perturbatively. There is no unique definition of
shower time $t$ as  explained in more detail in the upcoming Section
\ref{subsec:time}. 

The perturbative evolution  is described by an operator
$U(t_F,t_0)$. The expectation value of the observable $F$, including
shower effects, is
\begin{equation}
  \sigma[F] = (F|\rho(t_F)) = (F|U(t_F,t_0)|\rho(t_0))\;.
\end{equation}
$U(t_F,t_0)$ is assumed to be unitary\footnote{A non-unitary evolution
can be used to resum soft gluon effects.} 
in the sense that the total cross section $\sigma_T$ is not affected by 
evolution
\begin{equation}
  \sigma_T=(1|\rho(t_F)) = (1|U(t_F,t_0)|\rho(t_0))  = (1|\rho(t_0))\;.
\end{equation}
The evolution operator is the solution of the equation
\begin{equation}
  \frac{dU(t,t_0)}{dt} =[\mathcal{H}_I(t) - \mathcal{V}(t)]U(t,t_0)\;.
  \label{diffeq}
\end{equation}
Here, $\mathcal{H}_I(t)$ describes the transformation of a state 
$\{p,f,s^\prime,c^\prime,s,c\}_m$ to another state \\ 
$\{\hat{p},\hat{f}, \hat{s}^\prime,\hat{c}^\prime, \hat{s},\hat{c}\}_{m+1}$
by the emission of a resolved particle. This transition between the states is 
of course constrained by overall momentum conservation. On the other hand, 
$\mathcal{V}(t)$ describes the unresolved/virtual emission and, therefore,
 does not alter momentum or flavour configurations of the particles. 
Nevertheless, it can change colour configurations, which can have an effect on 
further emissions. It can be further decomposed into 
a colour diagonal, $\mathcal{V}_E(t)$, and a colour off-diagonal part,
$\mathcal{V}_S(t)$
\begin{equation}
\mathcal{V}(t) = \mathcal{V}_E(t) + \mathcal{V}_S(t)\;.
\end{equation}

Traditional parton showers correspond to the large $N_c$ limit. 
The colour structure is, therefore, always diagonal ($\mathcal{V}_S(t) \to 0$). 
In that case, Eq.~\eqref{diffeq} can be solved as
\begin{equation}
  U(t,t_0) = N(t,t_0) + \int_{t_0}^t d\tau~ U(t,\tau)\mathcal{H}_I(\tau)
  N(\tau,t_0)\;,
\end{equation}
where $N(t,t_0)$ is the Sudakov form factor (a number) defined as
\begin{equation}
  N(t,t_0) = \exp\left(-\int_{t_0}^t d\tau~\mathcal{V}(\tau)
  \right)\;.
\end{equation}

In case of a non-trivial colour evolution, the exponentiation of a 
non-diagonal matrix is cumbersome. Instead, only the colour diagonal part,
$\mathcal{V}_E(t)$, is exponentiated while the off-diagonal part, 
$\mathcal{V}_S(t)$ is treated perturbatively on the same footing as 
$\mathcal{H}_I(t)$
\begin{equation}
  U(t,t_0) = N(t,t_0) + \int_{t_0}^t d\tau~ U(t,\tau)\left[\mathcal{H}_I
  (\tau)-\mathcal{V}_S(\tau)\right]N(\tau,t_0) \;, 
  \label{idea-perturbative}
\end{equation}
with
\begin{equation}
  N(t,t_0) = \exp\left(-\int_{t_0}^t d\tau~\mathcal{V}_E(\tau)
  \right)\;.
\end{equation}
This is the final evolution equation of the Nagy-Soper parton shower.
Notice, however, that the current implementation in \deductor{} 
involves some approximations which are documented in Section 
\ref{subsec:colourevolution}.

In the following we provide the definitions of the real splitting 
operator $\mathcal{H}_I(t)$ and the virtual operator $\mathcal{V}(t)$. 
As we will see, the 
unitarity condition on $U(t,t_F)$ will allow to determine $\mathcal{V}(t)$ in 
terms of $\mathcal{H}_I(t)$.

%------------------------X------------------------X------------------------X---%
\subsection{$\mathcal{H}_I(t)$ - Real splitting operator}
\label{subsec:Realsplittingoperator}
%------------------------X------------------------X------------------------X---%
The real splitting operator describes the transition of an $m$-particle 
ensemble to an $(m+1)$-particle one, by splitting a particle into two,
\begin{equation}
  \{p,f,s^\prime,c^\prime,s,c\}_m \to \{\hat{p},\hat{f},\hat{s}^\prime,
  \hat{c}^\prime,\hat{s},\hat{c}\}_{m+1}\;.
\end{equation} 
This splitting is constrained by flavour and momentum conservation. 
The parameters of the transition require further specifications, because they 
depend on a particular choice of momentum mappings, shower time and splitting 
functions. The behaviour of the splitting operator is fixed in the infrared 
regime in order to reproduce the correct singular limits of QCD amplitudes.

After emitting a particle, one has to modify the momenta in the event to 
preserve momentum conservation and the on-shellness of all particles. This 
is achieved by momentum mappings $R_l$, which require
 three additional variables parametrizing the momentum of 
the emitted particle, $\Gamma_l$, and a variable,  $\chi_l$, specifying its
flavour $\hat{f}_{m+1}$. 
Thus, the new momenta and flavours are given by
\begin{equation}
  \{ \hat{p},\hat{f} \}_{m+1} = R_l(\{p,f\}_m, \Gamma_l,\chi_l)\;,
  \label{MomentumOperator}
\end{equation}
where $l \in \{a,b,1,...,m\}$. The inverse of this transformation is denoted 
by $Q_l$
\begin{equation}
  Q_l(\{ \hat{p},\hat{f} \}_{m+1}) = \{p,f\}_m\;.
  \label{MomentumOperatorInverse}
\end{equation}

The Nagy-Soper parton shower makes use of a global momentum mapping 
\cite{Nagy:2007ty}, i.e. all final state particles' momenta are modified 
to account for the momentum of the emitted particle.
The freedom inherent in the definition of $R_l$ may be exploited to improve
the resummation for certain observables. For instance, the study presented 
in Ref.~\cite{Nagy:2009vg} has shown, that the $p_T$ spectrum for Drell-Yan 
$Z$-boson production depends strongly on the momentum mapping for 
initial state parton splittings.
The original $R_l$ from Ref.~\cite{Nagy:2007ty} has been found inadequate
and subsequently modified in Ref.~\cite{Nagy:2014nqa}. The modified version
has been adopted in \deductor{}. 

One can define a momentum mapping operator $\mathcal{P}_l$ satisfying two 
conditions, the first one being
\begin{equation}
  \frac{1}{m!}\int [d\{p^\prime,f^\prime \}_m]\,(\{\hat{p},\hat{f}\}_{m+1}|
  \mathcal{P}_l|\{p^\prime,f^\prime\}_m)\,g(\{p^\prime,f^\prime\}_m) 
  = g(\{p,f\}_m)\;, 
  \label{hidden_delta0}
\end{equation}
where $g(\{p,f\}_\lambda)$ is an arbitrary test function and $\{p,f\}_m$ 
is determined by Eq.~\eqref{MomentumOperatorInverse}. The second condition is
\begin{equation}
\begin{split}
  \frac{1}{(m+1)!}\int & [d\{\hat{p},\hat{f}\}_{m+1}]\, 
  g(\{\hat{p},\hat{f}\}_{m+1})\, (\{\hat{p},\hat{f}\}_{m+1}|\mathcal{P}_l
  |\{p,f\}_m)  \\
  & = \frac{1}{(m+1)}\sum_{\hat{f}_{m+1} \in \chi_l} \int d\Gamma_l\,
  g(\{\hat{p}^\prime,\hat{f}^\prime\}_{m+1})\;,
  \label{hidden_delta}
\end{split}
\end{equation}
where $\{\hat{p}^\prime,\hat{f}^\prime\}_{m+1}$ is determined by 
Eq.~\eqref{MomentumOperator}. $\mathcal{P}_l$ contains a $\delta$-function,
which links the $(m+1)$-particle kinematics to the $m$-particle kinematics with
$R_l$. The two conditions imply that the Jacobians resulting from the 
integration over the $\delta$-function in Eqs.~\eqref{hidden_delta0}~and
\eqref{hidden_delta} have been absorbed in $\mathcal{P}_l$.

The evolution of the density matrix can be determined by studying the
factorization of QCD amplitudes in the soft and collinear limits.
We remind that in the limit when two partons become 
collinear, $\hat{p}_l \parallel \hat{p}_{m+1}$, the amplitude
factorizes as 
\begin{equation}
  \ket{\mathcal{M}(\{\hat{p},\hat{f}\}_{m+1})} \approx T_l^\dagger(f_l\to 
  \hat{f}_l+\hat{f}_{m+1}) V_l^\dagger(\{\hat{p},\hat{f}\}_{m+1}) 
  \ket{\mathcal{M}(\{p,f\}_m)}\;,
  \label{Mfactorization}
\end{equation}
where $T_l^\dagger(f_l\to \hat{f}_l+\hat{f}_{m+1})$ is an operator in colour 
space (see Ref.~\cite{Nagy:2007ty}). $V_l^\dagger(\{\hat{p},\hat{f}\}_{m+1})$ 
is the splitting operator in spin space. Contrary to 
traditional parton showers, the Nagy-Soper formulation does not use the 
Altarelli-Parisi splitting kernels~\cite{Altarelli:1977zs}. Instead, the 
required functions are derived directly from matrix elements without taking the
collinear limit first~\cite{Nagy:2007ty}. This allows, for example, for a 
direct access to polarization information. Furthermore, the functions are not 
singular in the soft limit~\cite{Nagy:2008ns}. 
In the soft limit, on the other hand, when the momentum of a gluon 
$\hat{p}_{m+1}$ vanishes, a similar approximation is valid
\begin{equation}
  \ket{\mathcal{M}(\{\hat{p},\hat{f}\}_{m+1})} \approx \sum_{l} T_l^\dagger(f_l\to 
  \hat{f}_l+\hat{f}_{m+1}) V_l^{\mathrm{soft}\,\dagger}(\{\hat{p},\hat{f}\}_{m+1}) 
  \ket{\mathcal{M}(\{p,f\}_m)}\;,
  \label{Mfactorization2}
\end{equation}
where $l$ runs over all partons and $V_l^{\mathrm{soft}}$ is an eikonal factor.

Combining these approximations, we may write the following evolution equation
for the quantum density matrix
\begin{equation}
  |\rho_{m+1}) = \sum_l \mathcal{S}_l\,|\rho_m)\;.
\end{equation}
Consequently $\mathcal{H}_I(t)$ is defined by the splitting operators 
$\mathcal{S}_l$ at a fixed shower time $\mathcal{T}_l(\{p,f \}_m)$ 
(see Section~\ref{subsec:time})
\begin{equation}
  \mathcal{H}_I(t) = \sum_l \mathcal{S}_l \, \delta\big( t -\mathcal{T}_l
  (\{p,f \}_m) \big)\;.
  \label{def:realsplittingoperator}
\end{equation}

We shall not reproduce the exact form of $\mathcal{S}_l$ in the most general
case, as it can be found in Ref.~\cite{Nagy:2007ty}. Nevertheless, we would like
to point out that there is an ambiguity in distributing the soft limit among
the contributions with different momentum mappings.
In order to illustrate this issue, we give the form of $\mathcal{H}_I(t)$ 
in the slightly less involved spin averaged case
\begin{equation}
\begin{split}
  (\{\hat{p},\hat{f},\hat{c}^\prime, \hat{c}\}_{m+1}|&\mathcal{H}_I(t)|
  \{p,f,c^\prime,c\}_m) \\[0.2cm]
  = &  \, (m+1) \, \sum_{l,k} \delta\big(t-\mathcal{T}_l(\{p,f\}_m)\big) \,
  (\{\hat{p},\hat{f}\}_{m+1}|\mathcal{P}_l|\{p,f\}_m) \\
  &\times \frac{n_c(a)n_c(b) \eta_a \eta_b}{n_c(\hat{a}) n_c(\hat{b}) 
  \hat{\eta}_a \hat{\eta}_b} \frac{f_{\hat{a}/A} (\hat{\eta}_a,\mu_F^2) 
  f_{\hat{b}/B} (\hat{\eta}_b,\mu_F^2)}{f_{a/A} (\eta_a,\mu_F^2) f_{b/B} 
  (\eta_b,\mu_F^2)}  \\
  &\times \frac{1}{2} \, \left[ \delta_{kl} \big(1-\delta_{\hat{f}_{m+1},g}\big) 
  \overline{w}_{ll}
  (\{\hat{p},\hat{f}\}_{m+1})  \right. \\
  & +\delta_{kl} \delta_{\hat{f}_{m+1},g} \left[\overline{w}_{ll}
  (\{\hat{p},\hat{f}\}_{m+1})  - \overline{w}_{ll}^{eikonal}(\{\hat{p},
  \hat{f}\}_{m+1})\right] \\
  & \left. -(1-\delta_{kl}) \delta_{\hat{f}_{m+1},g} A_{lk}
  (\{\hat{p}\}_{m+1}) \overline{w}_{lk}^{dipole}(\{\hat{p},\hat{f}\}_{m+1})
   \right] \\
  &\times \left[ (\{\hat{c}^\prime,\hat{c}\}_{m+1}|T_l^\dagger(f_l\to 
  \hat{f}_l + \hat{f}_{m+1}) \otimes T_k(f_k \to \hat{f}_k + \hat{f}_{m+1})
  |\{c^\prime,c\}_m) \right. \\
  &\quad \left. + (\{\hat{c}^\prime,\hat{c}\}_{m+1}|T_k^\dagger(f_k\to 
  \hat{f}_k + \hat{f}_{m+1}) \otimes T_l(f_l \to \hat{f}_l + \hat{f}_{m+1})
  |\{c^\prime,c\}_m) \right]\;,
\end{split}
  \label{spinav-splittingoperator}
\end{equation}
where the sum runs over two partons $l$ and $k$, not necessarily
different, and  $l$ is the emitter. The function $\overline{w}_{ll}$
is the spin averaged squared splitting function,  while
$\overline{w}^{eikonal}_{ll}$ is its soft approximation (see
Refs.~\cite{Nagy:2008ns,Nagy:2012bt}).  Furthermore, 
\begin{equation}
  \overline{w}_{lk}^{dipole}(\{\hat{p},\hat{f}\}_{m+1}) = - 4\pi \alpha_s
  \frac{((\hat{p}_{m+1}\cdot\hat{p}_l)\hat{p}_k - (\hat{p}_{m+1}\cdot
  \hat{p}_k)\hat{p}_l)^2}{(\hat{p}_{m+1}\cdot\hat{p}_l)^2(\hat{p}_{m+1}
  \cdot\hat{p}_k)^2}\;,
  \label{wlkdipole}
\end{equation}
which, for massless $l$ and $k$, reduces to the well known squared eikonal 
factor \cite{Bassetto:1984ik}. 
The function $A_{lk}(\{\hat{p}\}_{m+1})$ removes the singularity
when the partons $m+1$ and $k$ become collinear. 
\begin{equation}
  A_{lk}(\{\hat{p}\}_{m+1}) = \frac{(\hat{p}_{m+1}\cdot \hat{p}_k)(\hat{p}_l
  \cdot \hat{Q})}{(\hat{p}_{m+1}\cdot \hat{p}_{k})(\hat{p}_l\cdot \hat{Q}) 
  +  (\hat{p}_{m+1}\cdot \hat{p}_{l})(\hat{p}_k\cdot \hat{Q})}\;,
  \label{Alk_def}
\end{equation}
where $\hat{Q}$ is the total final state momentum. 
The ambiguity we have mentioned is due to the fact that the product
$A_{lk}\,\overline{w}_{lk}^{dipole}$ must only satisfy two conditions:
1) it may not have a singularity in the collinear limit 
$\hat p_{m+1} \parallel \hat p_k$; 2) it should correctly reproduce
the singularity at $\hat p_{m+1} \to 0$.
Formulae~\eqref{wlkdipole}~and~\eqref{Alk_def} only correspond to 
one specific choice.

%------------------------X------------------------X------------------------X---%
\subsection{$\mathcal{V}(t)$ - Virtual splitting operator}
\begin{figure}[h!]
\begin{center}
  \includegraphics[width=12cm]{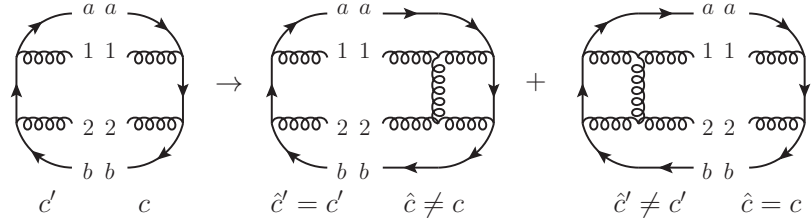}\;
  \caption{\textit{Modification on the colour structure by the inclusion
  of the virtual operator $\mathcal{V}(t)$.}}
  \label{virtualcolour}
\end{center}
\end{figure}

\begin{figure}[h!]
\begin{center}
  \includegraphics[width=13cm]{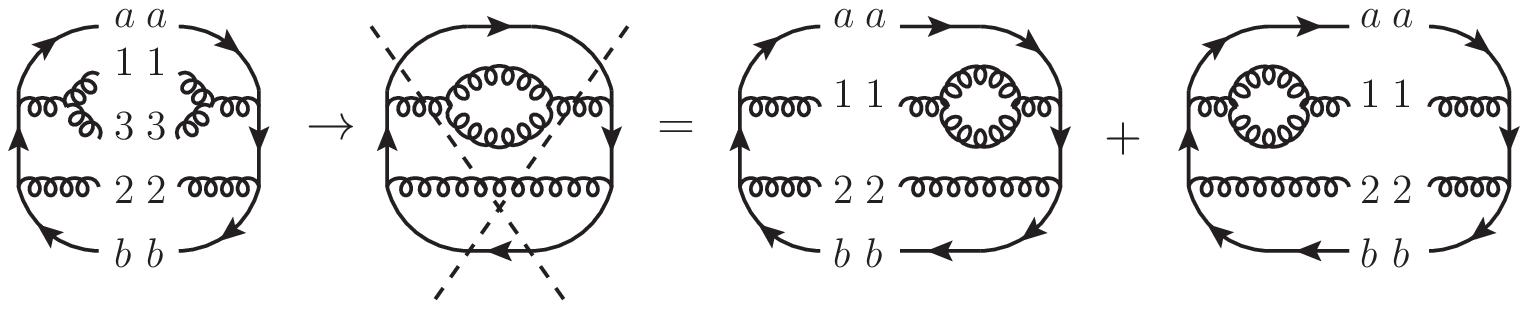}\\
  \hspace{-0.4cm}\includegraphics[width=13cm]{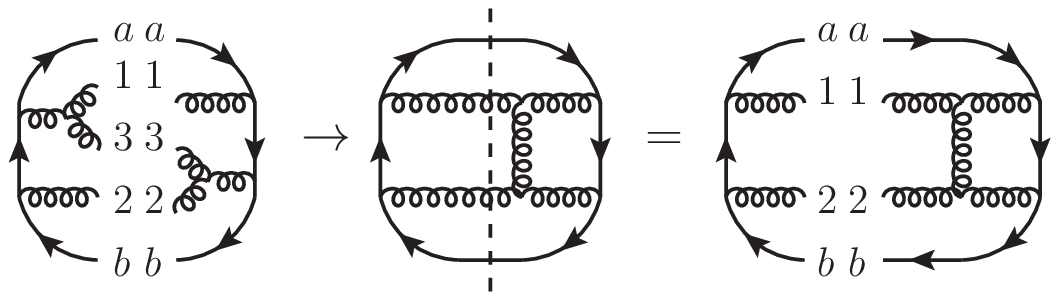}
  \caption{\textit{Correspondence between a real radiation
contribution as given by $\mathcal{H}_I(t)$ and its counterpart in
$\mathcal{V}(t)$, for a direct splitting (upper line) and an
interference contribution  (lower line).  In both cases the left part
of the first graph corresponds to $l=1$. 
For the right part $k=1$ (upper graph) or $k=2$ (lower graph).  
In the case of the interference contribution, by convention,
we attach the virtual  correction to the amplitude of the parton
$k$.}}
  \label{colour-virtual}
\end{center}
\end{figure}

%------------------------X------------------------X------------------------X---%
As advertised, the virtual splitting operator, $\mathcal{V}(t)$, 
can be determined to a large extent from the real splitting operator
$\mathcal{H}_I(t)$ by the requirement of unitarity
\begin{equation}
  \big(1\big|\mathcal{H}_I(t) - \mathcal{V}(t)\big|\rho(t)\big) = 0\;,
\end{equation}
which should be valid for any $|\rho(t))$. This can be interpreted 
with the help of the optical theorem. For example, we can schematically write
\begin{equation}
\underbrace{\sum_{\text{Triple cuts}} \;
  \raisebox{-0.5cm}{\includegraphics[width=1cm]{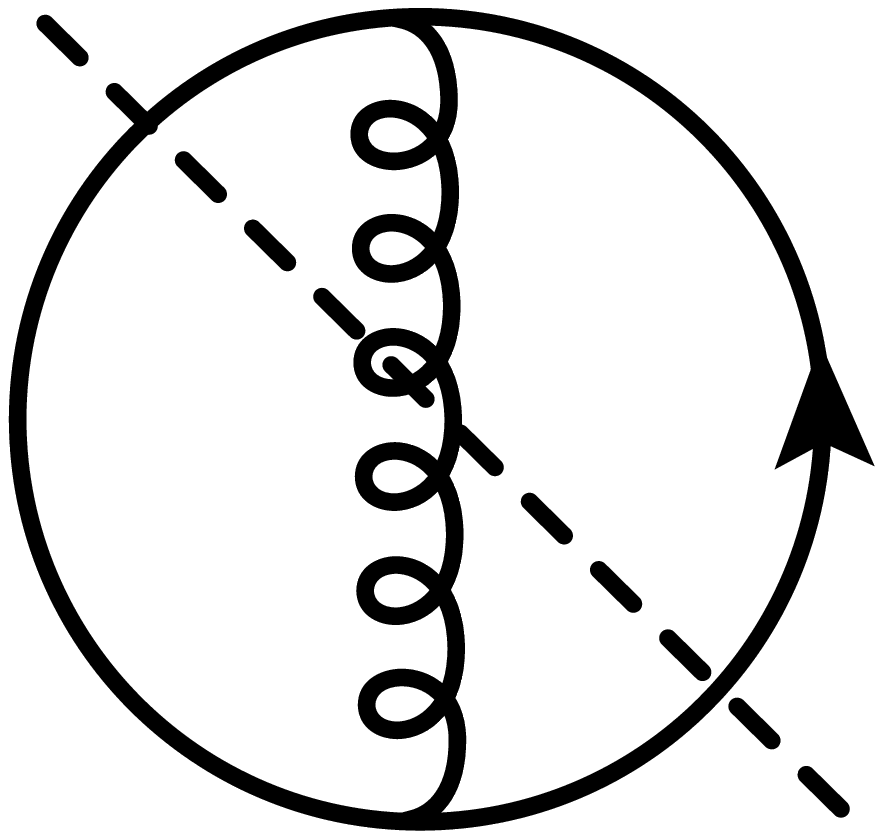}}
}_{(1|\mathcal{H}_I(t)|\rho)}
\; - \;
\underbrace{
  \Bigg[\mathrm{Im}\;
  \raisebox{-0.5cm}{\includegraphics[width=1cm]{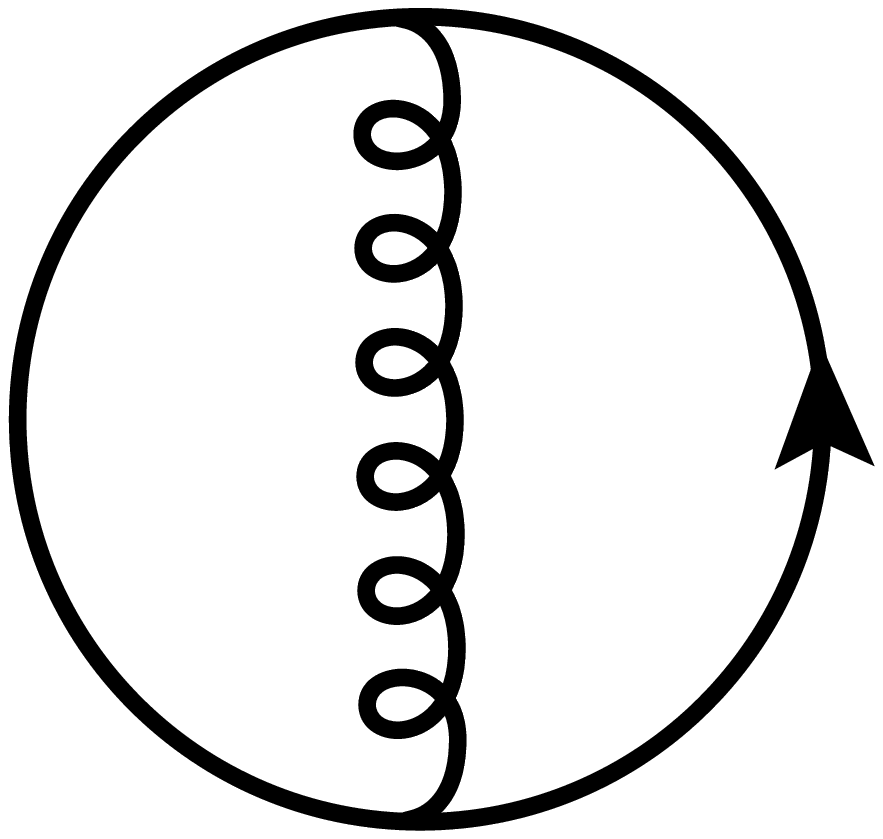}} \;- 
  \sum_{\text{Double cuts}}\;
  \raisebox{-0.5cm}{\includegraphics[width=1cm]{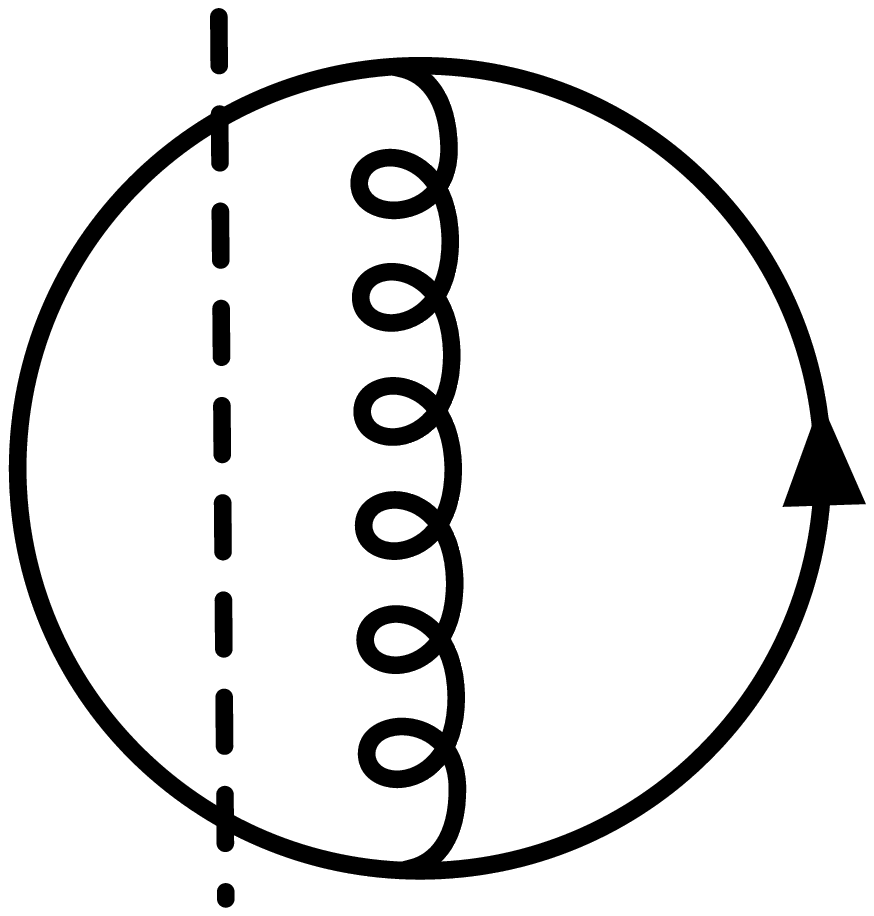}}\;\Bigg]
}_{(1|\mathcal{V}(t)|\rho)} = 0\;.
  \label{opticaltheorem}
\end{equation}
Notice that $\mathcal{V}(t)$ does not contain the exact virtual corrections.
Furthermore, it turns out that 
$(1|\mathcal{H}_I(t)|\{p,f,s^\prime,c^\prime,s,c\}_m)$ does
not contain spin correlations due to the integration over the 
azimuthal angle of the emitted partons \cite{Nagy:2007ty}. This allows us
to write
\begin{equation}
  (1|\mathcal{H}_I(t)|\{p,f,s^\prime,c^\prime,s,c\}_m) \equiv 2\braket{
  \{s^\prime\}_m|\{s\}_m} \braket{\{c^\prime\}_m|h(t,\{p,f\}_m)|\{c\}_m}\;.
  \label{Hinclusive}
\end{equation}
In consequence, the freedom in defining 
$\mathcal{V}(t)$ is restricted to the colour structure. In principle,
a virtual correction may be applied either on the $c^\prime$ or $c$ indices
of the $|\{c^\prime,c\}_m)$ state, see Fig.~\ref{virtualcolour}.
This leads to the following
\begin{equation}
  \mathcal{V}(t) = (h+i\phi) \otimes 1 + 1 \otimes (h^\dagger -i\phi)\;,
\end{equation}
where $\phi$ is a colour dependent phase, which can not be determined from
real radiation corrections. Its presence is a consequence of a Coulomb gluon 
exchange~\cite{Nagy:2012bt}
\begin{equation}
  \phi(t, \{p,f\}_m) = -2\pi \sum_{l \neq k} \frac{\alpha_s}
  {4\pi} \frac{1}{v_{kl}} \mathbf{T}_k \cdot \mathbf{T}_l\;,
  \qquad v_{kl} = \sqrt{1 - \frac{m^2_{f_k}m^2_{f_l}}{(p_k\cdot p_l)^2}}\;,
\end{equation}
where the sum runs over $l$ and $k$, both either in the initial or final
state. Examples of the relationship between $\mathcal{H}_I(t)$ and 
$\mathcal{V}(t)$ are depicted in Fig.~\ref{colour-virtual}.
The exact specification of $\mathcal{V}(t)$ can be found 
in Refs.~\cite{Nagy:2007ty,Nagy:2012bt}.

%
%------------------------X------------------------X------------------------X---%
\subsection{Colour and spin evolution}
\label{subsec:colourevolution}
%------------------------X------------------------X------------------------X---%

An essential part of the action of $\mathcal{H}_I(t)$ is the change of the
colour state. Starting from the two types of colour strings 
Eqs.~\eqref{openstring}~and~\eqref{closedstring}, the application of
$T_l^{\dagger}(f_l\to \hat{f}_l + \hat{f}_{m+1})$ on the $c^\prime$ index
and of $T_k(f_k\to \hat{f}_k + \hat{f}_{m+1})$ on the $c$ index of the
$|\{c^\prime,c\}_m)$ state results in a linear combination of the same 
basis vectors. This follows from the relations
\begin{align}
  \label{commutator}
  if^{abc}T^c &= [T^a,T^b] = T^aT^b - T^bT^a\;, \\
  \label{fierz}
  T^a_{ij}T^a_{kl} &= \frac{1}{2}\left[ \delta_{il}\delta_{jk} - 
  \frac{1}{N_c} \delta_{ij}\delta_{kl}\right]\;,
\end{align}
which can be translated into similar relations for 
$T^{(\dagger)}(f_l\to \hat{f}_l +\hat{f}_{m+1})$,
see Ref.~\cite{Nagy:2007ty}. 

\begin{figure}[h!]
\begin{center}
  \includegraphics[width=8cm]{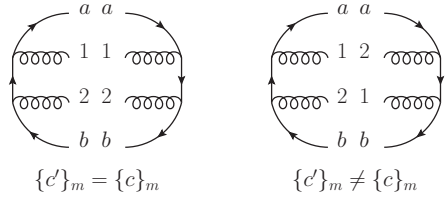}
  \vspace{-0.5cm}
  \caption{\textit{Example colour configurations for the 
  $\bar q(a) q(b) g(1) g(2)$ state.}}
  \label{fig:qqgg_colour}
\end{center}
\end{figure}

\begin{figure}[h!]
\begin{center}
  \includegraphics[width=10cm]{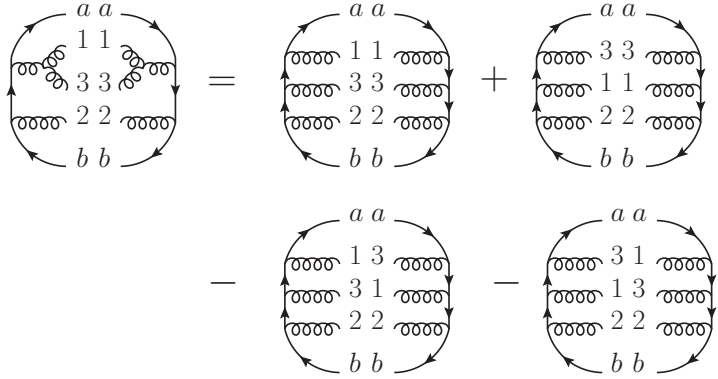}
  \vspace{-0.5cm}
  \caption{\textit{Example of a direct splitting $g\to gg$
  starting from a diagonal colour configuration. The left part
  of the first graph corresponds to $l=1$, while the right part
  to $k=1$. Moreover, $m+1=3$.}}
  \label{fig:qqggg_direct_1}
\end{center}
\end{figure}

\begin{figure}[h!]
\begin{center}
  \includegraphics[width=10cm]{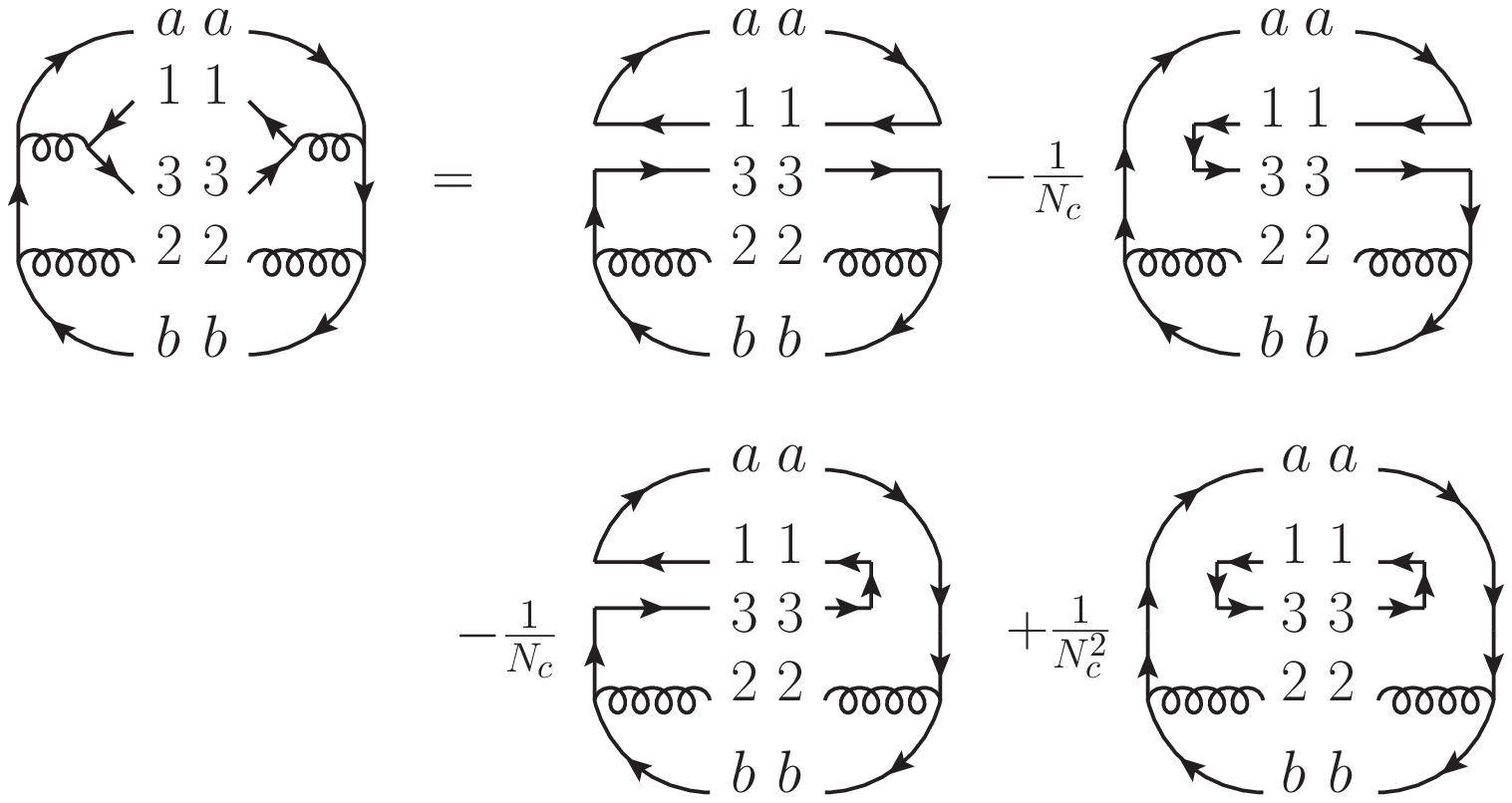}
  \vspace{-0.5cm}
  \caption{\textit{Example of a direct splitting $g\to q\bar q$
  starting from a diagonal colour configuration. The left part
  of the first graph corresponds to $l=1$, while the right part
  to $k=1$. Moreover, $m+1=3$.}}
  \label{fig:qqggg_direct_2}
\end{center}
\end{figure}

\begin{figure}[h!]
\begin{center}
  \includegraphics[width=10cm]{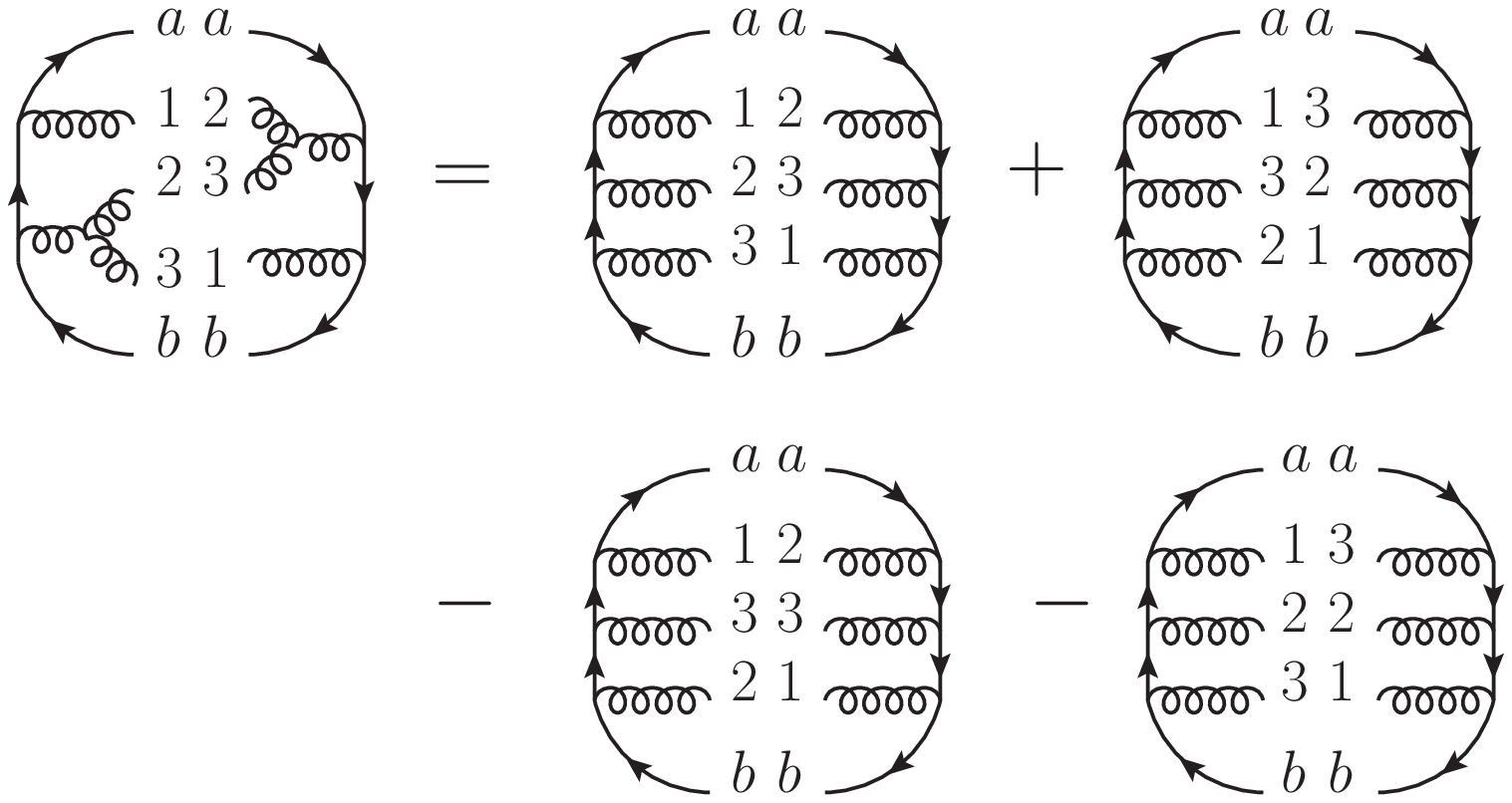}
  \vspace{-0.5cm}
  \caption{\textit{Example of a direct splitting $g\to gg$ starting 
  from an off-diagonal colour configuration. The left part
  of the first graph corresponds to $l=2$, while the right part
  to $k=2$. Moreover, $m+1=3$.}}
  \label{fig:direct_subleading}
\end{center}
\end{figure}

\begin{figure}[h!]
\begin{center}
  \includegraphics[width=10cm]{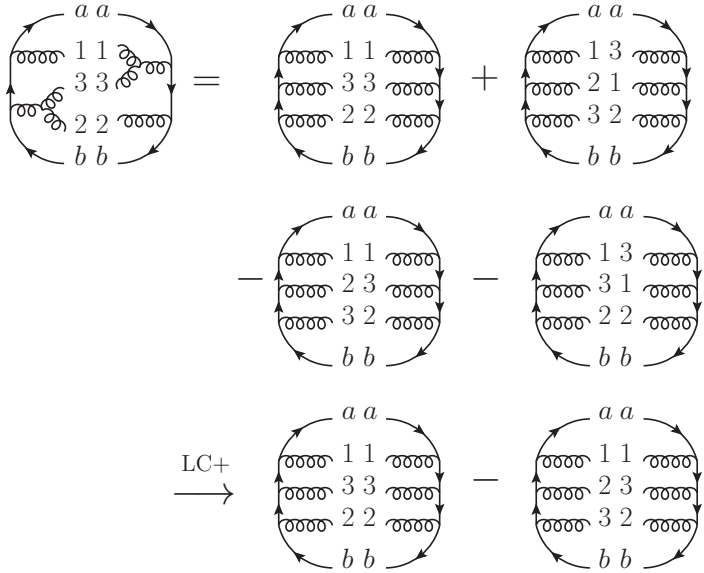}
  \vspace{-0.5cm}
  \caption{\textit{Example of an interference contribution of two 
  $g\to gg$ splittings starting from a diagonal colour configuration.
  The last line represents the LC+ approximation of the original colour
  structure. The left part
  of the first graph corresponds to $l=2$, while the right part
  to $k=1$. Moreover, $m+1=3$.}}
  \label{fig:qqggg_interference_1}
\end{center}
\end{figure}

\begin{figure}[h!]
\begin{center}
  \includegraphics[width=10cm]{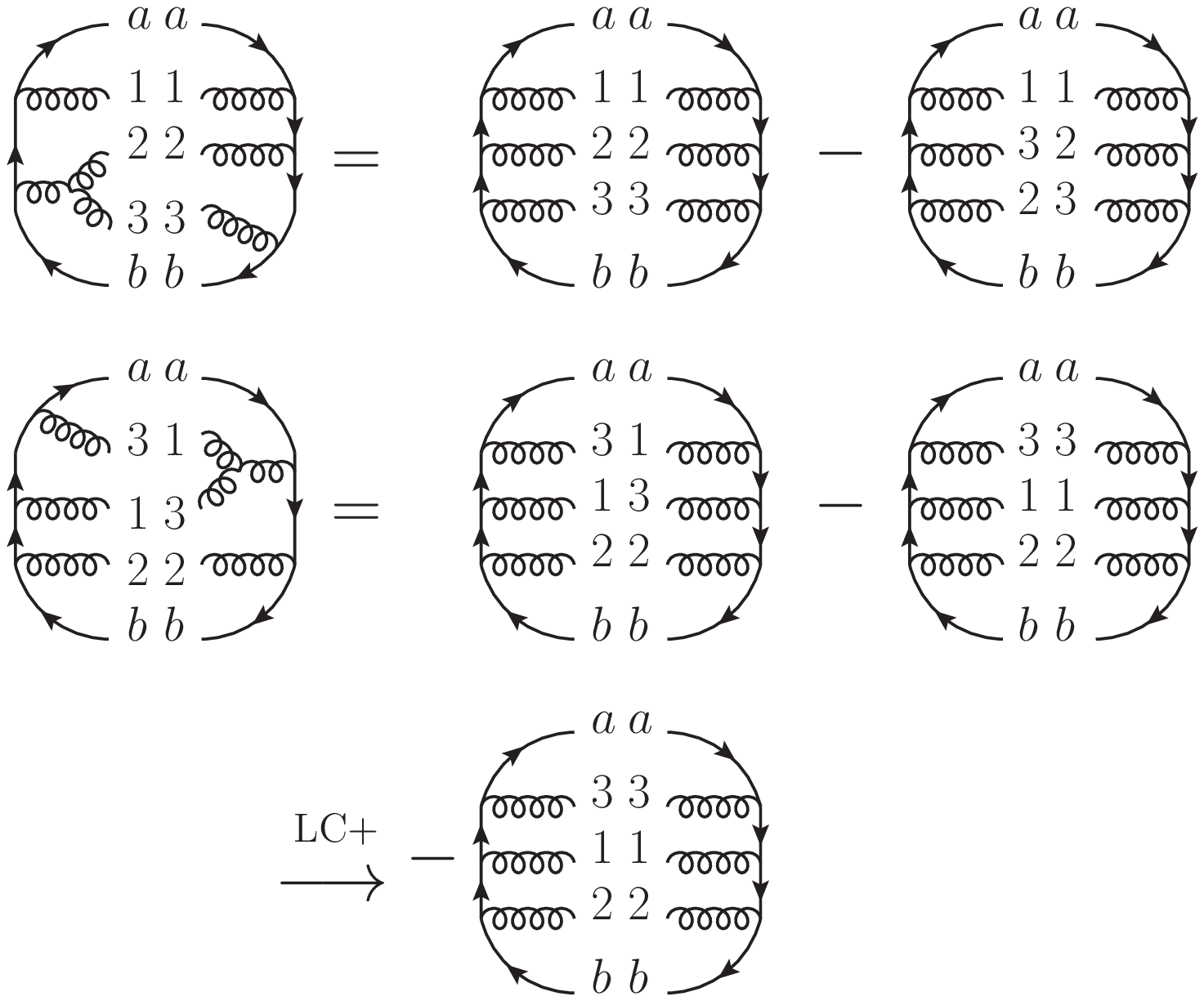}
  \vspace{-0.5cm}
  \caption{\textit{Example of interference contributions of 
  $g\to gg$ and $q\to qg$ splittings starting from a diagonal 
  colour configuration.  Upper graph: the left part
  of the first graph corresponds to $l=2$, while the right part
  to $k=b$. Lower graph: the left part
  of the first graph corresponds to $l=a$, while the right part
  to $k=1$. Moreover, $m+1=3$ in both cases. The last line represents 
  the LC+ approximation of the original colour structure of the lower 
  graph.}}
  \label{fig:qqggg_interference_2}
\end{center}
\end{figure}

In principle we have to consider diagonal, $\{c^\prime\}_m = \{c\}_m$, 
and off-diagonal, $\{c^\prime\}_m \neq \{c\}_m$, colour configurations,
see Fig.~\ref{fig:qqgg_colour} for illustration. 
Notice that the colour diagonal contributions contain the leading colour
contribution, while colour off-diagonal contributions are subleading.
The latter can be identified by crossing of
gluon lines. Traditional parton showers are restricted to the
leading colour approximation (or at best to colour diagonal contributions).
On the other hand, the Nagy-Soper parton shower allows for the evolution of
subleading colour configurations.
In Figs.~\ref{fig:qqggg_direct_1}, \ref{fig:qqggg_direct_2}, 
\ref{fig:direct_subleading}, \ref{fig:qqggg_interference_1}, 
and \ref{fig:qqggg_interference_2}, we give several examples of colour
evolution induced by direct splitting (application of $T_l^\dagger
\otimes T_l$) and interference contributions (application of
$T_l^\dagger \otimes T_k$, where $l\neq k$). 


The solution of the evolution equation, Eq.~\eqref{idea-perturbative},
requires a decomposition of the virtual operator into a diagonal and an 
off-diagonal colour part. This can be achieved with the help of the LC+ 
approximation introduced in Ref.~\cite{Nagy:2012bt}. The main features of this
approximation are 
\begin{enumerate}
\item Exact colour treatment of the collinear and soft-collinear limits;
\item Leading colour approximation for the pure soft limit (a part of 
      the subleading contributions is kept nevertheless). 
\end{enumerate}

According to the previous Section, the virtual splitting operator is defined
in terms of the real splitting operator. For the latter, the LC+ approximation
amounts to the introduction of a projector $C(l,m+1)$ acting on the colour 
states $\ket{\{c\}_{m+1}}$ in Eq.~\eqref{spinav-splittingoperator} as
\begin{align}
  \label{colourcorrmodA}
  T_k^\dagger(f_k\to \hat{f}_k + \hat{f}_{m+1})\ket{\{c\}_m} &\to C(l,m+1)
  T_k^\dagger(f_k\to \hat{f}_k + \hat{f}_{m+1})\ket{\{c\}_m}\;, \\
  \label{colourcorrmodB}
  \bra{\{c^\prime\}_m}T_k(f_k\to \hat{f}_k + \hat{f}_{m+1}) &\to
  \bra{\{c^\prime\}_m}T_k(f_k\to \hat{f}_k + \hat{f}_{m+1})C^\dagger(l,m+1)
  \;,
\end{align}
where 
\begin{equation}
  C(l,m+1)\ket{\{c\}_{m+1}} = \begin{cases}
    \ket{\{c\}_{m+1}} & \text{partons $l$ and $m+1$ form a $q\bar{q}$ pair}\;,\\
    \ket{\{c\}_{m+1}} & \text{partons $l$ and $m+1$ are colour connected}\;,\\
    0 & \text{otherwise}\;.
  \end{cases}
  \label{colourprojector}
\end{equation}
Colour connected partons are partons, which are direct neighbours on a colour
string as defined in Section \ref{subsec:confspace}. 
The definition Eq.~\eqref{colourprojector} guarantees that direct splittings 
are treated exactly. This is also true for the evolution of off-diagonal colour
configurations, as depicted in Fig.~\ref{fig:direct_subleading}.
In other cases, a part of the contributions is removed,
as can be seen in 
Figs.~\ref{fig:qqggg_interference_1}~and~\ref{fig:qqggg_interference_2}.

The colour projector $C(l,m+1)$ defines a decomposition  
\begin{equation}
  \mathcal{H}_I(t) \equiv \mathcal{H}_I^{LC+}(t) + \Delta\mathcal{H}_I(t)\;.
\end{equation}
All corrections to the LC+ approximation are of order $\mathcal{O}(1/N_c^2)$ 
and are treated perturbatively. It turns out that the virtual splitting 
operator $\mathcal{V}^{LC+}(t)$ defined in terms of $\mathcal{H}_I^{LC+}(t)$ 
is colour diagonal~\cite{Nagy:2012bt}. Therefore, we define
\begin{equation}
\mathcal{V}_E(t) \equiv \mathcal{V}^{LC+}(t)\;, \qquad 
\mathcal{V}_S(t) \equiv \Delta\mathcal{V}(t) = \mathcal{V}(t) - 
\mathcal{V}^{LC+}(t)\;.
\end{equation}
We can also define an approximate evolution operator
\begin{equation}
  U^{LC+}(t,t_0) = N^{LC+}(t,t_0) + \int_{t_0}^t d\tau~ U^{LC+}(t,\tau)
  \mathcal{H}_I^{LC+}(\tau)N^{LC+}(\tau,t_0)\;,
  \label{LCplusEvo}
\end{equation}
where
\begin{equation}
  N^{LC+}(t,t_0) = \exp\left( - \int_{t_0}^t d\tau~\mathcal{V}^{LC+}(\tau)
  \right)\;.
\end{equation}
On the other hand, the exact evolution operator is
\begin{equation}
  U(t,t_0) = U^{LC+}(t,t_0) + \int_{t_0}^t d\tau~ U(t,\tau) \left[\Delta
  \mathcal{H}_I(\tau) - \Delta\mathcal{V}(\tau)\right]U^{LC+}(\tau,t_0)\;.
  \label{fullEvo}
\end{equation}
The practical solution of these equations is described in 
Ref.~\cite{Nagy:2012bt}. The current implementation of 
\deductor{} relies on Eq.~\eqref{LCplusEvo}.

\begin{figure}[h!]
\begin{center}
  \includegraphics[width=9cm]{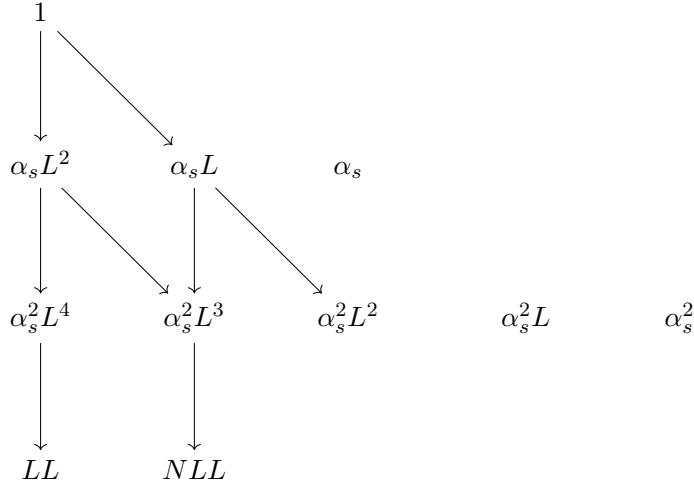}\;
  \caption{\textit{Illustration of the logarithm counting. One step in the 
  vertical direction is given by an insertion of $U^{LC+}$ and a diagonal step 
  is given by an insertion of $\Delta \mathcal{H}_I(t) - \Delta \mathcal{V}
  (t)$. One can see, that two insertions of $\Delta \mathcal{H}_I(t) - \Delta 
  \mathcal{V}(t)$ only contributes to the coefficient of $\alpha_s^n 
  L^{2n-2}$.}}
  \label{logcounting}
\end{center}
\end{figure}

Let us comment on the logarithmic accuracy of the Nagy-Soper parton shower.
Consider an observable $\mathcal{O}$, given by the expansion
\begin{equation}
  \langle \mathcal{O} \rangle = \sum_n c(n,2n)\alpha_s^n L^{2n} + \sum_n 
  c(n,2n-1)\alpha_s^n L^{2n-1} + \cdots\;,
\end{equation}
where $L$ denotes a potentially  large logarithm,
e.g. $L=\log(s/p_T^2)$, where $s$  is the scale of the hard process,
and $p_T$ is the transverse momentum of some final state, for example
a pair of oppositely charged leptons from the Drell-Yan process. The
evolution operator  $U(t,t_0)$ reproduces the coefficients $c(n,2n)$
and $c(n,2n-1)$ without  any approximations in colour. On the other
hand, $U^{LC+}(t,t_0)$ misses  part of the $c(n,2n-1)$ coefficient due
to subleading colour contributions from wide angle soft radiation.
For this reason, the LC+ approximation is only exact at leading
logarithm (LL), see Fig.~\ref{logcounting}. Inserting a factor of
$\left[\Delta\mathcal{H}_I(\tau) - \Delta\mathcal{V}(\tau)\right]$
generates the remaining contribution to the coefficient $c(n,2n-1)$.
In consequence, the evolution operator 
\begin{equation}
  U(t,t_0) = U^{LC+}(t,t_0) + \int_{t_0}^t d\tau~ U^{LC+}(t,\tau) \left[
  \Delta\mathcal{H}_I(\tau) - \Delta\mathcal{V}(\tau)\right]U^{LC+}
  (\tau,t_0)\;,
\end{equation}
is accurate at next-to-leading logarithm (NLL).

As pointed out before (see discussion around Eq.~\eqref{Hinclusive}), 
the virtual splitting operator is diagonal in spin. 
Therefore, spin dependent effects are 
only induced by the action of $\mathcal{H}_I(t)$. It has been proposed in 
Ref.~\cite{Nagy:2008eq} to include them on top of the spin averaged shower.
To this end, we note that the necessary modification in $\mathcal{H}_I(t)$, 
Eq.~\eqref{spinav-splittingoperator}, amounts to the replacement
\begin{equation}
\Phi_{lk}(\{\hat{p},\hat{f}\}_{m+1}) \; \longrightarrow \;
(\{\hat{s}^\prime,\hat{s}\}_{m+1}|Y_{lk}(\{\hat{p},\hat{f}\}
  _{m+1})|\{s^\prime,s\}_m) \; \Phi_{lk}(\{\hat{p},\hat{f}\}_{m+1}) \;,
\end{equation}
where
\begin{equation}
\begin{split}
  \Phi_{lk}(\{\hat{p},\hat{f}\}_{m+1}) &= \delta_{kl} 
  \; \big(1-\delta_{\hat{f}_{m+1},g}\big) \;
  \overline{w}_{ll}(\{\hat{p},\hat{f}\}_{m+1} \\
  &+\delta_{kl} \; \delta_{\hat{f}_{m+1},g} \; \left[\overline{w}_{ll}(\{\hat{p}
  ,\hat{f}\}_{m+1}  - \overline{w}_{ll}^{eikonal}\right] \\
  &-(1-\delta_{kl}) \; \delta_{\hat{f}_{m+1},g} \; A_{lk}(\{\hat{p}_{m+1}\}) \; 
  \overline{w}_{lk}^{dipole}(\{\hat{p},\hat{f}\}_{m+1})\;,
\end{split}
\end{equation}
and $Y_{lk}(\{\hat{p},\hat{f}\}_{m+1})$ is just the ratio between spin dependent 
and spin averaged splitting functions, see Ref.~\cite{Nagy:2008eq}. 

The parton shower evolution starts with a quantum density matrix
\begin{equation}
  (\{p,f,s^\prime,c^\prime,s,c\}_m|\rho(t_0)) = (\{p,f,c^\prime,c\}_m
  |\rho(t_0)) \otimes 	(\{s^\prime,s\}_m|\rho^s(\{p,f,c^\prime,c\}_m))\;,
\end{equation}
with a factorized spin dependence, $|\rho^s)$, with respect to the spin 
averaged quantum density $|\rho(t_0))$. The latter subsequently undergoes
spin averaged evolution. On the other hand, after $(n-m)$ real emission steps 
\begin{equation}
  |\rho^s_{n}) = Y_{l_n,k_n}(\{\hat{p},\hat{f}\}_{n}) \cdots Y_{l_{m+2},
  k_{m+2}}(\{\hat{p},\hat{f}\}_{m+2})Y_{l_{m+1},k_{m+1}}(\{\hat{p},\hat{f}\}
  _{m+1})|\rho^s(\{p,f,c^\prime,c\}_m))\;.
\end{equation}
Assuming that the final observable does not depend on the spin configuration, 
we obtain
\begin{equation}
  \sigma[F] = \sum_{\lambda}\frac{1}{\lambda!}\int [d\{p,f,c^\prime,c\}
  _\lambda] (F|\{p,f,c^\prime,c\}_\lambda) (\{p,f,c^\prime,c\}_\lambda
  |\rho(t_F))(1_{spin}|\rho^s_\lambda)\;,
\end{equation}
where
\begin{equation}
  (1_{spin}|\rho^s_\lambda) = \sum_{\{s^\prime,s\}_\lambda} (1_{spin}|
  \{s^\prime,s\}_\lambda)(\{s^\prime,s\}_\lambda|\rho^s_\lambda) = 
  \sum_{\{s\}_\lambda} (\{s,s\}_\lambda|\rho^s_\lambda)\;.
\end{equation}
%

%------------------------X------------------------X------------------------X---%
\subsection{Shower time}
\label{subsec:time}
%------------------------X------------------------X------------------------X---%

The forward evolution in the shower time is used to enforce an ordering 
in some chosen kinematic variable. 
This is necessary in order to correctly resum leading logarithms of infrared
sensitive quantities. In traditional 
parton shower programs the following ordering variables are used:
\begin{description}
\item[virtuality:] the invariant mass of two daughter partons produced through 
the splitting (e.g. \pyQ{}~\cite{Sjostrand:2006za}),
\item[transverse momentum:] the transverse momentum of the daughter 
partons with respect to the mother parton 
(e.g. \pythia{}~\cite{Sjostrand:2007gs}),
\item[angle:] the angle between the momenta of the daughter partons
(e.g. \herwig{}~\cite{Bahr:2008pv,Corcella:2000bw}).
\end{description}
The construction of a new parton shower allows to re-evaluate the advantages
of these variables, and replace them if necessary. In Ref.~\cite{Nagy:2014nqa},
the following variable has been proposed
\begin{equation}
  \Lambda^2_l = \frac{|(\hat{p}_l \pm \hat{p}_{m+1})^2-m^2_l|}
  {2p_l\cdot Q_0} Q_0^2\;,
  \label{lambdadef}
\end{equation}
where $\hat{p}_l$ is the emitter momentum after emission (daughter parton), 
$\hat{p}_{m+1}$ the emitted parton momentum (daughter parton), 
$p_l$ the emitter momentum before emission (mother parton with mass $m_l$) 
and $Q_0$ is
the total final state momentum. 
The minus sign between $\hat{p}_l$ and $\hat{p}_{m+1}$ in Eq.~\eqref{lambdadef}
applies to an initial state splitting.
The shower time with $\Lambda^2_l$ ordering
is then given by
\begin{equation}
  \mathcal{T}_l(\{p,f\}_m) \equiv -\log\left(\frac{\Lambda_l^2}{Q_0^2}\right)\;.
\end{equation}

The particular form of $\Lambda_l^2$ may be justified as 
follows~\cite{Nagy:2014nqa}. Consider a splitting of a final state parton
with momentum $p_0$ into two partons with momenta $p_1$ and $p_2$. 
After splitting
the mother parton has non-vanishing additional virtuality 
$v_0^2$ ($= p_0^2-m_0^2$),
 while both daughter partons are on-shell. In subsequent steps, 
both daughter partons will split as well, and
acquire additional virtualities $v_1^2$ and $v_2^2$ ($= p_i^2-m_i^2$). Clearly,
$v_0^2$ will be modified by this procedure. The validity of the on-shell
approximation used in the first step necessitates the effect on $v_0^2$
to be negligible. This can be translated into the following conditions
\cite{Nagy:2014nqa}
\begin{equation}
  \frac{v_1^2}{2 p_1\cdot Q_0} \ll \frac{v_0^2}{2p_0\cdot Q_0} \quad 
  \text{ and }\quad   \frac{v_2^2}{2 p_2\cdot Q_0} \ll 
  \frac{v_0^2}{2p_0\cdot Q_0}\;.
\end{equation}
Up to normalization, the terms $\frac{v_l^2}{2 p_l\cdot Q_0}$ are equal 
to $\Lambda^2_l$. A similar discussion holds for the initial state shower.
We note that one of the consequences of $\Lambda_l^2$ ordering is an
enlarged phase space  for initial state splittings compared to $p_T$
ordering.

%------------------------X------------------------X------------------------X---%
\subsection{Consequences of backward evolution}
\label{subsec:backwardevolution}
%------------------------X------------------------X------------------------X---%

Initial state splittings are generated by the backward evolution 
formalism first introduced by 
Sj\"ostrand in Refs.~\cite{Sjostrand:1985xi,Bengtsson:1986gz}. 
In this approach, the Sudakov form factor contains a ratio of parton 
distribution functions to guide the evolution of the momentum fraction 
during splitting. 
Nevertheless, as long as mass effects are neglected and the splitting functions
used in the shower and PDF evolution are the same,
one can factor out the non-perturbative part completely. This matches 
the treatment of final state radiation, where hadronization only
happens after showering. Unfortunately, the splitting functions used in the
Nagy-Soper parton shower are very different from those of Altarelli-Parisi
away from the collinear limit. Thus, non-perturbative information will
inevitably affect the evolution. 

In the presence of mass effects, however, the non-perturbative input can not
be factorized, independently of the approach to the parton shower
(traditional or otherwise), unless the evolution of PDFs is modified. 
We describe the required modification after Ref.~\cite{Nagy:2014oqa},
which has been implemented in \deductor{}. 

In principle we would like to decompose the density matrix as
\begin{equation}
  |\rho) = \mathcal{L}(t) |\rho_{\text{pert}})\;,
  \label{rhopert}
\end{equation}
where $|\rho_{\text{pert}})$ would contain no non-perturbative information
if splittings were restricted to the quasi-collinear limit. Here,
the luminosity operator $\mathcal{L}(t)$ is
\begin{equation}
  \mathcal{L}(t)|\Phi_m) \equiv \frac{f_{a/A}(\eta_a,\mu_A^2e^{-t}) f_{b/B}
  (\eta_b,\mu_B^2e^{-t})}{4n_c(a)n_c(b) 4\eta_a \eta_b p_A\cdot p_B}|\Phi_m)\;,
\end{equation}
where we used a shorthand notation for the statistical states, 
$\Phi_\lambda = \{p,f,s^\prime, c^\prime, s,c\}_\lambda$.
The scale $\mu_i^2$ is given by $\mu_i^2 = 2p_i\cdot Q_0$, 
where $p_i$ are the hadron momenta and $Q_0$ the total final state momentum. 
Therefore, the PDFs are evaluated at the scale 
$\mu_i^2e^{-t} = |(\hat{p}_i-\hat{p}_{m+1})^2-m_i^2|$ 
which is the virtuality of the splitting. 
Let us define a perturbative real splitting operator, 
$\mathcal{H}_I^{\text{pert}}(t)$, through
\begin{equation}
  \mathcal{H}_I(t) = \mathcal{L}(t) \mathcal{H}_I^{\text{pert}}(t) 
  \mathcal{L}^{-1}(t)\;.
  \label{Hpert}
\end{equation}
Unfortunately, $\mathcal{V}(t)$ contains a convolution of
splitting functions with PDFs, because it is an integral of 
Eq.~\eqref{Hpert}. It is thus impossible to trivially factor
out the non-perturbative contribution. Nevertheless, we can define 
a quasi-perturbative virtual splitting operator, $\mathcal{V}^{\text{pert}}(t)$,
via the evolution equation. Indeed, substituting Eq.~\eqref{rhopert} in
Eq.~\eqref{diffeq} we obtain
\begin{equation}
  \left(\frac{d}{dt} \mathcal{L}(t)\right) |\rho_{\text{pert}}) + 
  \mathcal{L}(t)\left(\frac{d}{dt} |\rho_{\text{pert}}) \right) = 
  \left[\mathcal{H}_I(t) - \mathcal{V}(t)\right]\mathcal{L}(t)
  |\rho_{\text{pert}})   \;.
\end{equation}
This allows us to write
\begin{equation}
  \frac{d}{dt} |\rho_{\text{pert}}) = \left[\mathcal{H}_I^{
  \text{pert}}(t) - \mathcal{V}^{\text{pert}}(t)\right]|\rho_{\text{pert}})
  \;,
\end{equation}
with
\begin{equation}
  \mathcal{V}^{\text{pert}}(t) = \mathcal{V}(t) + \mathcal{L}^{-1}(t)  
  \left(\frac{d}{dt} \mathcal{L}(t)\right)\;,
\end{equation}
where we used the fact that $\mathcal{L}(t)$ commutes with $\mathcal{V}(t)$ 
since it does not change the momenta. 

To ensure that $\mathcal{V}^{\text{pert}}(t)$ is independent of PDFs
in the quasi-collinear limit, 
the splitting functions obtained by deriving $\mathcal{L}(t)$ must match
the quasi-collinear limit of $\mathcal{V}(t)$.
The respective functions are given in Ref.~\cite{Nagy:2014oqa}.
As expected, they only differ from the Altarelli-Parisi kernels
at non-zero mass. They also imply a modified evolution of the PDFs.
The effect is mostly visible on the $b$-quark PDF. Clearly, only processes
predominantly generated from initial state $b$-quarks will be influenced.

Modifying the evolution of parton distribution functions may be used for other
purposes as well. For instance, adding a higher order term $P^{(2)}_{a\hat{a}}$ 
as follows
\begin{equation}
\begin{split}
  \frac{df_{a/A}(\eta_a,\mu^2)}{d\log(\mu^2)} &= \sum_{\hat{a}} \int 
  \frac{dz}{z} \frac{\alpha_s(\mu^2/z)}{2\pi} P_{a\hat{a}}(z,\mu^2/z)f_{
  \hat{a}/A}(\eta_a/z,\mu^2) \\
  &+ \sum_{\hat{a}} \int \frac{dz}{z} \left(\frac{\alpha_s(\mu^2/z)}{2\pi}
  \right)^2 P^{(2)}_{a\hat{a}}(z,\mu^2/z)f_{\hat{a}/A}(\eta_a/z,\mu^2) \;,
\end{split}
\end{equation}
with 
\begin{equation}
  P^{(2)}_{a\hat{a}}(z,\mu^2/z) = -2\pi\beta_0 \log(\lambda_R) P_{a\hat{a}}
  (z,\mu^2/z)\;,
\end{equation}
where $\beta_0 = (33-2n_f)/(12\pi)$ and
\begin{equation}
  \lambda_R = \exp\left( - \frac{C_A(67-3\pi^2)-10n_f}{3(33-2n_f)}\right)\;,
\end{equation}
where $n_f$ is the number of light active flavours,
allows to incorporate the cusp anomalous dimension at NLL for 
processes with less then three coloured particles \cite{Catani:1990rr}.

%------------------------X------------------------X------------------------X---%
\subsection{A summary of ambiguities}
%------------------------X------------------------X------------------------X---%

Let us conclude this Section with a list of ambiguous components of the 
Nagy-Soper parton shower:
\begin{enumerate}
  \item  Momentum Mappings
  \item Splitting functions  
  \item Soft partition function  
  \item Colour treatment  
  \item Spin treatment  
  \item Shower time 
  \item PDF evolution
\end{enumerate}
We stress that this shower is a rather recent construction still under
development.  It is to be expected that there will be changes to any
of the particular solutions discussed in the previous Subsections.

%------------------------X------------------------X------------------------X---%
\section{Matching NLO matrix elements to the parton shower}
%------------------------X------------------------X------------------------X---%

Matching NLO calculations with parton showers is a widely explored subject and 
there exist several matching schemes, the most popular being 
\powheg{}~\cite{Nason:2004rx,Frixione:2007vw} and 
\mcnlo{}~\cite{Frixione:2002ik,Frixione:2003ei} 
(for other proposals see e.g.~\cite{Kramer:2003jk,Soper:2003ya,Kramer:2005hw,
Nagy:2005aa,Giele:2007di,Torrielli:2010aw,Hoche:2010pf,Platzer:2011bc,
Hartgring:2013jma}). 
A general comparison between these two can be found in 
e.g.~\cite{Hoeche:2011fd}. 
In order to benefit from the recently implemented subtraction scheme
based on the Nagy-Soper parton shower splitting 
kernels~\cite{Bevilacqua:2013iha},
we chose the \mcnlo{} formalism. Notice, that the implementation of \mcnlo{} 
in \sherpa{}~\cite{Hoeche:2011fd} also benefits from the consistency between 
the parton shower (Catani-Seymour)
and subtraction terms in the matching. However, once the implementation
of the Nagy-Soper parton shower will be made accurate at subleading terms
in the colour expansion, our matching procedure will allow for a 
similar accuracy in the matched sample. This may also be possible in the 
future within \sherpa{}, see Ref.~\cite{Platzer:2012np}. Before we discuss 
the challenges of parton shower matching, we give a brief overview of 
the general objectives.

\begin{description}
  \item \textbf{Preserving the NLO cross section normalization:} When
    considering an inclusive observable $F$, the fixed order
    normalization of the cross section should be preserved. We thus require
  \begin{equation}
    (F|U(t_F,t_0)|\rho(t_0)) = \sigma^{NLO}[F]\;.
  \end{equation}

  \item \textbf{Defining event samples at NLO level:} Matching to parton shower
  is the only way to define events at NLO. As we will see, matching
  {\it \`a la}
  \mcnlo{} imposes a shower specific subtraction scheme. Without matching, the 
  weights of the real matrix element and the subtraction terms are not
  coupled kinematically unless in a strict limit, and diverge
  separately. Due to the matching scheme those weights are combined
  and one obtains real emission phase space configurations with a finite, but 
  not necessarily positive, weight. The matching renders the virtual
  corrections finite as well.
  
  \item \textbf{Next-to-Leading Logarithmic accuracy for infrared sensitive 
  observables:} Infrared sensitive observables, which are affected by large
  logarithms, $L$, of some kinematic variable at fixed-order, 
  are replaced by resummed predictions. Traditional parton showers 
  only resum the leading logarithms (LL) 
  $\alpha_s^nL^{2n}$, whereas the Nagy-Soper shower also allows to resum the 
  next-to-leading logarithms (NLL) $\alpha_s^n L^{2n-1}$. Therefore, we require
  differential distributions to be accurate at the NLO$+$NLL level.

  \item \textbf{High $p_T$ emissions according to matrix elements:} The parton 
  shower is valid in the soft and collinear regimes, the description of
  high $p_T$ emissions is, therefore, not reliable. 
  On the other hand, NLO calculations are valid in 
  this region. It is desirable to recover the NLO prediction for high $p_T$ 
  emissions despite applying the parton shower.

  \item \textbf{Connection to low energy physics:} The parton evolution 
  down to a low scale allows to include non-perturbative phenomena due to 
  hadronization and multiple interactions.
\end{description}
%
%
%------------------------X------------------------X------------------------X---%
\subsection{The quantum density matrix at next-to-leading order}
%------------------------X------------------------X------------------------X---%

For a generic $2\to m$ process at NLO, one can write the quantum density 
matrix in a perturbative expansion in $\alpha_s$, according to
\begin{equation}
  |\rho) = \underbrace{|\rho_m^{(0)})}_{\text{Born} , \; \mathcal{O}(1)} + 
  \underbrace{|\rho_m^{(1)})}_{\text{Virtual}, \; \mathcal{O}(\alpha_s)} + 
  \underbrace{|\rho_{m+1}^{(0)})}_{\text{Real}, \; \mathcal{O}(\alpha_s)} +
  \mathcal{O}(\alpha_s^2)\;.
  \label{rho_defNLO}
\end{equation}
Note that we count the leading order contribution as order $1$ in the strong
coupling $\alpha_s$. $|\rho_m^{(0)})$ and $|\rho_{m+1}^{(0)})$ are tree
level matrix elements, whereas  $|\rho_m^{(1)})$ is the one-loop amplitude. 
The definitions of these densities are analogous to the definition given in Eq. 
\eqref{rho_def}. So far Eq.~\eqref{rho_defNLO} suffers from infrared 
divergences, however, the expansion of $|\rho)$ in $\alpha_s$ is defined within 
dimensional regularization. Based on this quantum density matrix, the 
expectation value of the observable $F$ including shower effects reads 
\begin{equation}
  \sigma [F]^{PS}  = (F|U(t_F,t_0)|\rho) = \sum_{\lambda = m}^{\infty} 
  \frac{1}{\lambda!} \int [d\Phi_\lambda] (F|\Phi_\lambda)
  (\Phi_\lambda|U(t_F,t_0)|\rho)\;,
  \label{NLOobs}
\end{equation}
where $\Phi_\lambda = \{p,f,s^\prime, c^\prime, s,c\}_\lambda$. The quantum
density $|\rho)$ accounts for the hard matrix elements for $\lambda = m$ and
$\lambda =m+1$. This naive description of the cross section suffers 
from double counting, as we will show using the iterative solution to the
evolution equation expanded to $\mathcal{O}(\alpha_s)$. 
Indeed, there is
\begin{equation}
  |\rho(t_F))= U(t_F,t_0)|\rho) \approx |\rho) + \int_{t_0}^{t_F} d\tau 
  \left[ \mathcal{H}_I(\tau) - \mathcal{V}(\tau)\right]|\rho_m^{(0)})+
  \mathcal{O}(\alpha_s^2)\;.
  \label{PSexpand}
\end{equation}
As we can see from the unitarity condition $(1|\left[ \mathcal{H}_I(\tau) - 
\mathcal{V}(\tau)\right] = 0$, the total cross section $(1|\rho(t_F))$ is 
conserved. On the other hand, one does not recover the NLO 
prediction for inclusive observables $F$ because in general 
$(F|\left[ \mathcal{H}_I(\tau) - \mathcal{V}
(\tau)\right] \neq 0$. Even without this requirement the result is not
correct, since it contains the first emission contributions twice, once from
the real emission quantum density $|\rho^{(0)}_{m+1})$, and once from the parton 
shower approximation $\mathcal{H}_I(\tau)|\rho^{(0)}_m)$. We will now
show how to overcome this problem.
%
%
%------------------------X------------------------X------------------------X---%
\subsection{Matching fully inclusive processes to parton shower}
\label{sec:matchinginclusive}
%------------------------X------------------------X------------------------X---%

Let us begin with a class of processes, which have a well defined total cross 
section at leading order, e.g. $pp \to t\bar{t}$ or $pp \to 
W^+W^-$. Working along the line of the \mcnlo{} scheme,
we notice that the additional parton shower contribution
in Eq.~\eqref{PSexpand} can be cancelled by including appropriate 
counterterms. Let us, therefore, introduce the following modified quantum 
density matrix
\begin{equation}
  |\bar{\rho}) \equiv |\rho) - \int_{t_0}^{t_F} d\tau \left[ \mathcal{H}_I
  (\tau) - \mathcal{V}(\tau)\right]|\rho_m^{(0)})+ \mathcal{O}(\alpha_s^2)
  \;.
  \label{RhoBar}
\end{equation}
The total cross section $(1|\bar{\rho}) = (1|\rho) = 
\sigma^{NLO}$ is unchanged, due to the unitarity condition. On the other hand,
considering $U(t_F,t_0)|\bar{\rho})$ and expanding the evolution equation shows
that the undesired parton shower contributions are cancelled up to 
$\mathcal{O}(\alpha_s)$. This cancellation is non-trivial, because the 
modified quantum density matrix $|\bar{\rho})$, now depends explicitly on the 
parton shower splitting kernels, the momentum mappings, the ordering of the 
emissions and the choice of the starting shower time $t_0$.

For an infrared safe observable $F$, we have
\begin{equation}
\begin{split}
  \bar{\sigma} [F] &=\frac{1}{m!}\int  [d\Phi_m] (F|U(t_F,t_0) |\Phi_m) 
  \left[  (\Phi_m|\rho_m^{(0)})  + (\Phi_m|\rho_m^{(1)}) + \int_{t_0}^{t_F} 
  d\tau (\Phi_m|\mathcal{V}(\tau)|\rho_m^{(0)})	 \right]  \\
  &+\frac{1}{(m+1)!} \int [d\Phi_{m+1}] (F|U(t_F,t_0) |\Phi_{m+1}) \left[
   (\Phi_{m+1}|\rho_{m+1}^{(0)}) - \int_{t_0}^{t_F} d\tau 
   (\Phi_{m+1}|\mathcal{H}_I(\tau)|\rho_m^{(0)}) \right]\;.
\end{split}
  \label{XsecMatched}
\end{equation}
In the \mcnlo{} approach, the parton shower splitting kernels are used to 
provide subtraction terms for the infrared divergences in $|\rho^{(1)}_m)$ and
$|\rho^{(0)}_{m+1})$. Thus, the infrared cutoff $t_F$ may be removed by 
taking the 
limit $t_F \to \infty$. Even though this is a source of a mismatch between 
the fixed order and the shower calculation, it is numerically small due to
the exponential damping by the Sudakov form factor as discussed in 
Ref.~\cite{Sjostrand:1985xi}. In the real subtracted cross section, 
described in the 
second line of Eq.~\eqref{XsecMatched}, we use the definition of the real 
splitting operator from Eq.~\eqref{def:realsplittingoperator} and write
\begin{equation}
  \int_{t_0}^\infty d\tau~\mathcal{H}_I(\tau) = \sum_l \mathbf{S}_l
  \int_0^\infty d\tau~\delta(\tau-t_l)\Theta(\tau - t_0) 
  = \sum_l \mathbf{S}_l \Theta(t_l - t_0)\;.
  \label{H_sum}
\end{equation}
Here the sum runs over all external legs and $\mathbf{S}_l$ is the total 
splitting kernel for a given external leg $l$. We want to emphasize that 
$\mathbf{S}_l$ also contains non-singular contributions like the massive 
$g \to Q\bar{Q}$ splitting. Therefore, the operator $\mathcal{H}_I(t)$ contains
more than is needed for a subtraction scheme. The parameter $t_l$ is the shower 
time defined in Section~\ref{subsec:time}. Hence, $\Theta(t_l-t_0)$ represents 
the ordering of the emissions and the $t_0$ dependence introduces a dynamical 
restriction of the subtraction phase space. The choice of $t_0$ will be 
discussed at the end of this chapter, 
because it has non-trivial consequences. The real 
subtracted cross section is finite in $d=4$ dimensions, as $t_l$ is allowed 
to approach infinity.

Integrating the virtual operator $\mathcal{V}(\tau)$ without an infrared cutoff 
is more complex, considering that there is an explicit integration over the 
splitting variables. Hence, we have to integrate this part in $d=4-2\epsilon$ 
dimensions analytically to extract the $1/\epsilon^2$ and $1/\epsilon$ poles. 
The integrated virtual operator takes the form
\begin{equation}
  \int_{t_0}^\infty d\tau~ \mathcal{V}(\tau) = \sum_l \int d\Gamma_l\; 
  \mathbf{S}_l \Theta(t_l-t_0) \equiv \mathbf{I}(t_0) + \mathbf{K}(t_0)\;,
  \label{Vtheta}
\end{equation}
where $d\Gamma_l$ is the phase space integration of the additional parton.
The decomposition of the integrated $\mathcal{V}(\tau)$ into two operators 
$\mathbf{I}(t_0)$ and $\mathbf{K}(t_0)$ is arbitrary. However, we choose 
$\mathbf{I}(t_0)$ to match the divergencies of the virtual amplitude, 
as it is customary. We emphasize this 
structure to illustrate that the parton shower naturally incorporates a 
subtraction scheme similar to the Catani-Seymour 
framework~\cite{Catani:1996vz,Catani:2002hc}. 
In the case of initial state partons it is necessary to include additional
collinear counterterms, denoted by $\mathbf{P}$, needed for the 
renormalization of the parton distribution functions. 
Thus, the matched cross section reads
\begin{equation}
\begin{split}
  \bar{\sigma} [F] &=\int \frac{[d\Phi_m]}{m!}(F|U(t_F,t_0) |\Phi_m)\left[ 
  (\Phi_m|\rho_m^{(0)}) + (\Phi_m|\rho_{m}^{(1)})+(\Phi_m|[\mathbf{I}(t_0) + 
  \mathbf{K}(t_0) + \mathbf{P}]|\rho_m^{(0)})	 \right] \\
  &+ \int \frac{[d\Phi_{m+1}]}{(m+1)!}  (F|U(t_F,t_0) |\Phi_{m+1}) \left[ 
  (\Phi_{m+1}|\rho_{m+1}^{(0)}) -  \sum_l (\Phi_{m+1}|\mathbf{S}_l
  |\rho_m^{(0)}) \Theta(t_l-t_0)\right]\;.
\end{split}
  \label{Masterformula}
\end{equation}
For future reference, let us define the shorthands, as in 
Ref.~\cite{Frixione:2002ik}:
\begin{align}
	&(\Phi_m|S) \equiv (\Phi_m|\rho_m^{(0)}) + (\Phi_m|\rho_{m}^{(1)})  + 
(\Phi_m|[\mathbf{I}(t_0) + \mathbf{K}(t_0) + \mathbf{P}]|\rho_m^{(0)})\;, 
\label{Sevent0}\\
	&(\Phi_{m+1}|H) \equiv (\Phi_{m+1}|\rho_{m+1}^{(0)}) -  \sum_l 
(\Phi_{m+1}|\mathbf{S}_l|\rho_m^{(0)}) \Theta(t_l-t_0)\;.
\label{Hevent0}
\end{align}
The total cross section is then given by
\begin{equation}
  \bar{\sigma}^{NLO} [1] =\frac{1}{m!}\int [d\Phi_m](1|\Phi_m) (\Phi_m|S) + 
  \frac{1}{(m+1)!} \int [d\Phi_{m+1}]  (1|\Phi_{m+1}) (\Phi_{m+1}|H)\;,
  \label{GenNLOSample}
\end{equation}
whereas including parton shower evolution amounts to the integrals
\begin{equation}
\begin{split}
  \bar{\sigma} [F]^{PS} &=\frac{1}{m!}\int [d\Phi_m]  (F|U(t_F,t_0)|\Phi_m) 
  (\Phi_m|S) \\
  &+ \frac{1}{(m+1)!} \int [d\Phi_{m+1}]  (F|U(t_F,t_0)|\Phi_{m+1}) 
  (\Phi_{m+1}|H)\;.
\end{split}
  \label{Finaleventgeneration}
\end{equation}
Matching is a two step procedure, which consists of first generating the samples
according to Eq.~\eqref{Sevent0}~and~\eqref{Hevent0}, followed by the 
application  of $U(t_F,t_0)$.
%
%

%------------------------X------------------------X------------------------X---%
\subsection{Matching in the presence of singularities in the born approximation}
\label{subsec:jetxsec}
%------------------------X------------------------X------------------------X---%

For processes with massless partons at leading order, the matching prescription
as described in Eq.~\eqref{Masterformula} must be slightly modified by the 
inclusion of generation cuts, as discussed for example in 
Refs.~\cite{Frederix:2011ig,Alwall:2014hca}. A naive modification to the 
matching prescription would be to make the following replacements
\begin{align}
 \label{Sevent}
 (\Phi_m|S) &\to (\Phi_m|S)F_I(\{\hat{p},\hat{f}\}_m)\;, \\
 \label{Hevent}
 (\Phi_{m+1}|H) &\to (\Phi_{m+1}|H)F_I(\{p,f\}_{m+1})\;,
\end{align}
where $F_I(\{p,f\}_\lambda)$ is a jet function applied during the
generation of events, on the momenta and flavours of $\Phi_\lambda$.
Applying the parton shower to these ensembles shows that double
counting is not removed. Indeed, substituting Eq.~\eqref{Sevent} and
Eq.~\eqref{Hevent} in  Eq.~\eqref{Finaleventgeneration} we obtain
\begin{equation}
\begin{split}
  \bar{\sigma}[F]^{PS} &= \frac{1}{m!}\int [d\Phi_m] (F|U(t_F,t_0)|\Phi_m)
  (\Phi_m|S)F_I(\{\hat{p},\hat{f}\}_m) \\
  &+\frac{1}{(m+1)!}\int [d\Phi_{m+1}] (F|U(t_F,t_0)|\Phi_{m+1})(\Phi_{m+1}
  |H)F_I(\{p,f\}_{m+1})\;.
\end{split}
\end{equation}
Expanding the evolution operator, as given by Eq.~\eqref{PSexpand}, yields
\begin{equation}
\begin{split}
  \bar{\sigma}[F]^{PS} 
  &\approx\frac{1}{m!}\int [d\Phi_m](F|\Phi_m)(\Phi_m|\left[|\rho_m^{(0)})
  +|\rho_m^{(1)}) +\mathbf{P}|\rho^{(0)}_m) \right]
  F_I(\{\hat{p},\hat{f}\}_m) \\
  &+\frac{1}{(m+1)!} \int [d\Phi_{m+1}] (F|\Phi_{m+1})(\Phi_{m+1}|\rho^{(0)}
  _{m+1})F_I(\{p,f\}_{m+1}) \\
  &+\int \frac{[d\Phi_m]}{m!}\frac{[d\Phi_{m+1}]}{(m+1)!}\int_{t_0}^{t_F}
  d\tau~(F|\Phi_{m+1})(\Phi_{m+1}|\mathcal{H}_I(\tau)|\Phi_m)\\
  &\quad\times(\Phi_m|\rho^{(0)}_m) \Big[F_I(\{\hat{p},\hat{f}\}_m)-
  F_I(\{p,f\}_{m+1})\Big] + \mathcal{O}(\alpha_s^2)
  \;,
  \label{naivejetmatching}
\end{split}
\end{equation}
where the $\mathbf{I}(t_0)+\mathbf{K}(t_0)$ contribution of $(\Phi_m|S)$ has 
been cancelled by the linear expansion of the Sudakov form factor. 
The offending term is present in the 3rd and 4th lines of 
Eq.~\eqref{naivejetmatching}.
It does not vanish because $F_I(\{\hat{p},\hat{f}\}_m)$ is not equal to
$F_I(\{p,f\}_{m+1})$ for non-singular configurations, despite the fact that
their momenta and flavours are related according to 
$\{p,f\}_{m+1} = R_l(\{\hat{p},\hat{f}\}_m,\Gamma_l,\chi_l)$.
The mismatch can be cured by 
enforcing the subtraction terms to fulfill $F_I(\{\hat{p},\hat{f}\}_m)$, i.e. 
by modifying the real subtracted cross section according to
\begin{multline}
  (\Phi_{m+1}|H) \longrightarrow \\[0.2cm] (\Phi_{m+1}|\tilde{H}) 
  \equiv (\Phi_{m+1}|\rho_{m+1}^{(0)}) -  \sum_l 
  (\Phi_{m+1}|\mathbf{S}_l|\rho_m^{(0)}) \Theta(t_l-t_0)
  F_I(Q_l(\{p,f\}_{m+1}))\;,
  \label{modSub}
\end{multline}
where we make use of the inverse momentum mapping $Q_l$, as defined in 
Eq.~\eqref{MomentumOperatorInverse}. Thus,
\begin{equation}
  F_I(Q_l(\{p,f\}_{m+1})) = F_I(\{\hat{p},\hat{f}\}_m)\;.
\end{equation}
This modification allows us to compute cross sections with massless partons, at 
the same time introducing restrictions on the functional form of $F_I$. 
Using Eq.~\eqref{modSub} in addition to Eq.~\eqref{Hevent} in the definition of 
the cross section in Eq.~\eqref{Finaleventgeneration} and expanding the shower
evolution yields
\begin{equation}
\begin{split}
  \bar{\sigma}[F]^{PS} 
  &\approx\frac{1}{m!}\int [d\Phi_m](F|\Phi_m)(\Phi_m|\left[|\rho_m^{(0)})
  +|\rho_m^{(1)}) + \mathbf{P}|\rho^{(0)}_m)\right]
  F_I(\{\hat{p},\hat{f}\}_m) \\
  &+\frac{1}{(m+1)!} \int [d\Phi_{m+1}] (F|\Phi_{m+1})(\Phi_{m+1}|\rho^{(0)}
  _{m+1})F_I(\{p,f\}_{m+1}) \\
  &+\int \frac{[d\Phi_m]}{m!}\frac{[d\Phi_{m+1}]}{(m+1)!}\int_{t_0}^{t_F}
  d\tau~(F|\Phi_{m+1})(\Phi_{m+1}|\mathcal{H}_I(\tau)|\Phi_m)\\
  &\quad\times(\Phi_m|\rho^{(0)}_m) \Big[1-F_I(\{p,f\}_{m+1})\Big]
  F_I(\{\hat{p},\hat{f}\}_m)+\mathcal{O}(\alpha_s^2)
  \;. 
  \label{good_matching}
\end{split}
\end{equation}
The double counting is removed from Eq.~\eqref{good_matching}, 
if the following condition is satisfied
\begin{equation}
  \Big[1-F_I(\{p,f\}_{m+1})\Big]F(\{p,f\}_{m+1})=0\;,
  \label{mismatch}
\end{equation}
where we have used the fact that $(F|\Phi_{m+1}) \sim F(\{p,f\}_{m+1})$.
This is achieved if
\begin{equation}
  F_I(\{p,f\}_{m+1}) = 1\;, \text{ for } F(\{p,f\}_{m+1})\neq 0 \;,
\end{equation}
and can be understood as the generation cuts $F_I(\{p,f\}_{m+1})$ being more
inclusive than cuts that are applied on the final observable 
$F(\{p,f\}_{m+1})$.

Our previous discussion of double counting relied on the fact that the 
splitting functions were the same in the NLO subtraction scheme 
and in the parton shower.
However, in the case of initial state splittings, this is not the case due
to mass effects and the presence of different PDFs. 
The parton shower PDFs are evolved as explained
in Section~\ref{subsec:backwardevolution} using modified leading order
splitting kernels. On the other hand, the fixed order calculation employs
NLO PDFs. We point out that the implied mismatch is of order 
$\mathcal{O}(\alpha_s^2)$ as long as both PDF sets are equal at some scale.
This means that we use an NLO PDF parameterization as an input for the parton 
shower PDF evolution. A second source of mismatch is the treatment of 
mass effects in the initial state. In the case of the parton shower, 
the splitting kernels are modified as explained 
in Section~\ref{subsec:backwardevolution}, 
whereas the specifics of the treatment of mass thresholds depends on 
the PDF collaboration~\cite{Thorne:2014toa}. 

Finally, let us note that a last source of mismatch is due to scales in the 
strong coupling constant in the subtraction terms and in the shower. However, 
just like in the case of PDF evolution, it is of higher order.
%
%
%------------------------X------------------------X------------------------X---%
\subsection{On-shell projection}
\label{subsec:onshell}
%------------------------X------------------------X------------------------X---%

As mentioned before, the Nagy-Soper parton shower allows for massive initial 
state partons and treats the charm and bottom quarks as massive. In an 
NLO calculation the charm quark is usually assumed massless, whereas the bottom 
quark is only present in the initial state if it is massless as well, which 
corresponds to the 5-flavour scheme. The practical application of the matching 
procedure requires, therefore, an on-shell projection for charm 
and bottom quarks. We use different procedures in the initial and 
final state cases.

\subsubsection*{Initial state quarks}
We proceed iteratively by making one quark massive at a time.
For definiteness, we will modify $p_a$ which is assumed to have $p_a^2 = 0$,
whereas the mass, $m_b$, of $p_b$ is arbitrary. 
The total momentum is given by
\begin{equation}
  Q = p_a + p_b\;,
\end{equation}
where
\begin{equation}
  p_a = \eta_a p_A\;, \qquad p_b = \eta_bp_B + \frac{m^2_b}{\eta_bs}p_A\;,
\end{equation}
with $\eta_a$ and $\eta_b$ the momentum fractions, $p_{A/B}$ the 
hadron momenta and $s=2(p_A\cdot p_B)$ the hadronic center of mass energy 
squared. This can be written as
\begin{equation}
  Q = \left(\eta_a + \frac{m^2_b}{\eta_bs}\right)p_A + \eta_bp_B\;, \qquad 
  Q^2 = s\eta_a\eta_b + m^2_b\;.
\end{equation}
We shall determine $\hat{\eta}_a$, such that 
\begin{equation}
  \hat{p}_a = \hat{\eta}_a p_A + \frac{m^2_a}{\hat{\eta}_a s}p_B\;,
  \quad \hat{p}_a^2 = m_a^2 \neq 0\;.
\end{equation}
The new total momentum $\hat{Q}=\hat{p}_a+p_b$ and its invariant mass 
then read 
\begin{equation}
\begin{split}
  \hat{Q} &= \left(\hat{\eta}_a + \frac{m^2_b}{\eta_bs}\right)p_A + 
  \left(\eta_b+\frac{m^2_a}{\hat{\eta}_as}\right)p_B\;, \\
  \hat{Q}^2 &= s\eta_b\hat{\eta}_a + \frac{m^2_b m^2_a}{s\eta_b\hat{
  \eta}_a} +m^2_b+m^2_a\;.
\end{split}
\end{equation}
The requirement, $\hat{Q}^2 = Q^2$, yields 
\begin{equation}
  \hat{\eta}_a = \eta_a\left(\frac{1}{2} - \frac{m^2_a}{2Q^2} + 
  \frac{\sqrt{(m^2_a-Q^2)^2-4m^2_am^2_b}}{2Q^2}\right)\;.
\end{equation}
To ensure total momentum conservation, the complete final state has to be 
boosted in the $z$-direction. The rapidity of the boost, $\omega$, is
\begin{equation}
  e^\omega = \frac{Q^2+m^2_b-m^2_a}{2m^2_b} - 
  \frac{\sqrt{\left(Q^2+m^2_b-m^2_a\right)^2-4Q^2m^2_b}}{2m^2_b}\;,
  \label{eWmamb}
\end{equation}
which for $m_b=0$ reduces to
\begin{equation}
  e^\omega = \frac{Q^2}{Q^2-m^2_a}\;.
  \label{eWma}
\end{equation}

\subsection*{Final state quarks}
In the case of a final state massless quark with momentum $p_l$, 
we exploit the momentum mapping procedure of the Nagy-Soper 
parton shower to generate a non-zero mass $m_l$. The new momentum is given by
\begin{equation}
  \hat{p}_l = \lambda p_l + \frac{1-\lambda+y}{2a_l}Q\;,
  \qquad \hat{p}_l^2 =m^2_l\;,
\end{equation}
where $Q$ is the total final state momentum and
\begin{equation}
  a_l = \frac{Q^2}{2 p_l \cdot Q}\;, \qquad b_l = \frac{m^2_l}{2 p_l\cdot Q}
  \;.
\end{equation}
The parameter $y$ is determined by requiring that the invariant mass of 
the total recoiling momentum, $K=Q-p_l$, be preserved, i.e. $K^2=\hat{K}^2$,
with $\hat{K}=Q-\hat{p}_l$. The result is
\begin{equation}
  y=b_l\;.
\end{equation}
As a consequence, $K$ and $\hat{K}$ are connected 
by a boost~\cite{Nagy:2007ty} given by
\begin{equation}
  \mathcal{B}^{\mu\nu} = g^{\mu\nu} - 2 \frac{(K+\hat{K})^\mu(K+
  \hat{K})^\nu} {(K+\hat{K})^2} + 2\frac{\hat{K}^\mu K^\nu}{K^2}\;.
  \label{CSboost}
\end{equation}
The parameter $\lambda$ is determined by the on-shell condition 
$\hat{p}_l^2 = m_l^2$, which yields 
\begin{equation}
  \lambda = \sqrt{(1+y)^2-4a_lb_l}\;,
%  \label{lambdadef}
\end{equation}
which is well-defined if
\begin{equation}
  p_l \cdot Q > m_l \bigg( \sqrt{Q^2} - \frac{m_l}{2} \bigg) \;.
\end{equation} 
In the center-of-mass frame of $Q$, this becomes
\begin{equation}
  E_l > m_l \bigg( 1 - \frac{m_l}{2\sqrt{Q^2}} \bigg) \;.
\end{equation} 
Clearly, arbitrarily soft quarks cannot be projected on the mass-shell. 

In case of $2\to 2$ processes, $a_l =1$ and the 
boost in Eq.~\eqref{CSboost} is singular. In that case one can use an 
alternative boost~\cite{Nagy:2007ty}
\begin{equation}
  \mathcal{B}^{\mu\nu} = g^{\mu\nu} + \left(\frac{K\cdot n}{\hat{K}\cdot 
  n}-1\right)n^\mu \bar{n}^\nu + \left(\frac{\hat{K}\cdot n}{K\cdot 
  n}-1\right)\bar{n}^\mu n^\nu\;,
\end{equation}
with 
\begin{equation}
  n = \sqrt{\frac{2}{Q^2}}\;p_l\;,\qquad
  \bar{n} =\sqrt{\frac{2}{Q^2}}\;\big(Q-p_l\big)\;.
\end{equation}
%
%
%------------------------X------------------------X------------------------X---%
\subsection{Initial conditions for the parton shower}
\label{subsec:initime}
%------------------------X------------------------X------------------------X---%

As described in Section~\ref{subsec:time}, the shower evolution emissions are 
strongly ordered in the parameter $\Lambda^2_l$, which is related to the shower 
time $t$ as follows,
\begin{equation}
  e^{-t} = \frac{\Lambda^2_l}{Q_0^2}\;, \qquad \Lambda^2_l = 
  \frac{|(\hat{p}_l\pm \hat{p}_{m+1})^2-m^2_l|}{2p_l\cdot Q_0} Q_0^2\;,
  \label{timel}
\end{equation}
where $p_l$ is the emitter momentum and $Q_0$ is the total final state
momentum  before emission. The emitter momentum after emission is
$\hat{p}_l$, while  $\hat{p}_{m+1}$ is the momentum of the emitted
particle. The parton shower evolution starts at an initial time $t_0$,
which has to be determined by the user. Since the choice of $t_0$ has
a noticable influence on the differential distributions, it is
necessary to provide a sensible prescription, which is the  purpose of
this Section.

Consider first a process at leading order, e.g. $pp \to t\bar{t}$.
The parton shower will generate additional radiation. However, the
resulting hard  jets are described very poorly, because the parton
shower description is only  valid close to infrared limits. As a
consequence, the $p_T$ distribution of the  top quark pair is not
reliably reproduced for large values of the transverse momentum.  Since
the initial time $t_0$ restricts the available phase space for parton
emissions, it will have a strong influence on the distribution, as
reported,  for example, in Ref.~\cite{plehn:2005cq}. Notice that the
situation may be improved with merging of different multiplicity
samples, see Ref.~\cite{Mangano:2006rw}.

Matching at NLO has the advantage of providing one hard emission
exactly through the matrix element. According to our modified quantum
density matrix, we removed all parton shower contributions to the
first emission. Therefore, corrections to the $p_T$ distribution are
formally of higher order, even though they can be a priori
large~\cite{Hoeche:2011fd,Alioli:2008tz}.  The ultimate goal is to
choose $t_0$ such that  the description of the high $p_T$ regime is
close to that of the NLO fixed order approximation. This can be
achieved by restricting  the parton shower phase space to allow only
splittings with virtualities  lower than the splittings already
present in the hard matrix element.

We choose $t_0$ to approximate the virtuality of the hard process on
an event-by-event basis. Inspired by
Refs.~\cite{Frixione:2003ei,Hoeche:2014qda}, we define
\begin{equation}
  e^{-t_0} = \min_{i\neq j} \left\{ \frac{2p_i\cdot p_j}{Q_0^2} \right\}\;,
  \label{initime}
\end{equation}
where $p_i$ and $p_j$ are pairwise different external momenta of the process 
considered. 

For the real radiation contribution, we use the subtraction terms 
Eq.~\eqref{modSub} to determine $t_0$ as follows
\begin{enumerate}
\item For each subtraction term
\begin{equation}
(\Phi_{m+1}|\mathbf{S}_l|\rho_m^{(0)}) \Theta(t_l-t_0)
  F_I(Q_l(\{p,f\}_{m+1}))\;,
\end{equation}
determine its individual $t_0$ according to Eq.~\eqref{initime}
while $t_l$ according to Eq.~\eqref{timel}. Retain $t_l$ if $t_l > t_0$.
\item If there is at least one $t_l$, then evaluate $t_0$ according to 
\begin{equation}
  e^{-t_0} = \min_{l} \left( e^{-t_l} \right)\;,
\end{equation}
where the minimum runs over all $t_l$ found in the previous step.
\item Otherwise, apply Eq.~\eqref{initime} to the $(m+1)$-particle kinematics. 
\end{enumerate} 
Let us note that the separate determination of $t_0$ in step 1 for each 
subtraction term, is necessary to avoid double counting.
%
%
%------------------------X------------------------X------------------------X---%
\subsection{A summary of ambiguities}
%------------------------X------------------------X------------------------X---%

In this Section, we recapitulate by listing the intrinsic uncertainties 
introduced by the \mcnlo{} matching formalism in combination with the 
Nagy-Soper parton shower.
\begin{description}
  \item \textbf{Parton distribution functions:} Parton distribution functions 
  are evolved differently in the NLO calculation and in the parton shower. 
  Nevertheless, we pointed out in Section~\ref{subsec:jetxsec}, 
  that the evolution itself is of higher order.
  Thus, the NLO accuracy is obtained as long as the evolutions share a common 
  point, e.g. at the low scale.
  \item \textbf{Parton masses:} A generic feature of the Nagy-Soper parton 
  shower is that it requires massive bottom and charm quarks in the initial 
  state. However, the fixed order calculation involves massless charm quarks 
  at least. Although the implied mismatch is power suppressed, it is necessary
  to introduce masses for the relevant quarks in the fixed order sample used
  as input for the shower. In Section~\ref{subsec:onshell}, we have provided 
  a possible algorithm. Nonetheless, other choices might influence the final 
  result.
  \item \textbf{Initial state shower time:} The choice of $t_0$ in the parton 
  evolution is, to a large extent, arbitrary. The only requirement is that 
  the NLO prediction be recovered for hard emissions. This can be achieved by 
  several different choices of $t_0$, with either fixed or configuration 
  dependent values. In Section~\ref{subsec:initime}, we have presented our 
  choice, but others are possible. Nevertheless, it is recommended to vary 
  $t_0$ by rescaling it by a common value within some reasonable range, 
  see Section~\ref{sec:setup}. 
\end{description}
%

%------------------------X------------------------X------------------------X---%
\section{Implementation}
%------------------------X------------------------X------------------------X---%
%
In this Section, we present a realization of the matching scheme
within the \helacnlo{} framework~\cite{Bevilacqua:2011xh}. Due to the
restricted functionality of the Nagy-Soper  parton shower present in
\deductor{} version 1.0.0~\cite{Nagy:2014mqa},  which we used as basis
for practical studies, our implementation is simplified as far as spin
and colour treatments are concerned. Since \deductor{} uses spin
averaged  splitting functions, we only provide unpolarized event
samples for showering.  On the other hand, we supply only leading
colour events even though \deductor{} works with the LC+ approximation
and may shower non-diagonal colour  configurations. While proper
colour matching does not present conceptual  challenges, it requires
some additional programming effort. The latter will be  necessary once
\deductor{} has full colour functionality. We believe that, for a
first study, our simplification is justified. More details on our
approach can be found in Appendix~\ref{App:NLOapprox}. Finally, we
note that,  at present, \deductor{} lacks unstable particle decays and
a hadronization model.
\subsection{Modifications in \dipoles{}}
The implementation of the matching scheme in the \helacnlo{} multi-purpose 
event generator only concerns 
\dipoles{}~\cite{Czakon:2009ss,Bevilacqua:2013iha}. 
We base our work on the previous implementation~\cite{Bevilacqua:2013iha}, 
where the majority of the parton shower operators $\mathcal{H}_I$ and 
$\mathcal{V}$ have already been included.
%
%------------------------X------------------------X------------------------X---%
\subsubsection*{Momentum mapping for initial state splitting}
%------------------------X------------------------X------------------------X---%
%
The momentum mapping implemented in \dipoles{} is based on 
Ref.~\cite{Nagy:2007ty}. For reasons, which we shortly discussed in 
Section~\ref{subsec:Realsplittingoperator}, the mapping implemented in 
\deductor{} is slightly different~\cite{Nagy:2009vg}. Here, we reproduce the
necessary formulae used in the new version of \dipoles{}.

We start from a set of momenta $\{\tilde{p}\}_m$ and consider the splitting 
$\tilde{p}_a \to p_a + p_{m+1}$. The incoming momenta are given by
\begin{equation}
  \tilde{p}_a = \tilde{\eta}_a P_A\;, \qquad 
  \tilde{p}_b =\tilde{\eta}_b P_B\;,
\end{equation}
where $P_A$ and $P_B$ are the hadron momenta. After the splitting, there is
\begin{equation}
  p_a = \frac{1}{z}\tilde{p}_a\;, \quad p_b = \tilde{p}_b\;,
\end{equation}
and the momentum $p_{m+1}$ of the emitted particle is parametrized by
\begin{equation}
  p_{m+1} = x_a \tilde{p}_a + x_b \tilde{p}_b + k_\perp\;,
\end{equation}
where
\begin{equation}
  |k_\perp|^2 = 2x_a x_b (\tilde{p}_a \cdot \tilde{p}_b)\;,\quad
  x_a = \frac{1}{z} -1 -y\;,\quad
  x_b = zy\;.
\end{equation}
The phase space variables $y$, $z$ and $\phi$ are defined as
\begin{equation}
\begin{split}
  &z = \frac{\tilde{\eta}_a}{\eta_a} = \frac{s\tilde{\eta}_a 
  \tilde{\eta}_b}{s\eta_a \eta_b} = \frac{\tilde{Q}^2}{(p_a+p_b)^2}\;,\\
  &y = - \frac{(p_a - p_{m+1})^2}{2(\tilde{p}_a\cdot \tilde{p}_b)} = 
  \frac{2p_a\cdot p_{m+1}}{\tilde{Q}^2}\;,\\
  &\phi \in [0,2\pi]\;,
\end{split}
\end{equation}
where $\tilde{Q} = p_a+p_b -p_{m+1} = \tilde{p}_a + \tilde{p}_b$. To ensure 
momentum conservation the remaining final state particles have to be boosted,
$p_i = \mathcal{B} \tilde{p}_i$, with the Lorentz transformation
\cite{Nagy:2014nqa}
\begin{equation}
\begin{split}
  \mathcal{B}^{\mu\nu}(\omega,v_\perp) &= g_\perp^{\mu\nu} + 
  \frac{e^\omega \tilde{p}_a^\mu \tilde{p}_b^\nu + e^{-\omega} 
  \tilde{p}_b^\mu\tilde{p}_a^\nu}{\tilde{p}_a\cdot\tilde{p}_b} \\[0.2cm]
  &+ \sqrt{\frac{2}{\tilde{p}_a\cdot \tilde{p}_b}}\left[e^\omega v_\perp^\mu 
  \tilde{p}_b^\nu - \tilde{p}_b^\mu v_\perp^\nu \right] - e^\omega v_\perp
  ^2 \frac{\tilde{p}_b^\mu\tilde{p}_b^\nu}{\tilde{p}_a\cdot \tilde{p}_b}\;,
\end{split}
\label{NewBoost}
\end{equation}
where
\begin{equation}
\begin{split}
  &g_\perp^{\mu\nu} = g^{\mu\nu} - \frac{\tilde{p}_a^\mu \tilde{p}_b^\nu + 
  \tilde{p}_b^\mu\tilde{p}_a^\nu}{\tilde{p}_a \cdot \tilde{p}_b}\;,\\
  &e^\omega = \frac{1}{z} -x_a = 1+y\;,\\
  &v_\perp = - \frac{e^{-\omega}}{\sqrt{2\tilde{p}_a\cdot\tilde{p}_b}}
  k_\perp\;.
\end{split}
\end{equation}
In the case, where $\tilde{p}_b$ is the emitter one simply exchanges 
$\tilde{p}_a$ and $\tilde{p}_b$ in the equations above. This momentum mapping 
is implemented in the $\mathbf{I}$ and $\mathbf{KP}$ operators of the 
\dipoles{} package, while the modification only concerns initial-final state 
interference contributions. For the real subtracted cross section, the 
inverse transformation has to be applied
\begin{equation}
\begin{split}
  &\tilde{p}_a = z p_a\;, \quad
  \tilde{p}_b = p_b\;, \\
  &\tilde{p}_i = \mathcal{B}^{-1} p_i\quad (i=1,...m)\;,
\end{split}
\end{equation}
with
\begin{equation}
\begin{split}
  \left(\mathcal{B}^{-1}\right)^{\mu\nu}(\omega,v_\perp) &= g_\perp^{\mu\nu} + 
  \frac{e^{-\omega} \tilde{p}_a^\mu \tilde{p}_b^\nu + e^{\omega} 
  \tilde{p}_b^\mu\tilde{p}_a^\nu}{\tilde{p}_a\cdot\tilde{p}_b} \\[0.2cm]
  &+\sqrt{\frac{2}{\tilde{p}_a\cdot \tilde{p}_b}}\left[e^\omega\tilde{p}_b^
  \mu v_\perp^\nu  - v_\perp^\mu \tilde{p}_b^\nu  \right] - e^\omega 
  v_\perp^2 \frac{\tilde{p}_b^\mu\tilde{p}_b^\nu}{\tilde{p}_a\cdot 
  \tilde{p}_b}\;.
\end{split}
\end{equation}
%
%------------------------X------------------------X------------------------X---%
\subsubsection*{Dynamical phase space restriction}
%------------------------X------------------------X------------------------X---%
%
A second modification to the existing subtraction scheme is the implementation 
of a dynamical cutoff $\Theta(t_l - t_0)$ on the available dipole phase space, 
which represents the ordering of the emissions in the parton shower evolution.
The new subtraction terms read (see Section~\ref{sec:matchinginclusive})
\begin{equation} 
  \int_{t_0}^\infty d\tau~\mathcal{H}_I(\tau) = \sum_l \mathbf{S}_l\Theta
  (t_l-t_0)\;,
\end{equation}
where the splitting operator $\mathbf{S}_l$ is already available in 
\dipoles{}. The shower time $t_l$ is given by (see 
Section~\ref{subsec:initime})
\begin{equation}
  e^{-t_l}=\frac{|(\hat{p}_l \pm \hat{p}_{m+1})^2 -m^2_l|}{2p_l\cdot Q_0}\;,
\end{equation}
and can be reconstructed from the $\{\hat{p},\hat{f}\}_{m+1}$ kinematics. 
The numerator is given by real radiation momenta, whereas the denominator 
$2p_l\cdot Q_0$ is given by the momenta before the splitting. However, the
denominator can also be reconstructed by making explicit use of the momentum
mappings. 

This phase space restriction is also implemented in the integrated dipoles. 
In order to use as much existing code as possible, we implement it
in the form of
\begin{equation}
  \int_{t_0}^\infty d\tau~V(\tau) = \sum_l \int d\Gamma_l\; \mathbf{S}_l 
  \Theta(t_l-t_0) = \sum_l \int d\Gamma_l\; \mathbf{S}_l 
  \left[1-\Theta(t_0-t_l)\right]\;.
\end{equation}
Therefore, we can calculate the finite remainder in $d=4$ dimensions 
and subtract it from the complete expression. We follow closely the outlined 
semi-numerical strategy presented in~\cite{Bevilacqua:2013iha}.

We note that, even though the restriction is necessary for matching, it can
be also used in fixed order NLO calculation as an effective cutoff on the 
subtraction phase space similar to $\alpha_{\mathrm{max}}$ in the 
Catani-Seymour subtraction  scheme 
\cite{Frixione:1995ms,Nagy:1998bb,Nagy:2003tz}.
%

%------------------------X------------------------X------------------------X---%
\subsection{Monte Carlo techniques}
%------------------------X------------------------X------------------------X---%
%
With the modifications described above, it is possible to use \helacnlo{} to
generate an event sample ready for showering with \deductor{}.
We produce events subprocess by subprocess. 
First, we use \OneLoop{} \cite{vanHameren:2009dr} 
to obtain a set of unweighted leading order events
with the virtual contributions
\begin{equation}
  \omega_i(\{p,f\}_m) = 1 + \frac{(\{p,f\}_m|\rho^{(1)}_m)}{(\{p,f\}_m
  |\rho^{(0)}_m)}\;, 
\end{equation}
where only the finite part of $|\rho^{(1)}_m)$ is included. At this point
\begin{equation}
 \sigma[\text{LO} + \text{V}] = \frac{\sigma[\text{LO}]}{N}\sum_{i=1}^N w_i
 \;.
\end{equation}
This set of weighted events is subsequently reweighted using \dipoles{}
in order to include the parton shower virtual operator. 
This corresponds to taking into account the integrated subtraction terms.
The weights become
\begin{equation}
  \omega_i(\{p,f\}_m) = 1 + \frac{(\{p,f\}_m|\rho^{(1)}_m)}{(\{p,f\}_m
  |\rho^{(0)}_m)} + \frac{(\{p,f\}_m|\mathbf{I}(t_0) + \mathbf{K}(t_0)
  +\mathbf{P}|\rho^{(0)}_m)}{(\{p,f\}_m|\rho^{(0)}_m)}\;,
\end{equation}
yielding
\begin{equation}
 \sigma[\text{LO} + \text{V} + \text{I} + \text{KP}] 
 = \frac{\sigma[\text{LO}]}{N}\sum_{i=1}^N w_i
 \;.
\end{equation}
Notice that the integrated subtraction terms are Monte Carlo integrals
on the phase space $d\Gamma_l$ of the additional unresolved parton.
In order to obtain a good approximate of this integral,
we sample several points for a fixed born phase space point $\{p,f\}_m$. 
This proved to be advantageous in case of large cancellations in 
$\mathcal{V} (\tau)$. 

The real radiation events are generated separately with \dipoles{},
which was extended to provide unweighted events with positive and
negative weights,  $\pm 1$~\cite{Corcella:2001pi}. Notice that
unweighting is possible, because both the real radiation and the
respective subtraction weights correspond to the same phase space
point as described in Section~\ref{sec:matchinginclusive}.  For each
accepted event, we choose the most probable diagonal colour flow  with
the colour weight
\begin{equation}
  C_w(\{p,f,c\}_{m+1}) = \frac{(\{p,f,c\}_{m+1}|\rho^{(0)}_{m+1})}
  {\sum_{\{c^\prime\}_{m+1}}(\{p,f,c^\prime\}_{m+1}
  |\rho^{(0)}_{m+1})}\;.
  \label{colorweight}
\end{equation}
The generated events are stored in a Les Houches
file~\cite{Alwall:2006yp}, with  the initial parton shower time $t_0$
assigned to the variable \texttt{SCALEUP}.

%
%
%------------------------X------------------------X------------------------X---%
\subsection{Interface to \deductor{}}
%------------------------X------------------------X------------------------X---%
%
Events generated by \helacnlo{} are transferred to \deductor{}. This requires
an on-shell projection for charm and bottom quarks 
(see Section~\ref{subsec:onshell}) and the determination of a starting colour
configuration for each event.
%
%------------------------X------------------------X------------------------X---%
\subsubsection*{On-shell projection}
%------------------------X------------------------X------------------------X---%
%
The on-shell projection of Section~\ref{subsec:onshell} is applied iteratively
to each charm and bottom quark. In the case of an initial state transformation 
for two massless quarks bound to become massive, the order of longitudinal 
boost is relevant. Indeed, the first boost has a lower rapidity than the 
second as can be proven by inspection of Eqs.~\eqref{eWmamb}~and~\eqref{eWma}.
We choose the order at random in order to reduce the systematics.
In case the projection fails, which is only possible for soft bottom and charm 
quarks in the real radiation contribution, the event is rejected. 
Vetoing such emissions may only induce a negligible modification of the cross
section. Otherwise, effects, which are not under control, would be substantial.
This problem has to be studied case-by-case.
%
%------------------------X------------------------X------------------------X---%
\subsubsection*{Colour configurations}
%------------------------X------------------------X------------------------X---%
A colour flow generated as described before in the Les 
Houches~\cite{Alwall:2006yp} format must be translated to the internal 
representation of \deductor{} in terms of colour strings, see 
Section~\ref{subsec:confspace}.
Notice that only one colour flow is needed in the leading colour approximation
per event. A colour flow is given by a list of two colour indices.
Thus, the $i$-th particle carries the following pair of indices
\begin{equation}
\begin{split}
  &q(i) \to (\texttt{colour1[i]}\;,\;0) \\
  &\bar{q}(i) \to (0\;,\;\texttt{colour2[i]}) \\
  &g(i) \to (\texttt{colour1[i]}\;,\;\texttt{colour2[i]})\;.
\end{split}
\end{equation}
In \deductor{}, on the other hand, quarks, anti-quarks and gluons are 
represented by
\begin{equation}
\begin{split}
  &q(i) \to [\texttt{Q}\;,\;\texttt{next[i]}) \\
  &\bar{q}(i) \to (\texttt{prev[i]}\;,\;\texttt{A}]\\
  &g(i) \to (\texttt{prev[i]}\;,\;\texttt{next[i]})\;,
\end{split}
\end{equation}
where \texttt{Q} denotes the beginning and \texttt{A} the end of an open
colour string. \texttt{prev[i]} refers to the particle index in the 
event, which is to the left of particle with index $i$. In the same way 
\texttt{next[i]} refers to the right partner on the colour string. 
The algorithm to translate colour flows into colour string configurations is
\begin{enumerate}
  \item Swap the flavour and the colour indices for initial state partons.
  \item Enumerate all partons starting from $-1$, where $-1$ and $0$ are
  reserved for the initial state partons. Colour-neutral particles are 
  enumerated with numbers less than $-1$.
  \item Iterate over all particles with index $i$ applying
  \begin{enumerate}
  \item If (\texttt{colour1[i]}$\neq 0$ and \texttt{colour2[i]} $=0$) then 
        \texttt{prev[i]} = \texttt{Q}.\\
        If (\texttt{colour1[i]}$= 0$ and \texttt{colour2[i]} $\neq 0$) then 
        \texttt{next[i]} = \texttt{A}.
  \item Iterate over all particles with index $k\neq i$ applying\\
  If (\texttt{colour1[i]}$\neq 0$ and \texttt{colour1[i]}$=$
  \texttt{colour2}[k]) then \texttt{next[i]}=k; \texttt{prev[k]}=i.
  \end{enumerate}
\end{enumerate}

%------------------------X------------------------X------------------------X---%
\section{$t\bar{t} j$ production at the LHC with next-to-leading order
matching} 
%------------------------X------------------------X------------------------X---%
In this Section, we present results for $pp\to t\bar{t}j+X$ production
at  next-to-leading order,  obtained with  \helacnlo{}, 
 matched with the Nagy-Soper parton shower as
implemented in \deductor{}. The NLO QCD corrections to the considered
process have been  previously computed in
Ref.~\cite{Dittmaier:2007wz,Dittmaier:2008uj,Melnikov:2010iu}.
Matching to a parton shower has been first considered in
Ref.~\cite{Kardos:2011qa,Alioli:2011as} using the \powheg{} method
\cite{Nason:2004rx,Alioli:2010xd}. 
%
%------------------------X------------------------X------------------------X---%
\subsection{Setup}
\label{sec:setup}
%------------------------X------------------------X------------------------X---%

The results for $t\bar{t} j$ production are presented for $pp$
collisions at the  LHC with a center-of-mass energy of 8 TeV. The top
quark is assumed to be  stable and its mass is set to $m_t = 173.5$
GeV, while the charm and bottom  quarks are considered to be massless
at fixed order. Results are obtained  using the \textsc{Mstw2008nlo}
PDF set~\cite{Martin:2009iq} with five active  flavours and the
corresponding two-loop running of the strong coupling. We set  the
renormalization and factorization scales to the top quark mass,  $
\mu_R = \mu_F = m_t $, and the starting shower time to
\begin{equation}
  e^{-t_0} = \min_{i\neq j}\left\{\frac{2 p_i \cdot p_j}
  {\mu_T^2 Q_0^2}\right\}\;,
  \label{eq:T0}
\end{equation}
where $p_i$ and $p_j$ are external momenta, $Q_0$ is the total final
state  momentum and $\mu_T=1$ for the central prediction as explained
below.  Partons with pseudorapidity $|\eta| < 5$ are clustered using
the anti-$k_T$  jet algorithm~\cite{Cacciari:2008gp}, with the
separation parameter $R=1$. The resulting jets are sorted in
decreasing order  of $p_T$. We require the tagged jets
to have transverse momentum of $p_T > 50$ GeV and rapidity in the
range of $|y| < 5$.  

We restrict our analysis to the perturbative parton shower evolution.
Decays of unstable particles, hadronization and multiple interactions
are not taken into account. The parton shower treats the charm and
bottom as massive particles with masses $m_c =1.4$ GeV and $m_b=4.75$
GeV.  We provide the \textsc{Mstw2008nlo} PDF set at $\mu_F=1$  GeV as
the starting point for the evolution in \deductor{}.  We also use the
corresponding two-loop running of $\alpha_s$,  and restrict the parton
shower to the leading colour approximation. Therefore, results 
presented in this section are accurate up to $\mathcal{O}(1/N_c^2)$.
\begin{figure}[t!]
\begin{center}
  \includegraphics[width=0.49\textwidth]{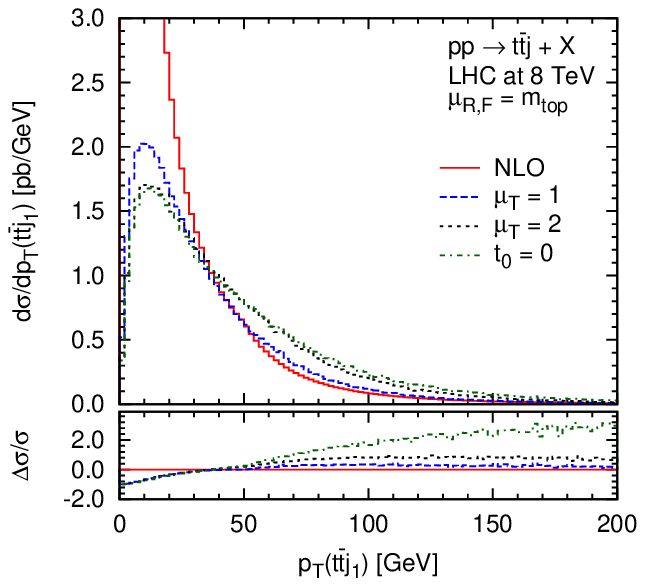}
  \includegraphics[width=0.49\textwidth]{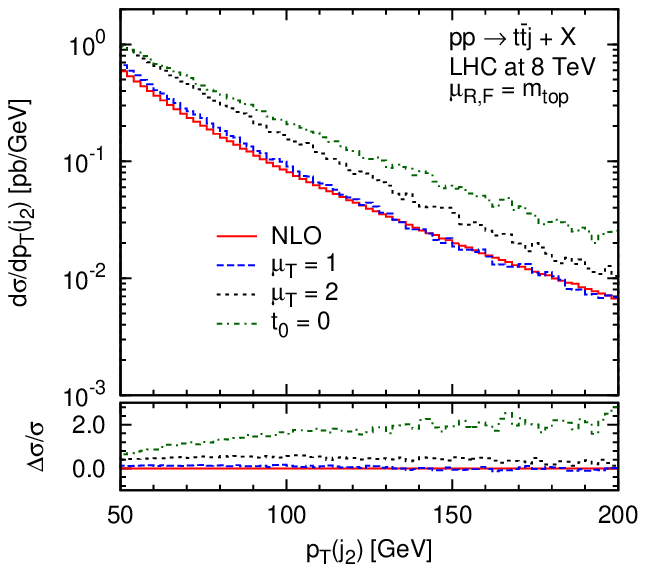}
\end{center}
  \caption{\textit{Differential cross section distributions as a
      function of the transverse momentum of the $t\bar{t}j_1$  system
      (left panel) and of the second hardest jet (right panel) for $pp
      \to t\bar{t}j+X$ at the LHC with $\sqrt{s} = 8$ TeV. Comparison between
      the NLO result obtained with \textsc{Helac-Nlo} and the results
      produced by matching NLO predictions to \textsc{Deductor} for
      three different choices of the starting time. 
      The scale choice is $\mu_F = \mu_R = m_t$. The
      lower panels display the relative  deviation from the fixed
      order result. Notice, that $t_0 = 0$ corresponds to unrestricted 
      shower radiation without relation to the kinematics of the underlying 
      event. This choice should not be used in practice and it is 
      shown for illustration purposes only.}}
  \label{fig:pTspectrum}
\end{figure}

In order to address the theoretical uncertainties, we investigate the
scale  dependence of cross sections and distributions on the
unphysical scales  $\mu_R$, $\mu_F$ and the rescaling parameter
$\mu_T$.  Here, $\mu_R$ is varied simultaneously with  $\mu_F$ between
$\mu_{R,F} = m_t/2$ and $\mu_{R,F} = 2m_t$.  On the other hand, both the
central value of $\mu_T$, $\mu_{T0}$, and its  variation range require
a more thorough discussion, because this parameter  is specific to the
Nagy-Soper parton shower.  Guidelines for a suitable choice of $t_0$
have already been presented in Section~\ref{subsec:initime}.  In
practice, we consider exclusive distributions, which are especially
sensitive to the parton shower effects, e.g. the transverse momentum,
$p_T$,  spectrum of the $t\bar{t}j_1$ system or of the second jet,
$j_2$, which are equivalent at NLO.  For large $p_T$ values, the fixed
order prediction is reliable and we would  not like the shower to
introduce substantial shape differences there.  This can be achieved
by a suitable choice of $\mu_{T0}$ as shown in
Fig.~\ref{fig:pTspectrum}, where the lower panel shows the relative
deviation from the fixed order result defined as
\begin{equation}
  \frac{\Delta \sigma}{\sigma} \equiv \frac{\sigma^\nlops-\sigma^\nlo}
  {\sigma^\nlo}\;.
\end{equation}

\begin{figure}[t!]
\begin{center}
  \includegraphics[width=0.49\textwidth]{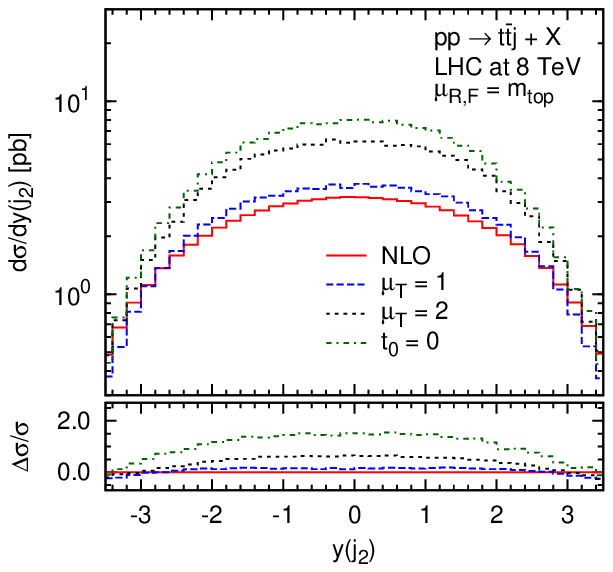}
  \includegraphics[width=0.49\textwidth]{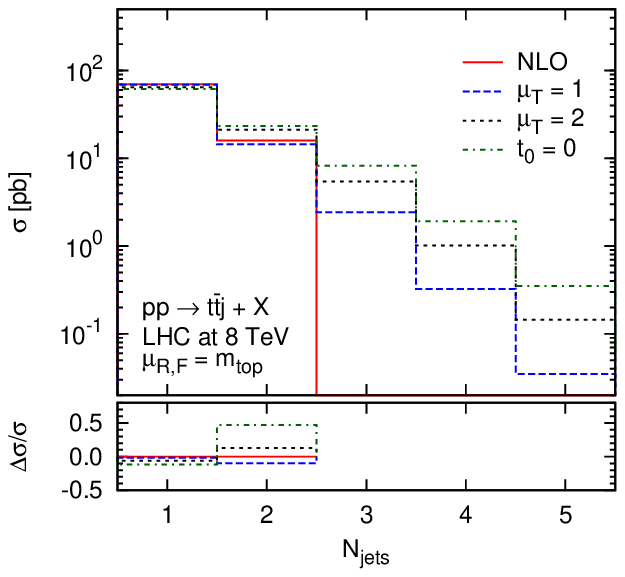}
\end{center}
  \caption{\textit{Differential cross section distribution as a
      function of the rapidity of the second hardest jet (left panel)
      and inclusive jet cross sections (right panel) for $pp \to t\bar{t}j+X$
      at the LHC with $\sqrt{s} = 8$ TeV. Comparison between the NLO
      result obtained with \textsc{Helac-Nlo} and the results produced
      by matching NLO predictions to \textsc{Deductor} for three
      different choices of the starting time.
      The scale choice is $\mu_F = \mu_R = m_t$. The
      lower panels display the relative  deviation from the fixed
      order result.}}
  \label{fig:Deltay_Njets}
\end{figure}

The $p_T$ spectra of $t\bar{t} j_1$ and $j_2$ coincide at NLO because
of  momentum conservation. After including shower effects,  they
deviate since the $t\bar{t}j_1$ system recoils against several
jets. In addition, the parton shower generates corrections to the
structure of the second jet.  We can see that for $t_0 = 0$ and
$\mu_T = 2$ the parton shower  overshoots the tail of the transverse
momentum spectra. The large higher  order corrections can be explained
by the exponentiation of non-singular  emissions, as already reported
in Ref.~\cite{Hoeche:2011fd} in the case of Higgs
production. Decreasing the value of $\mu_T$ helps to recover the NLO
predictions. We conclude that an appropriate choice of the central
value of  the starting time rescaling parameter is
\begin{equation}
\mu_{T0}=1\;.
\end{equation}
We expect that the choice of $\mu_T$ will affect all exclusive
distributions.  We present two examples in
Fig.~\ref{fig:Deltay_Njets}: the rapidity of the second jet, $y(j_2)$,
and the number of hard jets.
Finally, after fixing the central value of $\mu_T$, we choose to vary
it between $\mu_T = \mu_{T0}/2$ and  $\mu_T =
2\mu_{T0}$. This initial time variation is the  dimensionless
analogue of the resummation scale variation originally  introduced in
Refs.~\cite{Dasgupta:2001eq, Dasgupta:2002dc, Banfi:2004nk,
  Bozzi:2005wk,Bozzi:2010xn} and first used in the context of event
generators in Ref.~\cite{Hoche:2012wh}. 
%
%
%
%------------------------X------------------------X------------------------X---%
\subsection{Results}
%------------------------X------------------------X------------------------X---%

The fixed order cross section at NLO obtained from the \helacnlo{}
framework, including the theoretical error estimated via scale
variation, reads
\begin{equation}
  \sigma^{\nlo}_{pp \to t\bar{t}j+X} 
= 86.04^{~\,+5.10~(~+6\%)}_{-11.41~ (-13\%)} 
  \text{~pb}\;.
  \label{totxsecnlo}
\end{equation}
In order to obtain a reliable result including showering effects, we
study the  dependence of the cross section on the generation cut,
denoted by $p_T^{cut}$.  Table~\ref{tab:totalxsec} contains values
obtained at $\mu_F=\mu_R=m_t$ and  $\mu_T=1$. 
\begin{table}[t!]
\begin{center}
\begin{tabular}{ |c|c|c| }
  \hline \hline
&&\\
  $p_T^{cut}$ [GeV]  & $\sigma^\nlops_{pp \to t\bar{t}j+X}$ 
[pb] & $\epsilon$ [$\permil$]  \\
&&\\
  \hline\hline
  $5$   & $86.51 \pm 0.21 $ & $2.4$ \\
  $10$  & $86.26 \pm 0.17 $ & $2.0$ \\
  $15$  & $86.22 \pm 0.14 $ & $1.6$ \\
  $30$  & $86.11 \pm 0.13 $ & $1.5$ \\
  $40$  & $86.01 \pm 0.08 $ & $0.9$ \\
  $50$  & $84.58 \pm 0.07 $ & $0.8$ \\
  \hline \hline
\end{tabular}
  \caption{\textit{Total cross section for $pp \rightarrow t\bar{t} j + X$
  at  the LHC with $\sqrt{s}=8$~TeV, together with statistical and relative 
  errors, for different values of the generation cut. Results are produced 
   by matching \textsc{Helac-Nlo} predictions to \textsc{Deductor}. 
   The cross section is calculated for $\mu_F=\mu_R=m_t$ and $\mu_T=1$.}}
  \label{tab:totalxsec}
\end{center}
\end{table}
\begin{figure}[t!]
\begin{center}
  \includegraphics[width=0.49\textwidth]{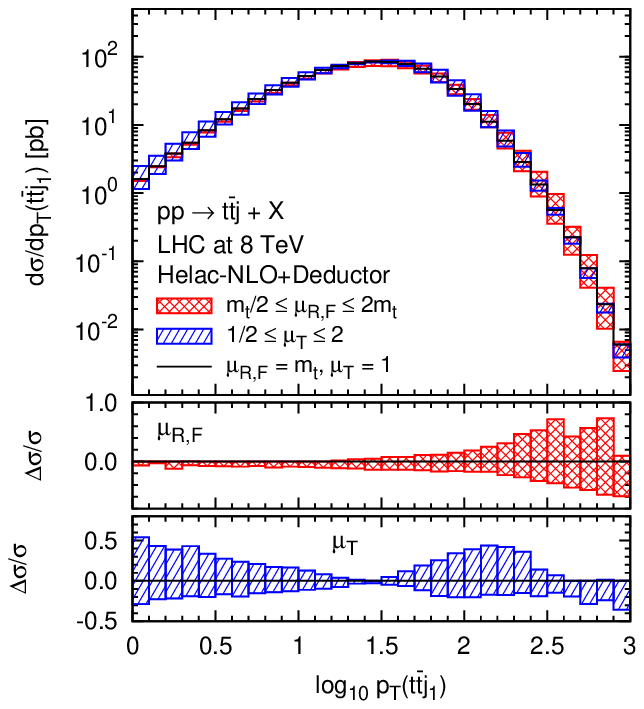}
  \includegraphics[width=0.49\textwidth]{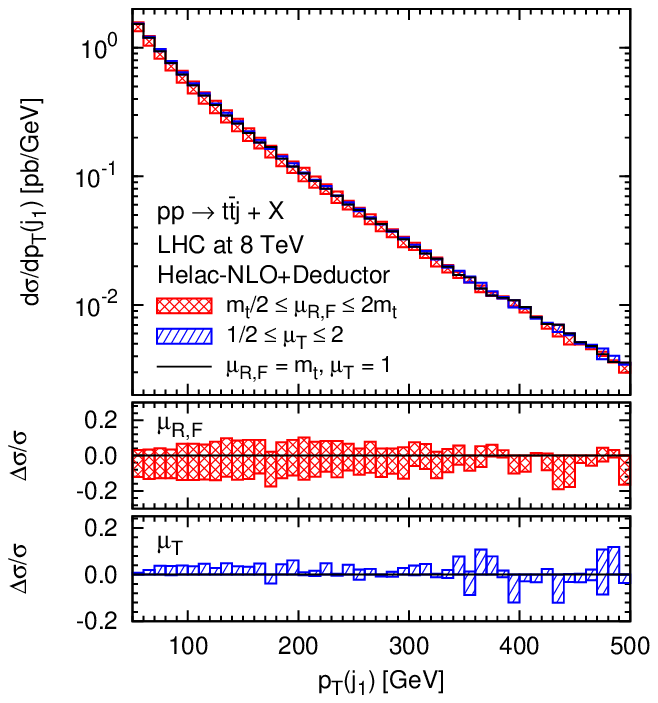}\\
\end{center}
  \caption{\textit{ Differential cross section distributions as a
      function of the transverse momentum of the $t\bar{t}j_1$  system
      (left panel) and of the  hardest jet (right panel) for $pp
      \to t\bar{t}j +X$ at the LHC with $\sqrt{s} = 8$ TeV. Results are
      produced by matching \textsc{Helac-Nlo} predictions to
      \textsc{Deductor}.  The uncertainty bands depict scale and
      initial shower time variation.  The lower panels display the
      corresponding relative deviation from the central value,
      separately for $\mu_{R,F}$ and $\mu_T$.}} 
  \label{fig:pT_scalevar}
\end{figure}
\begin{figure}[t!]
\begin{center}
  \includegraphics[width=0.49\textwidth]{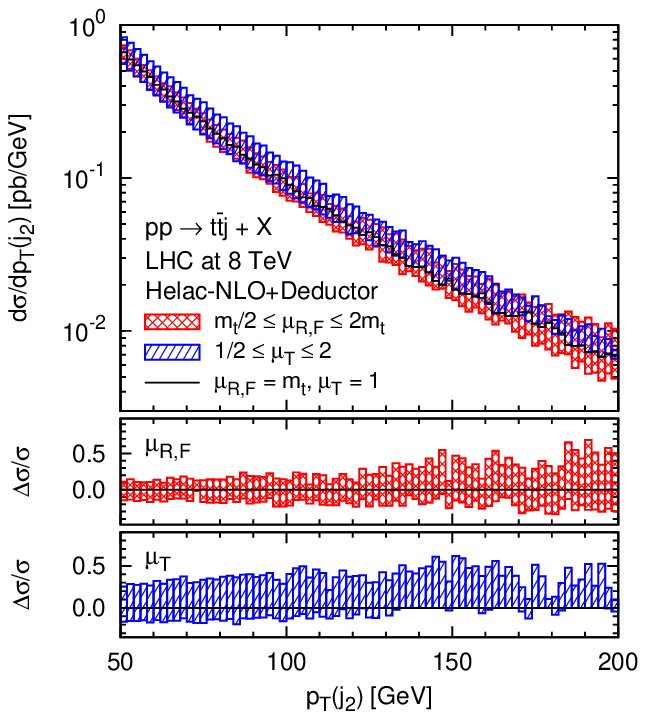}
  \includegraphics[width=0.49\textwidth]{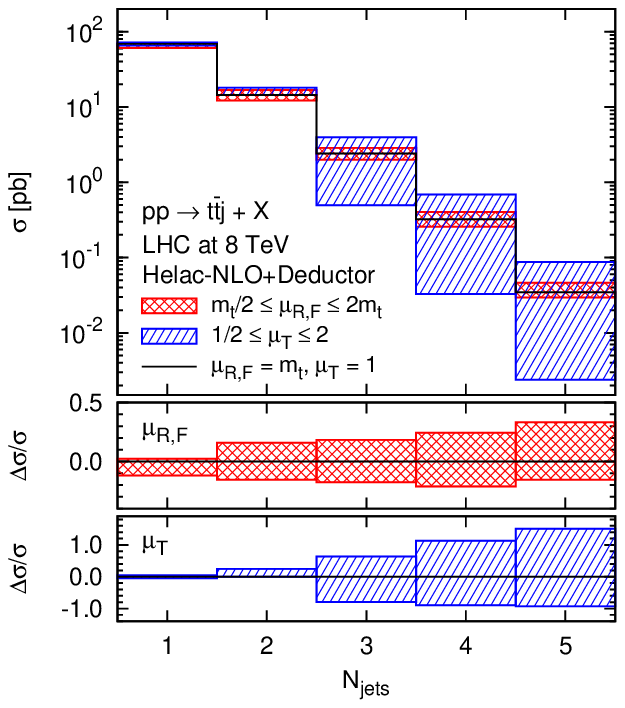}\\
\end{center}
  \caption{\textit{ Differential cross section distribution as a
      function of the transverse momentum of the  second jet
      (left panel) and inclusive jet cross sections (right panel)  for
      $pp \to t\bar{t}j+X$ at the LHC with $\sqrt{s} = 8$ TeV. Results are
      produced by matching \textsc{Helac-Nlo} predictions to
      \textsc{Deductor}.  The uncertainty bands depict scale and
      initial shower time variation.  The lower panels display the
      corresponding relative deviation from the central value,
      separately for $\mu_{R,F}$ and $\mu_T$.}}
  \label{fig:Njets_scalevar}
\end{figure}
We see that for all values of $p_T^{cut}$ but the last one, for which the 
generation cut is equal to the analysis cut, the total cross section is 
compatible with the fixed order prediction. 
For the following study, we choose a generation cut of $p_T(j_1)> 30$ GeV. 
Varying the renormalization/factorization scale between $m_t/2 \le
\mu_{R,F} \le 2m_t$ and the initial shower time rescaling parameter between 
$1/2 \le \mu_T \le 2$, the total cross section, together 
with its uncertainties, obtained after applying the parton shower is
\begin{equation}
  \sigma^{\nlops}_{pp \to t\bar{t}j+X}
  =86.11^{~\,+4.38~(~+5\%)}_{-10.88~ (-13\%)}
  \text{ [scales] } 
  ^{+0.80~(+1\%)}_{+2.17~(+3\%)}\text{ [PS time]~pb}\;,
  \label{totxsecnlops}
\end{equation}
where the upper, (lower) values are given for $\mu_{R,F}=m_t/2~(2m_t)$
and  $\mu_T = 1/2~(2)$.  The scale dependence of the
total cross section,  taken very conservatively as a maximum of the
upper and lower results, is $13\%$ or $9\%$ after symmetrization. The
dependence on the initial shower time is, by comparison,
negligible. The situation is quite different when it comes to
differential distributions as can be observed in
Figs.~\ref{fig:pT_scalevar}~and~\ref{fig:Njets_scalevar}, where  the
transverse momentum of the $t\bar{t}j_1$ system, the first and second
hard jets together with inclusive jet cross sections are given. The
variation bands for $\mu_{R,F}$ and $\mu_T$ in
Figs.~\ref{fig:pT_scalevar}~and~\ref{fig:Njets_scalevar} have been
obtained using the following sets of three parameter values: $\mu_{R,F}=\{
m_t/2, m_t, 2m_t \}$  and  $\mu_T = \{ 1/2,~1,~2 \,\}$,
respectively. The lower panels of
Figs.~\ref{fig:pT_scalevar}~and~\ref{fig:Njets_scalevar} display
corresponding  relative deviations from the central value, separately
for $\mu_{R,F}$ and $\mu_T$.

We start the discussion with the transverse momentum of the $p_T(t\bar{t}j_1)$
system, which is presented in  Fig.~\ref{fig:pT_scalevar} (left
panel).  At leading order, this observable  is zero due to momentum
conservation. When real emission contributions at the NLO level are
included, this observable diverges as the transverse momentum of the
entire system goes to zero. Therefore, it can only be reliably
described by the fixed order calculation in the high $p_T$
region. However, including the parton shower,  the low $p_T$ behavior
is altered strongly by the Sudakov form factor as can be seen in
Fig.~\ref{fig:pT_scalevar}. Indeed, for low values of the transverse momentum,
the distribution is generated mostly by the parton shower. The reason
is that the real radiation contribution, which is responsible for the divergent
behaviour at fixed order receives subtractions, which
match the singular behaviour for $p_T \to 0$. These subtractions belong
to the same bin as the real radiation events themselves, contrary to the fixed
order calculation, where they belong to the zero bin. Thus, the
subtracted real radiation sample has low weight contributions for low $p_T$.
In consequence, we observe a moderate dependence on $\mu_T$ in this
region, which reaches factors of 1.5 at the
lower end of the spectrum.  This dependence decreases down to just a
few percent around $30$ GeV,  whereas  for moderate values of
$p_T(t\bar{t}j_1)$  it is  at the level of $20\%-45\%$.  The presence
of a minimum dependence on $\mu_T$ is due to a crossing of the
distribution for $\mu_T = 1/2$, which dominates at low $p_T$
with the distribution for $\mu_T = 2$, which dominates at
high $p_T$. The reason for a larger cross section for low values of $\mu_T$ and
$p_T$ is that the parton shower generates low $p_T$ radiation only
barely shifting events with zero $p_T$ to non-zero values. On the
other hand, high $\mu_T$ results in radiation of high $p_T$ partons,
which shift events to high $p_T$ values. The situation
is reversed when it comes to the renormalization and factorization
scale dependence. Here, visible deviations from the central value
occur once the matrix element is present. They grow substantially  up
to almost $80\%$ at the end of the spectrum.  This can be explained 
by the fact that
the variation of $\mu_{R,F}$ is only implemented in the matrix
element, while the shower does not depend directly  on those scales
but rather on $t_0$. 

The $p_T(j_1)$ distribution of the hardest jet, which is given in
Fig.~\ref{fig:pT_scalevar} (right panel) shows a rather constant and
small dependence on both parameters, $\mu_{R/F}$ and $\mu_T$, as it
has NLO accuracy. The  $p_T(j_2)$ distribution of the second 
jet, which is presented in Fig.~\ref{fig:Njets_scalevar}  (left
panel), also shows a rather constant scale dependence with a somewhat
larger variation range, as it is only LO accurate.  
 
Finally, inclusive jet cross sections are shown in
Fig.~\ref{fig:Njets_scalevar}  (right panel). As expected, the  NLO
cross section with exactly one jet, which is given in the first bin,
is rather insensitive  to  $\mu_T$. Its theoretical error is at the
$12\%$ level.  The $\mu_T$ dependence is slightly larger in the second
bin, where the two jet cross section, correct only at the LO level,
is stored. Also here, the theoretical error increases up to $16\%$. 
Starting from the third bin, cross
sections are described via the  shower evolution alone, therefore,
fairly large variations can be noticed for both parameters, $\mu_T$
and $\mu_{R,F}$. For example, the scale dependence for the cross
section with five jets is found to be  around $35\%$. In addition,
cross sections for higher multiplicities are severely affected by the
cutoff on the parton shower emission phase space. This strong
dependence can be partially cured  when  matrix elements for higher
multiplicities are merged together and matched to  the shower evolution as
described e.g. in Refs.~\cite{Mangano:2006rw,Alwall:2007fs} for the LO
and in  Refs.~\cite{Hoeche:2014qda,Frederix:2012ps} for the NLO case.

%------------------------X------------------------X------------------------X---%
\subsection{Comparison with other Monte Carlo event generators}
%------------------------X------------------------X------------------------X---%

In this section we compare our implementation of the NLO matching  to
the Nagy-Soper parton shower from  \deductor{}, as implemented  in
\textsc{Helac-Dipoles}, with other matching procedures and shower
programs. To be more specific, we  use: 
\begin{enumerate}
\item the \powhegbox{}~\cite{Alioli:2010xd} implementation of $pp \to
  t\bar{t}j + X$~\cite{Alioli:2011as} in conjunction with the
  \textsc{Pythia 8.1}~\cite{Sjostrand:2007gs}  Monte Carlo program
  version 8.183, with  the transverse-momentum ordered shower (dubbed
  \powheg{}+\pythia{}),
\item the automatic solution of \amcnlo{}~\cite{Alwall:2014hca} including
  \textsc{Pythia 6.4}~\cite{Sjostrand:2006za}  (version 6.428) with
  the virtuality-ordered or  mass-ordered shower (dubbed
  \amcnlo{}+\pyQ{}), 
\item the automatic solution of \amcnlo{} together with  \textsc{Pythia
  8.1}, once more  with  the transverse-momentum ordered shower (dubbed
  \amcnlo{}+\pythia{}).  
\end{enumerate}
Once again, we do not include top quark decays, hadronization and
multiple interactions. Thus, we are only comparing the perturbative
evolution of different parton showers.  In each case the default setup
of the programs is used. 

The total cross sections together with their theoretical
errors are
\begin{align}
  &\sigma^{\nlops}_{pp \to t\bar{t}j+X} (\text{\amcnlopyQ}) 
= 84.85^{~\,+8.95~(+11\%)}_{
  -13.75~(-16\%)} \text{ [scales]~pb}\,, \\
  &\sigma^{\nlops}_{pp \to t\bar{t}j+X} (\text{\amcnlopy}) 
= 89.55^{~\,+8.44~(~+9\%)}_{
  -15.41~(-17\%)} \text{ [scales]~pb}\,, \\
  &\sigma^{\nlops}_{pp \to t\bar{t}j+X} (\text{\powhegpy}) 
= 89.12^{+26.22~(+29\%)}_{
  ~\,-8.96~(-10\%)} \text{ [scales]~pb}\,.
\end{align}
We observe that all three calculations, although based on different shower
ordering variables, give compatible results and agree within  $4\%$
with the \helacnlo{}+\deductor{} result from Eq.~\eqref{totxsecnlops}.
We note that,  in case of  \pythia{}, the central prediction for our
process is larger.  Moreover, the scale dependence after
symmetrization is below $13\%$ for all cases but the
\powheg{}+\pythia{}, where it is slightly larger i.e. of the order of
$20\%$.
\begin{figure}[t!]
\begin{center}
  \includegraphics[width=0.49\textwidth]{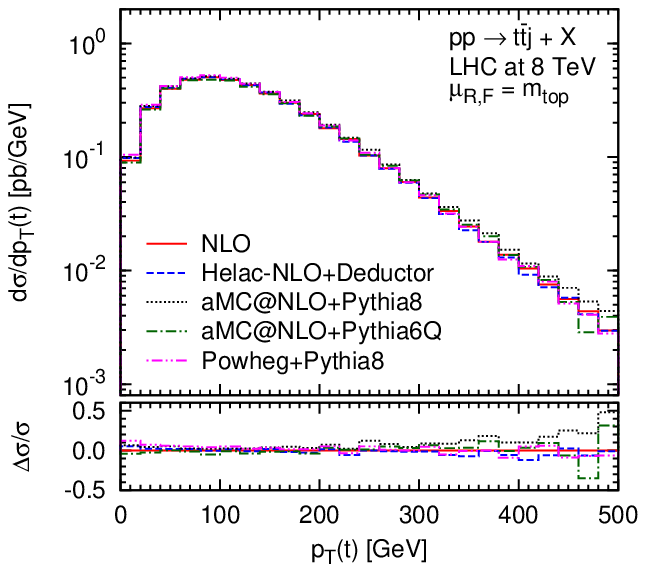} 
 \includegraphics[width=0.49\textwidth]{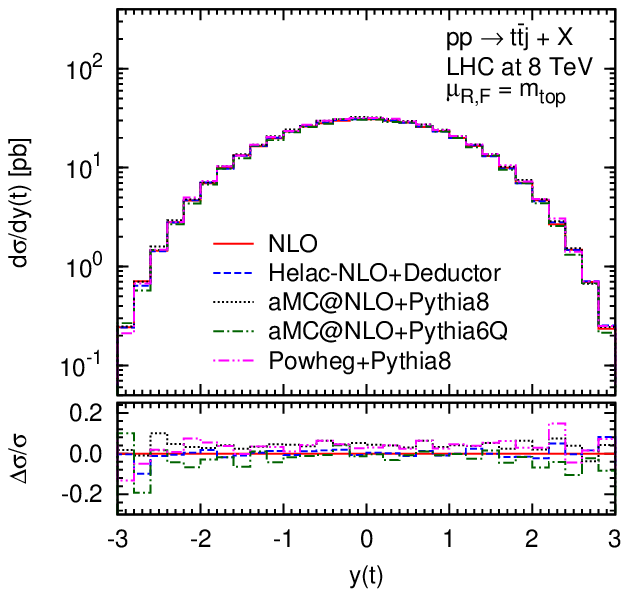}
  \includegraphics[width=0.49\textwidth]{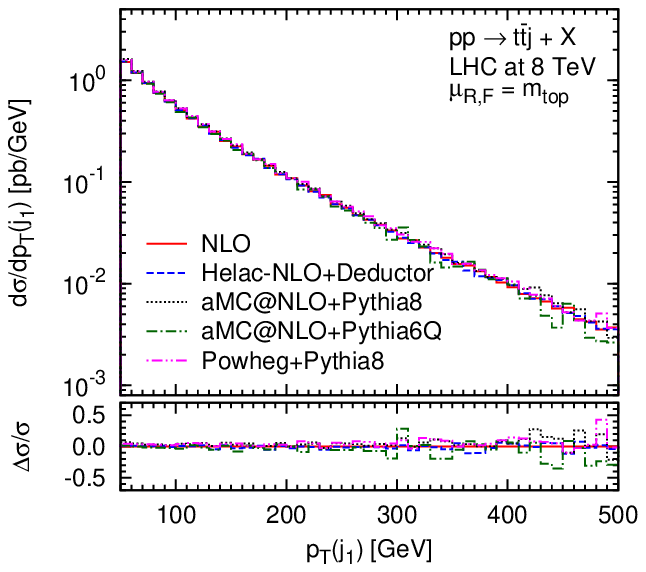} 
\includegraphics[width=0.49\textwidth]{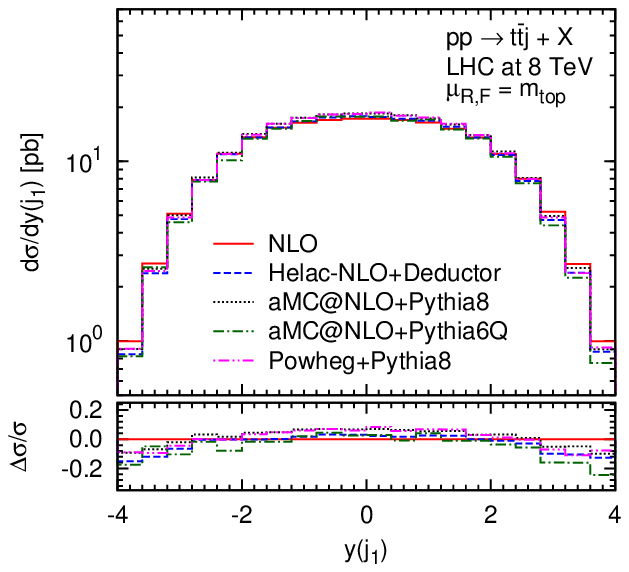}
\end{center}
  \caption{\textit{Differential cross section distributions as a
      function of the transverse momentum of the top quark  (left
      upper panel) and of the first jet (left lower  panel) as well as
      the rapidity of the top quark  (right upper panel) and of the
      first jet (right lower panel)   for $pp \to t\bar{t}j+X$ at the
      LHC with $\sqrt{s} = 8$ TeV.  Comparison between the NLO result
      obtained with \textsc{Helac-Nlo} and results produced by
      matching various NLO predictions to different parton showers.
      The scale choice is $\mu_F = \mu_R = m_t$.  The lower panels
      display the relative  deviation from the fixed order result.}}
  \label{fig:pt_y_com}
\end{figure}

In the next step, we extend our comparison  to   differential
distributions and start with observables that are rather  insensitive
to parton shower effects.  In Fig.~\ref{fig:pt_y_com}, the  transverse
momentum and rapidity distributions of the top quark and the first jet
are presented.  For each observable the NLO result obtained with
\textsc{Helac-Nlo}  is plotted together with  results produced by
matching various NLO predictions to different  parton showers. The
lower panels display the  relative deviation from the next-to-leading
order result.  All parton showers reproduce the corresponding NLO result
correctly, and slight deviations from the fixed order calculation can
only be seen  in the tails of the distributions due to smaller statistics.
\begin{figure}[t!]
\begin{center}
  \includegraphics[width=0.49\textwidth]{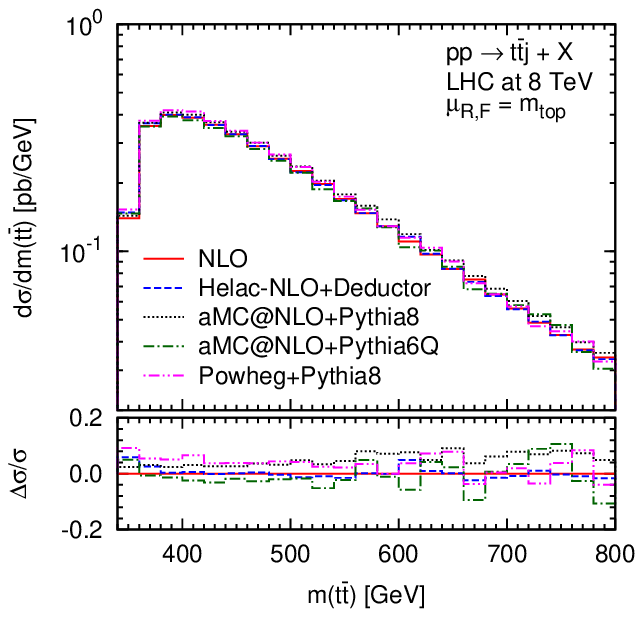}
  \includegraphics[width=0.49\textwidth]{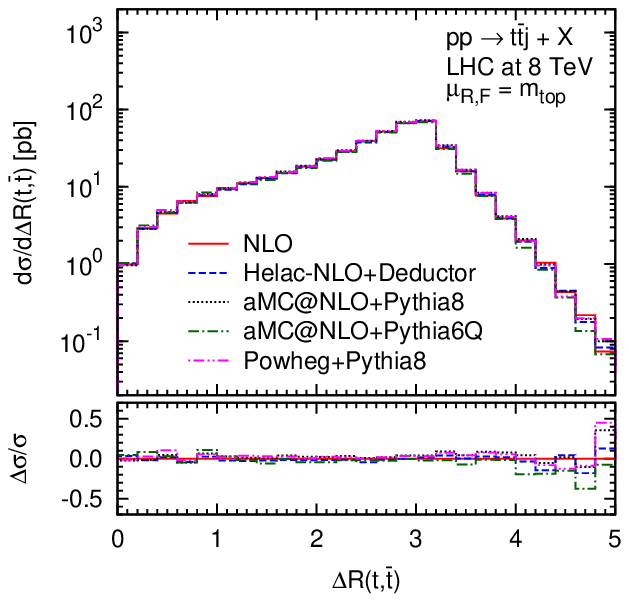}
\end{center}
  \caption{\textit{Differential cross section distributions as a
      function of the invariant mass  of the $t\bar{t}$ pair  (left
      panel) and of the $\Delta R_{t\bar{t}}$ (right panel)   for $pp
      \to t\bar{t}j+X$ at the LHC with $\sqrt{s} = 8$ TeV.  Comparison
      between the NLO result obtained with \textsc{Helac-Nlo} and
      results produced by matching various NLO predictions to
      different parton showers.  The scale choice is $\mu_F = \mu_R =
      m_t$.  The lower panels display the relative  deviation from the
      fixed order result.}}
  \label{fig:invariantmasses_com}
\end{figure}
\begin{figure}[t!]
\begin{center}
  \includegraphics[width=0.49\textwidth]{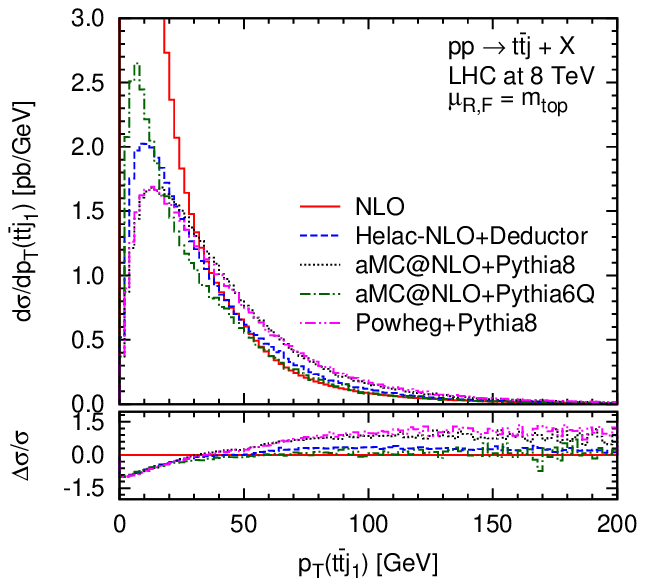}
  \includegraphics[width=0.49\textwidth]{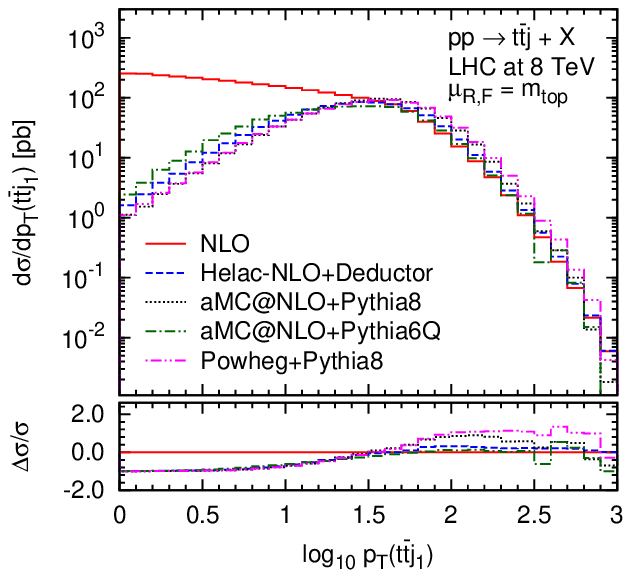}
\end{center}
  \caption{\textit{Differential cross section distributions as a
      function of the transverse momentum of the $t\bar{t}j_1$  system
      for $pp \to t\bar{t}j+X$ at the LHC with $\sqrt{s} = 8$ TeV.
      Comparison between the NLO result obtained with
      \textsc{Helac-Nlo} and results produced by matching various NLO
      predictions to different parton showers.  The scale choice is
      $\mu_F = \mu_R = m_t$.  The lower panels display the relative
      deviation from the fixed order result.}}
  \label{fig:pTspectrum_com}
\end{figure}
\begin{figure}[t!]
\begin{center}
  \includegraphics[width=0.49\textwidth]{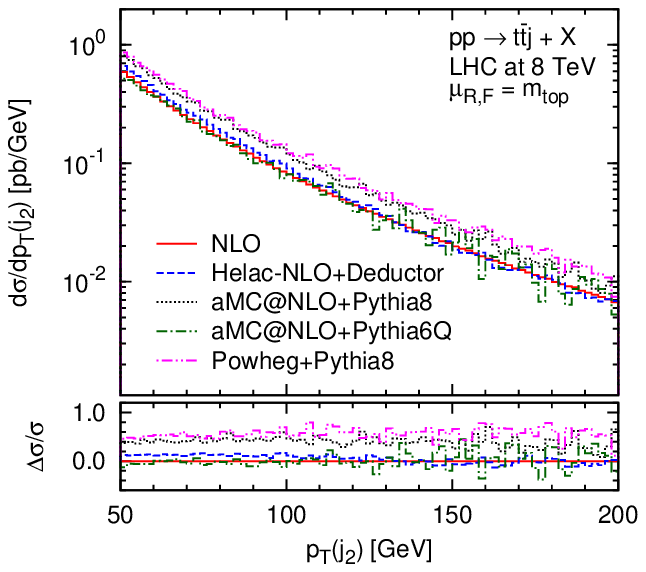}
  \includegraphics[width=0.49\textwidth]{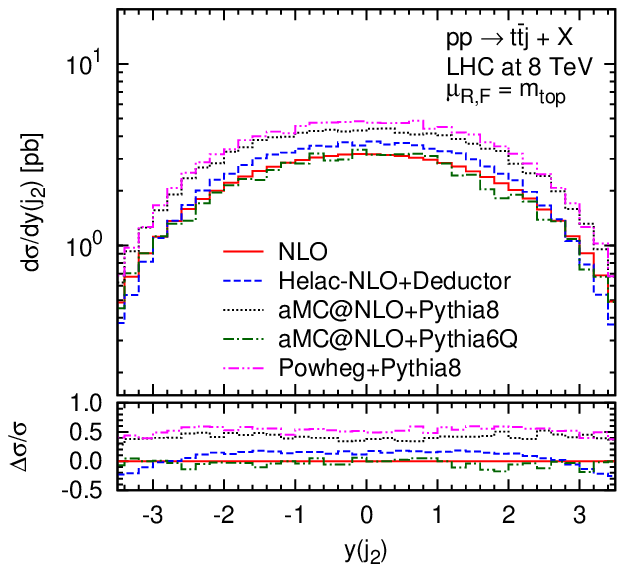}
\end{center}
  \caption{\textit{Differential cross section distributions as a
      function of the the transverse momentum (left panel)  and
      rapidity of  the second jet  (left panel) for $pp \to
      t\bar{t}j+X$ at the LHC with $\sqrt{s} = 8$ TeV.  Comparison
      between the NLO result obtained with \textsc{Helac-Nlo} and
      results produced by matching various NLO predictions to
      different parton showers.  The scale choice is $\mu_F = \mu_R =
      m_t$.  The lower panels display the relative  deviation from the
      fixed order result.}}
  \label{fig:Deltay_com}
\end{figure}
\begin{figure}[t!]
\begin{center}
  \includegraphics[width=0.49\textwidth]{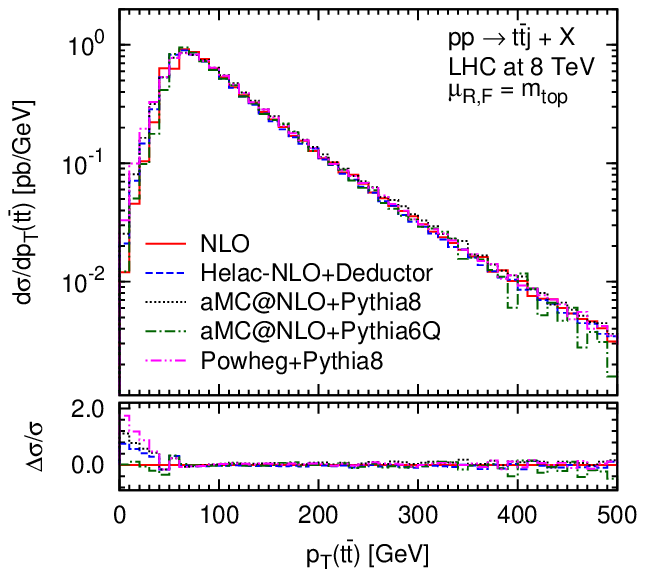} 
  \includegraphics[width=0.49\textwidth]{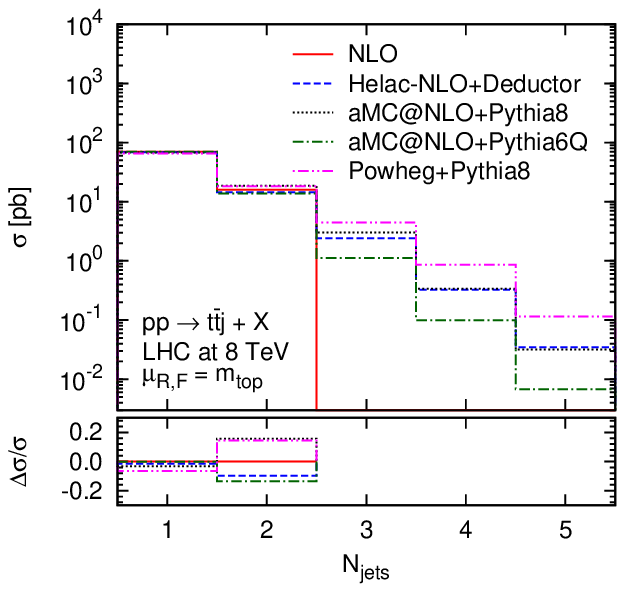}
\end{center}
  \caption{\textit{Differential cross section distribution as a
      function of the transverse momentum of the $t\bar{t}$ pair (left
      panel) and inclusive jet cross sections (right panel)   for $pp
      \to t\bar{t}j+X$ at the LHC with $\sqrt{s} = 8$ TeV.  Comparison
      between the NLO result obtained with \textsc{Helac-Nlo} and
      results produced by matching various NLO predictions to
      different parton showers.  The scale choice is $\mu_F = \mu_R =
      m_t$.  The lower panels display the relative deviation from the
      fixed order result.}}
  \label{fig:pt2_com}
\end{figure}

The same conclusions can be drawn from
Fig.~\ref{fig:invariantmasses_com}, where  we show the invariant mass
of the $t\bar{t}$ pair and the  angular separation in the
rapidity-azimuthal angle plane between the top and the anti-top quark  
\begin{equation}
\Delta R_{ij} = \sqrt{(y_i-y_j)^2+(\phi_i-\phi_j)^2} \,. 
\end{equation}
Overall, we observe that results obtained within the \heldeductor{}
framework are consistent with those of the other Monte-Carlo event
generators used in this study. Let us emphasize, that not only various
showers but also different matching schemes with distinct systematic
uncertainties are examined here. 

To assess potential differences among showers and  matching procedures
used in our analysis, we turn  to  observables, which are sensitive to
the initial  conditions of the parton evolution. As an example, we
present  in Fig.~\ref{fig:pTspectrum_com} the differential cross section
distribution as a function of the transverse momentum of the
$t\bar{t}j_1$ system. For clarity we give both a linear-scale (left
panel) and a log-scale (right panel) version of the plot.  We
observe that the matching with \pythia{} yields
similar results for both \mcnlo{} and \powheg{}, i.e. the hardest
spectra and a lack of agreement with
the fixed order  prediction in the region where hard, well separated
partons are produced. We have found discrepances even up to $120\%$.
This suggests, that for the process at hand, initial condition of the
transverse-momentum ordered  shower from \pythia{}  should  be
adjusted to restrict further the shower  phase space where emissions
become hard.  On the other hand,  \amcnlo{}+\pyQ{} nicely reproduces
the NLO differential cross section in the high $p_T$ region.   We
also note  that it produces a softer spectrum. As for the
\textsc{Helac-Nlo+Deductor} case, as  already explained, the parameter
$t_0$ was chosen to preserve the  NLO shape in the tail of the $p_T$
spectrum of the $t\bar{t}j_1$ system.

In Fig.~\ref{fig:Deltay_com},  the kinematics of the second jet is
analysed. To be more precise, distributions in the transverse momentum
and in rapidity are given. There is a good agreement between the
spectra of  \amcnlo{}+\pyQ{} and \helacnlo+\deductor{}, with a somewhat
narrower rapidity spectrum in the case of \helacnlo+\deductor{}.
For \amcnlo{}+\pythia{} and \powheg{}+\pythia{}, a difference in
normalization can be observed, but the differential $K$-factor remains flat.
This can be explained by  larger emission rates that increase the
amount of radiation.

In Fig.~\ref{fig:pt2_com}, the differential cross section as a function
of the transverse momentum of the $t\bar{t}$ pair  is presented.  We
observe a good agreement among all programs  for $p_T > 50$ GeV i.e.
above the analysis cut.  Below this value  the prediction is only
leading order accurate and thus strongly depends on the initial shower
conditions.  Also shown in Fig.~\ref{fig:pt2_com} are  inclusive jet
cross sections. In the first (second) bin the exclusive cross section
is  accurate at next-to-leading order (leading order).
Starting from the third bin on, cross sections are only described with
leading-logarithmic accuracy via the parton shower alone.  Results
for exclusive cross sections with $N_{\rm jets}=1$ and  
$N_{\rm jets}=2$ are  in agreement with our observations for other
observables.  The picture is vastly different for all showers,
however, when  $N_{\rm jets} \ge 3$. In this case, the  \powheg{}+\pythia{}
framework (\amcnlo{}+\pyQ{}) produces the highest (lowest) number of
hard jets. 
\begin{table}[t!]
\begin{center}
\begin{tabular}{ |c|c|c|c|c|c| }
  \hline \hline
&&&&&\\
  \textsc{Framework} 
& $\sigma^{[\ge 1 ~{\rm jet}]}$ 
& $\sigma^{[\ge 2 ~{\rm jets}]}$ 
& $\sigma^{[\ge 3 ~{\rm jets}]}$ 
& $\sigma^{[\ge 4 ~{\rm jets}]}$ 
& $\sigma^{[\ge 5 ~{\rm jets}]}$ \\
&&&&&\\
  \hline\hline
\helacnlo{}+\deductor{} & 86110 & 17204 & 2780 & 362 & 38 \\
\amcnlo{}+\pyQ{} & 84850 & 15030 & 1230  & 106 &  7 \\
\amcnlo{}+\pythia{} & 89556 & 21872 & 3377 &  372 & 32 \\
\powheg{}+\pythia{} & 89121 &  23744 & 5458 & 992 & 129  \\
  \hline \hline
\end{tabular}
  \caption{\textit{ Cross sections (in fb) for the inclusive jet rates
      at  the LHC with $\sqrt{s}=8$~TeV, according to the default
      settings of the various codes.}}
  \label{tab:inclusive_sigma}
\end{center}
\end{table}
\begin{table}[t!]
\begin{center}
\begin{tabular}{ |c|c|c|c|c| }
  \hline \hline
&&&&\\
  \textsc{Framework} & $\sigma^{[\ge 2]}/\sigma^{[\ge 1]}$ 
& $\sigma^{[\ge 3]}/\sigma^{[\ge 2]}$ 
& $\sigma^{[\ge 4]}/\sigma^{[\ge 3]}$ 
& $\sigma^{[\ge 5]}/\sigma^{[\ge 4]}$ 
\\
&&&&\\
  \hline\hline
\helacnlo{}+\deductor{} & 0.20 & 0.16 & 0.13 & 0.10  \\
\amcnlo{}+\pyQ{} & 0.18 & 0.08 & 0.09 & 0.07 \\
\amcnlo{}+\pythia{} & 0.24  & 0.15 & 0.11 & 0.09 \\
\powheg{}+\pythia{} & 0.27  & 0.23 & 0.18 & 0.13 \\
  \hline \hline
\end{tabular}
  \caption{\textit{ Cross section ratios for $(n + 1)/n$ inclusive jet rates 
      at  the LHC with $\sqrt{s}=8$~TeV, according to the default
      settings of the various codes.}}
  \label{tab:ratios}
\end{center}
\end{table}

Finally, cross sections (in fb) for inclusive
$\sigma(t\bar{t}+n  ~{\rm jets})$ rates are presented in
Table~\ref{tab:inclusive_sigma}. Table~\ref{tab:ratios} contains
cross section ratios, i.e.   $\sigma(t\bar{t}+n+1 ~{\rm
  jets})/\sigma(t\bar{t}+n ~{\rm jets})$. 

\section{Conclusions} 

In this publication, we have presented a next-to-leading order
matching scheme for the Nagy-Soper parton shower. We based our
construction on the original \mcnlo{} approach. Besides the general
formulation, we have performed real simulations for top-quark pair
production in association with a jet at the LHC, using an
implementation within the framework of the public codes \helacnlo{}
and \deductor{}.

Our general conclusion is that the combination
\helacnlo{}+\deductor{} is able to provide results for non-trivial
processes, which remain in reasonable agreement with other Monte Carlo
systems. Indeed, for observables, which are rather insensitive to showering
effects, the differential cross sections are in very good agreement
between different programs and the fixed order NLO
calculation. Inevitable differences for infrared sensitive
observables, on the other hand, seem to be justifiable in size in the
sense that all predictions have overlapping uncertainty bands. Of
course, further studies are needed here.

We note that our simulations are at the same level of logarithmic
precision as those of others. For now, we are only correct as far as
the leading behaviour is concerned. Future developments in \deductor{}
will allow us to include soft-gluon intereference effects yielding
next-to-leading logarithmic accuracy. This step still requires some
improvements of our implementation in \dipoles{}. In particular,
it will be necessary to transfer colour-configuration information
exactly, as opposed to the current leading-colour approximation. We
leave this to future work.

Finally, we expect that the Nagy-Soper parton shower will open new
opportunities for understanding parton shower systematics for
processes with non-trivial colour exchange. Here, the road is still
long, as we must remember that, ultimately, hadronization models must
be included. The latter, however, require tuning to the shower. There
are also interesting problems in merging different multiplicity
samples generated by our software.

\section*{Acknowledgements} 
We would like to thank M. Kr\"amer and Z. Nagy for discussions at
preliminary stages of this study.  The work of M. Czakon and
H. Hartanto was supported by the  German Research Foundation (DFG) via
the Sonderforschungsbereich/Transregio SFB/TR-9 ``Computational
Particle Physics''. M. Worek and  M. Kraus acknowledge support by the
DFG under Grant No.  WO 1900/1-1 (``Signals and Backgrounds Beyond
Leading Order.  Phenomenological studies for the LHC '').  In
addition, this research was supported in part by the Research Funding
Program ARISTEIA, HOCTools (co-financed by the European Union
(European Social Fund ESF) and Greek national funds through the
Operational Program "Education and Lifelong Learning" of the National
Strategic Reference Framework (NSRF)).

\appendix
%------------------------X------------------------X------------------------X---%
\section{Simplified matching scheme}
\label{App:NLOapprox}
%------------------------X------------------------X------------------------X---%
In this Appendix, we discuss the simplified matching prescription that
we have implemented starting from spin and colour averaged
amplitudes. We also prove that it yields NLO accuracy at leading colour, as
long as the observable is not sensitive to spin correlations.
The simplifications match the current \deductor{}
functionality and relate to a spin and colour averaged quantum density
\begin{equation}
  |\rho) = \sum_m \frac{1}{m!}\int [d\{p,f\}_m]|\{p,f\}_m)(\{p,f\}_m|\rho)
  \;,
\end{equation}
where
\begin{equation}
  (\{p,f\}_m|\rho) = \sum_{\{s\}_m,\{s^\prime\}_m} \sum_{\{c\}_m,
  \{c^\prime\}_m} \braket{\{s,c\}_m|\{s^\prime,c^\prime\}_m}
  \rho(\{p,f,s^\prime,c^\prime,s,c\}_m)\;.
\end{equation}
Averaging over spin and colour reduces tremendously the complexity of
the calculation. However, the parton shower requires a colour
configuration for each phase space point $\{p,f\}_m$ in order to
perform the evolution.  Starting from colour averaged weights
$(\{p,f\}_m|\rho)$ one can recover leading colour correlations,
$(\{p,f,c\}_m|\rho)$, by including a colour weight $C_w(\{p,f,c\}_m)$,
defined by
\begin{equation}
  C_w(\{p,f,c\}_m) = \frac{(\{p,f,c\}_m|\rho^{(0)}_m)}
  {\sum_{\{\hat{c}\}_m}(\{p,f,\hat{c}\}_m|\rho^{(0)}_m)}\;.
\end{equation}
The simplified matching prescription accounting for generation cuts is then
\begin{equation}
\begin{split}
  \sigma[F] &= \int\frac{[d\{p,f,c\}_m]}{m!}  (F|U(t_F,t_0)|\{p,f,c\}_m) 
  C_w(\{p,f,c\}_m)(\{p,f\}_m|\Omega|\rho^{(0)}_m) F_I(\{p,f\}_m) \\
  & + \int \frac{[d\{p,f,c\}_{m+1}]}{(m+1)!} (F|U(t_F,t_0)|\{p,f,c\}_{m+1})
  C_w(\{p,f,c\}_{m+1})\\
  &\quad\times(\{p,f\}_{m+1}|R) F_I(\{p,f\}_{m+1}) \\[0.2cm]
  &\equiv \sigma_m[F] + \sigma_{m+1}[F]\;,
\end{split}
  \label{app:NLOmatching}
\end{equation}
where the action of the operator $\Omega$ is given by
\begin{equation}
  (\{p,f\}_m|\Omega|\rho^{(0)}_m) = \omega(\{p,f\}_m)(\{p,f\}_m|
  \rho^{(0)}_m) \; ,
\end{equation}
with 
\begin{equation}
  \omega(\{p,f\}_m) = 1 + 
  \frac{(\{p,f\}_m|\rho^{(1)}_m)}{(\{p,f\}_m|\rho^{(0)}_m)} +
  \frac{(\{p,f\}_m|\mathbf{I}(t_0)+ \mathbf{K}(t_0)+ \mathbf{P}|
  \rho^{(0)}_m)}{(\{p,f\}_m|\rho^{(0)}_m)}\;.
  \label{app:NLOweight}
\end{equation}
The density matrix for the real subtracted cross section,
$(\{p,f\}_{m+1}|R)$, is defined as
\begin{equation}
\begin{split}
  (\{p,f\}_{m+1}|R) &\equiv (\{p,f\}_{m+1}|\rho^{(0)}_{m+1}) - 
  \int_{t_0}^\infty d\tau~\sum_{\{s,s^\prime,c,c^\prime\}}
  \braket{\{s^\prime,c^\prime\}_{m+1}|\{s,c\}_{m+1}} \\ 
  &\times\int \frac{[d\{\hat{p},\hat{f},\hat{s}^\prime, \hat{c}^\prime,
  \hat{s},\hat{c}\}_m]}{m!} (\{\hat{p},\hat{f},\hat{s}^\prime, 
  \hat{c}^\prime, \hat{s},\hat{c}\}_m|\rho^{(0)}_m) 
  \;F_I(\{\hat{p},\hat{f}\}_m) \\[0.2cm]
  &\times  (\{p,f,s^\prime,c^\prime,s,c\}_{m+1}|\mathcal{H}_I
  (\tau)|\{\hat{p},\hat{f},\hat{s}^\prime, \hat{c}^\prime,
  \hat{s},\hat{c}\}_m)\;,
\end{split}
  \label{app:NLOreal}
\end{equation}
where we specified explicitly the full quantum correlations in the subtraction
terms required for the removal of kinematic singularities.
$\mathcal{H}_I(t)$ is the Nagy-Soper real splitting
operator without any approximations. The inclusive jet
functions $F_I(\{p,f\}_m)$ define the generation cuts. 

Let us now study the accuracy of Eq.~\eqref{app:NLOmatching} after
expansion to next-to-leading order. The parton shower evolution
operator, currently available in \deductor{}, reads
\begin{equation}
  U(t_F,t_0) = N^{LC+}(t_F,t_0) + \int_{t_0}^{t_F}d\tau~U(t_F,\tau) 
  \overline{\mathcal{H}}_I^{LC+}(\tau)N^{LC+}(\tau,t_0)\;,
  \label{lcplusevo}
\end{equation}
where $\overline{\mathcal{H}}_I^{LC+}(\tau)$ denotes the spin averaged 
splitting operator in the LC+ approximation and $N^{LC+}(t_F,t_0)$ the 
corresponding Sudakov form factor, as discussed in 
Section~\ref{subsec:colourevolution}. Inserting Eq.~\eqref{lcplusevo}
in $\sigma_m[F]$ and expanding the evolution yields
\begin{equation}
\begin{split}
  \sigma_m[F] &= \frac{1}{m!}\int [d\{p,f,c\}_m] (F|\{p,f,c\}_m) 
  C_w(\{p,f,c\}_m)(\{p,f\}_m|\Omega|\rho^{(0)}_m)F_I(\{p,f\}_m) \\
  &+\frac{1}{m!}\int [d\{p,f,c\}_m] 
  \int_{t_0}^{t_F}d\tau~(F|\overline{\mathcal{H}}_I^{LC+}(\tau)-
  \mathcal{V}^{LC+}(\tau)|\{p,f,c\}_m) \\[0.2cm]
  &\times C_w(\{p,f,c\}_m)(\{p,f\}_m|\rho^{(0)}_m)F_I(\{p,f\}_m)
  +\mathcal{O}(\alpha_s^2)\;.
\end{split}
  \label{app:sigMstep1}
\end{equation}
Using the approximations
\begin{equation}
 \mathcal{V}(\tau) = \mathcal{V}^{LC+}(\tau) + \mathcal{O}(1/N_c^2) \; ,
\end{equation}
and 
\begin{equation}
  C_w(\{p,f,c\}_m)(\{p,f\}_m|\rho^{(0,1)}_m) = (\{p,f,c\}_m|\rho^{(0,1)}_m)
  +\mathcal{O}(1/N_c^2) \; ,
  \label{app:colourapprox}
\end{equation}
Eq.~\eqref{app:sigMstep1} reduces to
\begin{equation}
\begin{split}
  \sigma_m[F] &= \frac{1}{m!}\int [d\{p,f,c\}_m] (F|\{p,f,c\}_m) 
  (\{p,f,c\}_m|\left[|\rho^{(0)}_m)+|\rho^{(1)}_m) 
  +\mathbf{P}|\rho^{(0)}_m)\right] F_I(\{p,f\}_m)\\
  &+\frac{1}{m!}\int [d\{p,f,c\}_m] 
  \int_{t_0}^{\infty}d\tau~(F|\overline{\mathcal{H}}_I^{LC+}(\tau)|
  \{p,f,c\}_m)(\{p,f,c\}_m|\rho^{(0)}_m)F_I(\{p,f\}_m) \\
  &+\mathcal{O}(\alpha_s^2, 1/N_c^2, \Delta\sigma)\;.
\end{split}
  \label{app:sigMstep2}
\end{equation}
Here, $\Delta\sigma$ represents the error resulting from the 
limit $t_F \to\infty$.

Let us now turn to the $\sigma_{m+1}[F]$. Expanding the shower evolution
and making use of Eq.~\eqref{app:colourapprox} to recover the leading colour
correlated real matrix element, we find 
\begin{equation}
\begin{split}
  \sigma_{m+1}[F] &= \int\frac{[d\{p,f,c\}_{m+1}]}{(m+1)!} (F|\{p,f,c\}
  _{m+1}) F_I(\{p,f\}_{m+1})\Biggl[ (\{p,f,c\}_{m+1}|\rho^{(0)}_{m+1}) \\
  &-\int_{t_0}^\infty d\tau \sum_{\{\bar{s},\bar{s}^\prime,\bar{c}^\prime, 
  \bar{c}\}_{m+1}} \braket{\{\bar{s}^\prime, \bar{c}^\prime\}_{m+1}|\{
  \bar{s}, \bar{c}\}_{m+1}} C_w(\{p,f,c\}_{m+1}) \\
  &\times\int \frac{[d\{\hat{p},\hat{f},\hat{s}^\prime, \hat{c}^\prime,
  \hat{s},\hat{c}\}_m]}{m!} 
  (\{\hat{p},\hat{f},\hat{s}^\prime,\hat{c}^\prime, \hat{s},\hat{c}\}_m
  |\rho^{(0)}_m) \;F_I(\{\hat{p},\hat{f}\}_m)\\
  &\times (\{p,f,\bar{s}^\prime,\bar{c}^\prime,\bar{s},\bar{c}\}_{m+1}|
  \mathcal{H}_I(\tau)|\{\hat{p},\hat{f},\hat{s}^\prime, \hat{c}^\prime,
  \hat{s},\hat{c}\}_m) \Biggl]+ \mathcal{O}(\alpha_s^2,1/N_c^2) \\
  &\equiv \sigma^R_{m+1}[F] -\sigma^S_{m+1}[F]
  + \mathcal{O}(\alpha_s^2,1/N_c^2)\;.
\end{split}
  \label{app:sigMPstep1}
\end{equation}
In the following we will focus on the subtraction terms $\sigma^S_{m+1}[F]$.
The summation over $\{c\}_{m+1}$ can be eliminated by
\begin{equation}
\begin{split}
  \int [d\{p,f,c\}_{m+1}]&C_w(\{p,f,c\}_{m+1})(F|\{p,f,c\}_{m+1}) \\
  &=\int [d\{p,f\}_{m+1}](F|\{p,f\}_{m+1})+\mathcal{O}(1/N_c^2)\;.
\end{split}
\end{equation}
By a further approximation to the colour correlator
\begin{equation}
\begin{split}
  \braket{\{\bar{c}^\prime\}_{m+1}|\{\bar{c}\}_{m+1}} &= 
  \braket{\{\bar{c}\}_{m+1}|\{\bar{c}\}_{m+1}}
  \delta\left(\{\bar{c}^\prime\}_{m+1};\{\bar{c}\}_{m+1}\right) + 
  \mathcal{O}(1/N_c^2) 
  \;,
\end{split}
\end{equation}
we recover leading colour correlations
\begin{equation}
\begin{split}
  \int [d\{p,f\}_{m+1}]\sum_{ \{\bar{c}\}_{m+1}} (F|\{p,f\}_{m+1})  
  &\braket{\{\bar{c}\}_{m+1}|\{\bar{c}\}_{m+1}} \\ &= \int [d\{p,f, 
  \bar{c}\}_{m+1}](F|\{p,f,\bar{c}\}_{m+1})\;.
\end{split}
\end{equation}
In the following we rename $\bar{c} \to c$ and drop the second colour index
in $(\{p,f,\bar{s}^\prime,c, \bar{s},c\}_{m+1}|$. In addition, we use
\begin{equation}
  \mathcal{H}_I(\tau) =\mathcal{H}_I^{LC+}(\tau) +\mathcal{O}(1/N_c^2)
  \; ,
\end{equation}
to obtain the following form for $\sigma_{m+1}^S[F]$
\begin{equation}
\begin{split}
  \sigma_{m+1}^S[F] &= \frac{1}{(m+1)!m!}\int [d\{p,f,c\}_{m+1}] 
  [d\{\hat{p},\hat{f},\hat{s}^\prime, \hat{c}^\prime,\hat{s},\hat{c}\}_m]
  F_I(\{\hat{p},\hat{f}\}_m) F_I(\{p,f\}_{m+1})\\
  &\times(F|\{p,f,c\}_{m+1}) 
  \int_{t_0}^\infty d\tau \sum_{\{\bar{s},\bar{s}^\prime\}_{m+1}} 
  \braket{\{\bar{s}^\prime\}_{m+1}|\{\bar{s}\}_{m+1}} 
  (\{\hat{p},\hat{f},\hat{s}^\prime,\hat{c}^\prime, \hat{s},\hat{c}\}_m
  |\rho^{(0)}_m) \\
  &\times (\{p,f,\bar{s}^\prime,\bar{s},c\}_{m+1}|
  \mathcal{H}^{LC+}_I(\tau)|\{\hat{p},\hat{f},\hat{s}^\prime, 
  \hat{c}^\prime,\hat{s},\hat{c}\}_m) + \mathcal{O}(\alpha_s^2,1/N_c^2)
  \;.
\end{split}
  \label{app:sigMPstep3}
\end{equation}
It is now necessary to remove spin correlations. This can only be
achieved, if $(F|\{p,f,c\}_{m+1})$ and $F_I(\{p,f\}_{m+1})$ are
sufficiently inclusive to allow an azimuthal average. A typical case
would be an observable with NLO accuracy, where the azimuthal average
corresponds to the complete phase space integral over unresolved
jets. On the example of top-quark pair production with at least one
jet, this would be any observable insensitive to additional jets.
Once an azimuthal average has been performed, spin correlations vanish
as demonstrated in Ref.~\cite{Nagy:2007ty} (Chapter 12)
\begin{equation}
\begin{split}
  \int \frac{d\phi}{2\pi}&\sum_{\{\bar{s},\bar{s}^\prime\}_{m+1}} 
  \braket{\{\bar{s}^\prime\}_{m+1}|\{\bar{s}\}_{m+1}}(\{p,f,\bar{s}^\prime, 
  \bar{s},c\}_{m+1}|\mathcal{H}^{LC+}_I(\tau)|\{ \hat{p}, \hat{f}, 
  \hat{s}^\prime, \hat{c}^\prime, \hat{s}, \hat{c}\}_m)\\
  &= \braket{ \{\hat{s}^\prime\}_m|\{\hat{s}\}_m} 
  (\{p,f,c\}_{m+1}|\overline{\mathcal{H}}^{LC+}_I(\tau)|\{ \hat{p},\hat{f}, 
  \hat{c}^\prime, \hat{c}\}_m)\;,
\end{split}
  \label{app:sigMPspinav}
\end{equation}
which can be further transformed with
\begin{equation}
\begin{split}
  (\{p,f,c\}_{m+1}|\overline{\mathcal{H}}^{LC+}_I(\tau)|\{ \hat{p},\hat{f}, 
  \hat{c}^\prime, \hat{c}\}_m) &= 
  (\{p,f,c\}_{m+1}|\overline{\mathcal{H}}^{LC+}_I(\tau)|\{ \hat{p},\hat{f}, 
  \hat{c}\}_m) \\
  &\times \delta\left( \{ \hat{c}^\prime \}_m ; \{\hat{c} \}_m\right)+ 
  \mathcal{O}(1/N_c^2)\;.
\end{split}
  \label{app:sigMPH}
\end{equation}
Substituting Eq.~\eqref{app:sigMPspinav} and Eq.~\eqref{app:sigMPH} into Eq. 
\eqref{app:sigMPstep3} we obtain
\begin{equation}
\begin{split}
  \sigma_{m+1}^S[F] &= \frac{1}{(m+1)!m!} \int [d\{p,f, c\}_{m+1}] [d\{ 
  \hat{p}, \hat{f}, \hat{c}\}_m](F|\{p,f,c\}_{m+1}) \\
  & \times \int_{t_0}^\infty d\tau 
  (\{p,f,c\}_{m+1}|\overline{\mathcal{H}}^{LC+}_I(\tau)|\{ \hat{p},\hat{f}, 
  \hat{c}\}_m) F_I(\{\hat{p},\hat{f}\}_m)F_I(\{p,f\}_{m+1}) \\
  &\times  \sum_{\{\hat{s},\hat{s}^\prime\}_{m}} \braket{ 
  \{\hat{s}^\prime\}_m|\{\hat{s}\}_m}(\{ \hat{p}, \hat{f}, \hat{s}^\prime,
  \hat{s}, \hat{c}\}_m|\rho^{(0)}_m) + \mathcal{O}(\alpha_s^2, 1/N_c^2)\;.
\end{split}
  \label{app:sigMPstep4}
\end{equation}
The last line of Eq.~\eqref{app:sigMPstep4} corresponds to the leading
colour, spin averaged quantum density matrix
\begin{equation}
  \sum_{\{\hat{s},\hat{s}^\prime\}_{m}} \braket{ \{\hat{s}^\prime\}_m|
  \{\hat{s}\}_m}(\{\hat{p}, \hat{f}, \hat{s}^\prime, \hat{s}, \hat{c}\}_m
  |\rho^{(0)}_m)  = (\{ \hat{p}, \hat{f}, \hat{c}\}_m|\rho^{(0)}_m)\;.
\end{equation}
With these approximations the final expression for $\sigma^S_{m+1}[F]$ reads
\begin{equation}
\begin{split}
  \sigma^S_{m+1}[F]= & ~\frac{1}{(m+1)!m!}\int[d\{\hat{p},\hat{f},\hat{c}\}_{m+1}]
  [d\{p,f,c\}_m] 
  \int_{t_0}^{t_F}d\tau~(F|\{\hat{p},\hat{f},\hat{c}\}_{m+1}) \\
  &\times F_I(\{\hat{p},\hat{f}\}_{m+1})(\{\hat{p},\hat{f},\hat{c}\}_{m+1}|
  \overline{\mathcal{H}}_I^{LC+}(\tau)|\{p,f,c\}_m) \\
  &\times(\{p,f,c\}_m|\rho^{(0)}_m) F_I(\{p,f\}_m) +\mathcal{O}(\alpha_s^2,
  1/N_c^2, \Delta\sigma)\;.
\end{split}
  \label{app:sigMPstep5}
\end{equation}
Combining $\sigma_m[F]$ from Eq.~\eqref{app:sigMstep2} 
with $\sigma^R_{m+1}[F]$ and
$\sigma^S_{m+1}[F]$ from Eqs.~\eqref{app:sigMPstep1} and \eqref{app:sigMPstep5}
yields
\begin{equation}
\begin{split}
  \sigma[F] = & ~\sigma_m[F] + \sigma^R_{m+1}[F] -\sigma^S_{m+1}[F]\\
  = & ~ \sigma^{LC-NLO}[F] \\
  + & ~\frac{1}{(m+1)!m!}\int[d\{\hat{p},\hat{f},\hat{c}\}_{m+1}][d\{p,f,c\}_m] 
  \int_{t_0}^{t_F}d\tau~(F|\{\hat{p},\hat{f},\hat{c}\}_{m+1}) \\
  &\times (\{\hat{p},\hat{f},\hat{c}\}_{m+1}|\overline{\mathcal{H}}_I^{LC+}
  (\tau)|\{p,f,c\}_m)(\{p,f,c\}_m|\rho^{(0)}_m)  \\
  &\times F_I(\{p,f\}_m)\left[1-F_I(\{\hat{p},\hat{f}\}_{m+1}) \right]
  +\mathcal{O}(\alpha_s^2,1/N_c^2, \Delta\sigma)\;.
\end{split}
\end{equation}
The last term vanishes for suitable generation cuts as specified in
Section~\ref{subsec:jetxsec}.  In conclusion, we have shown that the
simplified matching approach yields a prescription
that is accurate at NLO to the level of leading colour.

\providecommand{\href}[2]{#2}\begingroup\raggedright\endgroup

\end{document}